\documentclass[11pt]{article}
\usepackage[]{amsmath,amssymb}
\usepackage{graphics,epsfig}
\usepackage{epic}
\usepackage{color}
\usepackage{lscape} 

\def\text#1{\mbox{\rm #1\ }}

\def\ie{{\rm i.e.,\/}\ }
\def\etc{{\rm etc.\/}\ }

\def \otimesdot {\stackrel{\cdot}{\otimes}}
\newcommand{\ud}[1]{\underline{#1}}

\newcommand{\ZZ}{\mathbb{Z}}

\newcommand{\CC}{\mathbb{C}}

\begin{document}

\title{Torus structure on graphs and twisted partition functions for
minimal and
affine models \vspace{0.8cm}}
\author{R. Coquereaux${}^{1}$ \thanks{%
~Email: Robert.Coquereaux@cpt.univ-mrs.fr}$\;$, M. Huerta${}^{1,}$ ${}^2$
\thanks{%
~Email: huertam@ictp.trieste.it} \\
\\
${}^1$ \textit{Centre de Physique Th\'eorique - CNRS} \\
\textit{Campus de Luminy - Case 907} \\
\textit{F-13288 Marseille - France} \\
\\
${}^2$ \textit{The Abdus Salam International Centre for Theoretical Physics}
\\
\textit{I-34100, Trieste, Italy}}
\date{}

\thispagestyle{empty}
\begin{titlepage}
\maketitle

\abstract{Using the Ocneanu quantum geometry of $ADE$ diagrams (and of other
diagrams belonging to higher Coxeter -- Dynkin systems), we discuss
the classification of twisted partition functions for affine and
minimal models in conformal field theory and study several examples
associated with the WZW, Virasoro and ${\cal W}_{3}$ cases.}

\vspace{3.cm}

\vspace{1.0cm}



\vspace{2. cm}

\noindent hep-th/0301215 \\
\noindent CPT-2003/P.4253 \\

\vspace*{0.3 cm}

\end{titlepage}



\section{Introduction}



\subsection{Purpose and structure of this article}

One of the purposes of our article is to present a discussion and a
classification of twisted partition functions for conformal field
theories associated with minimal models and affine models of type ADE,
as well as some of their generalizations associated with diagrams
belonging to higher Coxeter -- Dynkin systems.  The whole discussion
is based on the quantum geometry of these diagrams.  Since the graphs
themselves provide the necessary combinatorial data, we shall avoid as
much as possible to make any explicit use of the theory of affine Lie
algebras (or of their finite dimensional
counterparts).  Actually, we shall not use much information coming
from Conformal Field Theory, so that our presentation should be
understood by readers with  different backgrounds.

Many mathematical tools used in the study of the quantum geometry of graphs
were introduced by A. Ocneanu (in the context of operator algebras) and
later ``explained'' or adapted in various contexts (for instance CFT but not
only) by several authors;  this information is scattered in publications
of very different nature. Our presentation starts from very elementary
concepts and shows how one can calculate many (quantum) geometrical
quantities of interest by using rather straightforward algorithms.
 From the data encoded in $ADE$ diagram or their generalizations, we
remind the reader how the corresponding quantum geometry is related to
the (twisted or not) partition functions in affine models.  We then
move to minimal models in particular the unitary ones, discuss the
relation with graphs and give various examples (Ising, Potts and the
exceptional $E_{6}$ - ${\ A_{10}}$ model).  We also consider
twisted ${\cal W}_{3}$ minimal models.

Our discussion of twisted partition functions for minimal models can
be summarized as follows : to a pair $(G^{(1)},G^{(2)})$ of $ADE$ Dynkin
diagrams one can associate six types of sesquilinear forms on the
space of Virasoro characters.  These forms can be interpreted, in
terms of minimal models, as partition functions in boundary conformal
field theory with defects.  This classification rests on the
possibility of introducing several ``torus structures'' for the two
diagrams $G^{(1)}$ and $G^{(2)}$.  Torus structures are parametrized
by elements of a particular base in the Ocneanu algebra of quantum
symmetries; a torus structure may have a single twist, two twists, or
no twist at all.  The interpretation of what we call torus structures
in terms of defects (or twists) in a conformal field theory with
boundary was proposed by V.B. Petkova and J.-B. Zuber
\cite{PetkovaZuber:Oc}.  An application of these ideas to the
discussion of the different types of partition functions for minimal
models was presented in the publication\footnote{While finishing the
redaction of our paper, we received the recent preprint
\cite{PearceEtAl:integrable}; the authors use a concept of twisted
minimal model which is very similar to ours, they do not discuss the
same examples (besides the Potts model) and do not consider
generalized Coxeter-Dynkin systems, but they provide a nice lattice
realization of the twisted $SU(2)$ models.  The two papers share
therefore several features but focalize nevertheless on distinct
aspects of the same general theory.} \cite{PearceEtAl:minimalold}.  In
general, twisted partition functions are not modular invariant, and we
discuss what is left of this invariance in various cases.  We also
describe what happens when the $ADE$ Dynkin diagrams are replaced by
members of an higher Coxeter-Dynkin system (Di Francesco -- Zuber
diagrams in the case of $SU(3)$).

We want this article to be almost ``self contained'' and we shall have
therefore to remind the reader several facts or constructions that, in
principle, can be found in the literature.  For this reason we make
here a short list of several specific results of the present paper,
results that, to our knowledge, cannot be found elsewhere: the use of
induction/restriction matrices to obtain all twisted partition
functions (with one or two twists), the use of the multiplication
table of the algebra of quantum symmetries $Oc(G)$ in order to obtain
identities between toric matrices, the $12 \times 12$ multiplication
table of $Oc(E_{6})$, the list of toric matrices with two twists (and
the corresponding partition functions) for the affine $E_{6}$ model,
the behaviour of these functions with respect to the action of the
modular group, a general discussion of the various types of twisted
partition functions for minimal models (see however the previous
footnote), several explicit examples of twisted versions of Virasoro
minimal models, for instance $(A_{10}-E_{6})$ and several examples of
twisted ${\cal W}_{3}$ - minimal models, for instance $({\mathcal
A}_{4}-{\mathcal E}_{5})$.

Although many of the results and formulae that we mention
belong to the lore of CFT (in particular affine WZW models or minimal
models), we decide to adopt a presentation that uses graphs (or pairs
of graphs) as primary data, so that we can avoid, as much as possible,
to make use of results coming from the theory of Virasoro algebra or
of affine Lie algebras; we therefore hope that the reader will find
our presentation to be of independent interest.

\subsection{Torus structures of Dynkin diagrams and their generalizations}

Here is a brief presentation of the various structures that will be
discussed later in this paper.

\smallskip To a given Dynkin diagram $G$ (or to a member of an higher
Coxeter-Dynkin system) one associates the complex vector space (also
called $ G$) spanned by the vertices of this diagram.  In some cases (in
particular for all diagrams belonging to the $A$ series), this vector
space $G$ possesses an associative (and commutative) multiplication
law with positive \textsl{ integral} structure constants and it is
called the ``graph algebra''; one also says that the diagram (or the
corresponding vector space) admits ``self fusion''.  In the case of
ADE diagrams, whether or not the vector space of the diagram $G$ (with
Coxeter number $\kappa$) admits self fusion, it is anyway a module
over the graph algebra of the diagram $A_{\kappa -1}$, with the same
Coxeter number.  More generally, \textrm{i.e.,\/}\ for higher Coxeter
-- Dynkin systems, the vector space $G$ is a module over a particular
graph algebra that we call $\mathcal{A} (G)$.

\smallskip

Following Ocneanu \cite{Ocneanu:paths}, to every diagram $G$ (with or
without self - fusion) belonging to a Coxeter-Dynkin system, one can
associate a bi-algebra\footnote{
This bialgebra should be, technically, a weak Hopf algebra (or quantum
groupoid), but this structure, as far as we know, has only been checked in a
few cases, and we are not aware of any general proof (see however \cite
{CoqueTrinAr:weakhopf}).} $\mathcal{B}G$. By using a particular 
scalar product, it
is easier to think that $\mathcal{B}G$ is actually a di-algebra (a vector
space with two compatible associative algebra structures). There are two --
usually distinct -- block decompositions for this di-algebra (see later).
Blocks of the first type are labelled by points of a graph that we call $
\mathcal{A} (G)$. Blocks of the second type are labelled by points of a
graph that we call $Oc(G)$. The vector spaces spanned by the vertices of
these two graphs are themselves endowed with natural associative algebra
structures that we denote by the same symbol as the graphs themselves. The
algebra $\mathcal{A} (G)$, coincides, for $G$ of type ADE, with the
graph algebra of a particular member of the $A$ family, and it is a
commutative algebra, but $Oc(G)$, also called ``algebra of quantum
symmetries'' of $G$ is not always commutative.

The algebra of quantum symmetries $Oc(G)$, like the vector space $G$ itself,
comes with a particular basis and its multiplicative structure is encoded by
a graph called $Oc(G)$ whose vertices are in one to one correspondence with
the distinguished generators. In the particular case where $G$ is a member
of the $A$ series, the algebras $\mathcal{A} (G)$, $Oc(G)$ and $G$ coincide.

\smallskip

We call $i,j,\ldots$ the vertices of $\mathcal{A} (G)$, $a,b,\ldots$ the
vertices of $G$ and $x,y,\ldots$ the vertices of $Oc(G)$. Remember that
``vertices'' should be thought of as elements of the various (distinguished)
basis for the corresponding vector spaces. We denote by $\underline{0}$ the
identity of $Oc(G)$. The vector space $G$ is a module over $\mathcal{A} (G)$
, and the algebra $Oc(G)$ is a \textsl{bi-module} over $\mathcal{A} (G)$;
this bi-module structure is encoded by a set of matrices (toric
matrices) defined as follow:
\[
i.x.j = \Sigma_{y} (W_{x,y})_{ij} y
\]
A torus structure for the diagram $G$ is (by definition) specified by the
choice of a matrix $W_{x,y}$. If the dimension of $Oc(G)$ is $s$, the number
of independent toric structures is \textit{a priori} $s^{2}$, but very often
we may have  degeneracies, in the sense that we may obtain the
same toric matrix for different choices of the pair $(x,y)$.

It is convenient to introduce the following terminology: the undeformed
torus structure corresponds to the choice of the matrix $W_{\underline 0,
\underline 0}$, a deformed torus structure along one ``defect line''
specified by $x$ corresponds to the choice of the matrix $W_{\underline 0,
x} $ (or $W_{x,\underline 0}$) and a deformed torus structure along two
defect lines specified by $x$ and $y$ corresponds to the choice of the
matrix $W_{x,y}$. It is convenient to set $W_{x} \doteq W_{x,\underline 0}$
and in particular $W_{0} \doteq W_{\underline 0,\underline 0}$.

\subsection{Frustrated (or twisted) partition functions for affine models}
\subsubsection{Twisted partition functions for affine models}
For affine models characterized by the affine Kac - Moody algebra of
type $\widehat{su}(2)$ (chiral algebra), the classification of modular
invariant partition functions is well known \cite{CIZ}, and was shown
to be in one-to-one correspondence with $ADE$ Dynkin diagrams.  More
recently \cite{Ocneanu:paths}, it was shown that if the theory is
associated with the Dynkin diagram $G$, its modular invariant
partition function is given by $Z_{0} = \overline \chi W_{\underline
0, \underline 0} \chi$ where $ \chi$ is a vector of the complex vector
space\footnote{if $\kappa$ is the Coxeter number of $G$, $n$ denotes
the cardinality of the set of vertices of the diagram $A_{\kappa -1}$,
\ie $n = \kappa -1$ for a diagram $G$ of type ADE.} $\mathbb{C}^n$ and
$W_{\underline 0, \underline 0}$ is the toric matrix associated with
the origin of the Ocneanu graph of the diagram $G$.  This
characterization of partition functions uses only the (quantum)
geometry of the diagram $G$ and does not refer to the theory of affine
algebras; in this approach, for instance, the fact that $\chi$ could
be interpreted as a character of an affine Lie algebra is not used; in
particular, modular invariance is implemented by finite dimensional
matrices representing $SL(2,\mathbb{Z})$.

As shown in \cite{PetkovaZuber:twist} and \cite{PetkovaZuber:Oc}, the
other partition functions of type $Z_{x} = \overline \chi
W_{\underline 0,x} \chi$ , or more generally $Z_{x,y} = \overline \chi
W_{x,y} \chi$, can be interpreted as twisted partition functions in a
boundary conformal field theory (boundary ``of type'' $G$), in the
presence of defect lines of type $ x $ and $y$.  A simple algorithm
for the calculation of the matrices $ W_{\underline 0, x}$ was
presented in \cite{Coque:Qtetra} (where the example of $E_{6}$ was
chosen) and explicit results for all $ADE$ cases are given in
\cite{CoqueGil:ADE} (see also \cite{CoqueGil:Tmodular} for
generalizations to higher Coxeter-Dynkin systems).  The definition of
matrices $W_{x,y}$ in \cite{PetkovaZuber:Oc} looks different from ours
(we use the description of the bimodule structure of $Oc(G)$ over
$A_{\kappa -1}$) but it can be shown to be equivalent (see our comment in
section 4.2). The matrix $W_{0} \doteq W_{0,0}$ is a
modular invariant: it commutes with the generators $S$ and $T$,
representing $SL(2,\mathbb{Z})$ in the vector space spanned by the
vertices of the graph $\mathcal{A}_{\kappa-1}$.  The corresponding
sesquilinear form is the modular invariant partition function.  The
other matrices $W_{x,y}$ are associated with partition functions that
are not modular invariant.

\smallskip

For affine models characterized by the affine Kac - Moody algebra of type
$\widehat{su}(N)$, the story is very similar. Here $\chi$ is still a
vector of the complex vector space $\mathbb{C}^n$ but $n$ now denotes
the cardinality of the set of
vertices of a graph $\mathcal{A} (G)$ generalizing the $A_{\kappa -1}$ Dynkin
diagram. In the case of $SU(3)$ for instance, the $ADE$ diagrams are replaced
by the Di Francesco - Zuber diagrams, but we can again define the bialgebra $
\mathcal{B}G$ and the two related associative algebras $\mathcal{A} (G)$ and
$Oc(G)$. Torus structures on these diagrams and corresponding twisted
partition functions are defined as before.

\subsubsection{Twisted partition functions for minimal models and their higher
analogues}

\paragraph{Minimal models}

It has been known for quite a while (see for instance the book
\cite{Drouffe-Itzykson-book}) that the classification of modular
invariant partition functions for minimal models, unitary or not, also
follows a kind of $ADE$ classification, in the sense that every
partition function describing a minimal model can be associated with
\textsl{a pair} of Dynkin diagrams\footnote{This property received  
in \cite{KawaLongo} an 
interpretation  in the framework of the theory of local nets of von 
Neumann algebras.} $(G^{(1)},G^{(2)})$.  In our
set-up, this affirmation can be precisely formulated as follows: the
partition function of a minimal model of type $(G^{(1)},G^{(2)})$ can
be obtained as the sesquilinear form associated with the matrix
$W_{\underline 0, \underline 0}^{(1)}\otimes W_{\underline 0,
\underline 0}^{(2)}$ where these two matrices respectively describe
the undeformed torus structures of diagrams $G^{(1)}$ and $G^{(2)}$.
It is also well known that the obtained minimal model is unitary if
and only if the Coxeter numbers $ \kappa_{1}$ and $\kappa_{2}$ of the
two diagrams $G^{(1)}$ and $ G^{(2)}$ just differ by one unit.  The
usual situation for minimal models corresponds therefore to the choice
of the two trivial torus structures for the graphs $G^{(1)}$ and $
G^{(2)}$; the possibility of replacing these two torus structures by
more general ones (\textrm{i.e.,\/}\ matrices $W_{\underline 0,
\underline 0}^{(1)}$ and $W_{\underline 0, \underline 0}^{(2)}$ by
matrices $W_{x_{1},y_{1}}^{(1)}$ and $W_{x_{2},y_{2}}^{(2)}$) leads to
a natural classification of twisted partition functions for minimal
models.

\paragraph{Analogues of minimal models for general Coxeter-Dynkin systems
}

The general case of minimal models corresponds to the choice of two graphs
of type $SU(2)$ (\textrm{i.e.,\/}\ two arbitrary Dynkin diagrams of type
ADE) but one can also replace the two $ADE$ diagrams $G^{(1)}$ and $
G^{(2)}$ by members of an higher Coxeter-Dynkin system (for example
the Di Francesco - Zuber diagrams of type $SU(3)$) and obtain in this way
similar classifications.  Here the notion of ``minimal
model'' is generalized and the corresponding partition functions,
twisted or not, can be interpreted in terms of minimal models for
${\mathcal W}_{n}$ algebras (in particular ${\mathcal W}_{3}$ for the
Di Francesco - Zuber diagrams).

\subsection{A brief historical section}

Here we make a long story short and gather only a few references. Many
others can be found by looking at the quoted material.
Apologies for omissions.

The study of quantum geometry of $ADE$ graphs was, at the beginning,
presented as a nice example illustrating the general theory of
``paragroups'' and ``Ocneanu cells'' \cite{Ocneanu:paragroups}.
This class of examples and its generalizations turned out to be very
rich.  Much of the theory was developed by A. Ocneanu himself and
described (sometimes in a rather allusive way) at several meetings and
conferences during the years $ 95-2000$ ( for instance
\cite{Ocneanu:Marseille}).  As far as we know, the first published
material on this theory is \cite{Ocneanu:paths}.

    From the physical side, many relations existing between $ADE$ graphs and
physics (models of statistical mechanics) had been already observed and
investigated by V. Pasquier in his thesis (see \cite{Pasquier}). A
classification of modular invariant partition functions for conformal field
theories of $SU(2)$ type was obtained at the same time, \textrm{i.e.,\/}\ at
the end of the eighties, by \cite{CIZ} in a celebrated paper. Later, T.
Gannon (and collaborators) could obtain (\cite{Gannon}) similar results for
conformal field theories based on more general affine Kac -- Moody algebras.

Di Francesco and Zuber made the crucial observation \cite{DiFrancescoZuber}
that the $SU(3)$ classification could be related to a family of particular
graphs (that we call the Di Francesco -- Zuber diagrams), in a way similar
to the relation existing between the $SU(2)$ classification and the $ADE$
Dynkin diagrams. Several precisions concerning this classification were
brought by A. Ocneanu at the Bariloche school (\cite{Ocneanu:Bariloche}, see
also the lectures of J.-B. Zuber and D. Evans at the same school).

After the unpublished work by Ocneanu concerning the $ADE$ themselves,
it was more or less clear that the existence of modular invariant
partition functions associated with these diagrams (or their
generalizations) was only the tip of a theoretical iceberg.  For
instance, from the existence of several toric structures on $ADE$
diagrams, it was clear that the modular invariant partition function
was only describing a particular point of $Oc(G) $, and that other
``interesting'' partition functions claiming for a physical
interpretation existed in the theory.  A simple algorithm allowing one
to obtain the toric matrices $W_{x,\underline{0}}$ was explained in
\cite {Coque:Qtetra}, following the example of $E_{6}$, and, as
already mentioned, a physical interpretation of the $W_{x,y}$ in terms
of conformal field theory with a boundary and defects lines was given
in \cite{PetkovaZuber:Oc}.  Using the techniques explained in
\cite{Coque:Qtetra}, a systematic study of all $ADE$ cases was
performed in \cite{CoqueGil:ADE} and several interesting cases
belonging to the $SU(3)$ family were analyzed in
\cite{CoqueGil:Tmodular}. In  \cite{FuchsEtAl}, several properties of 
the twisted partition functions were interpreted  in terms of bimodules 
for Frobenius algebras.  More recently (see
\cite{PearceEtAl:integrable} and footnote 1), it was shown how to build
a lattice realization of these models.

\section{Quantum geometry on $ADE$ diagrams and their generalizations}

\subsection{From the classical to the quantum situation (in a nutshell)}

\paragraph{Classical situation.}

Representation theory of Lie groups ($SU(2)$, $SU(3)$, etc) and their
subgroups can be encoded by graphs. These graphs tell us how to decompose
the representations obtained by tensor multiplying irreducible
representations (irreps); actually it is enough to know what happens when
one tensor multiplies some irrep by the fundamental representations.
Representation theory of $SU(2)$ is encoded, in this way, by the graph $
A_{\infty}$ (it describes the coupling of an arbitrary spin with a spin $1/2$
). Representation theory of $SU(3)$ is characterized by two generalized $
\mathcal{A}_{\infty}$ diagrams differing only by orientation (multiplication
by the fundamentals $3 = (1,0)$ of $\overline 3=(0,1)$). Such a graph
defines an associative algebra (the ``graph algebra'') which is the the
Grothendieck ring spanned by the irreducible characters of the group. Notice
that the graph algebra of a subgroup is a module over the graph algebra of
the group and that the structure constants characterizing these
associative algebras, or modules, are positive integers.

\paragraph{Quantum situation.}

In the case of $SU(2)$, truncating the diagram $A_{\infty}$ leads to
the usual $A_{n}$ Dynkin diagrams.  In the case of $SU(3)$, truncating
one of the two  diagrams $\mathcal{A}_{\infty}$ leads to the Di
Francesco -- Zuber diagrams of type $\mathcal{A}$.  This can be
generalized to $SU(N)$ \cite{Zuber:generaldynkin}.  The vector space
spanned by the vertices of any $\mathcal{A}$ diagram, for a given
$SU(N)$ system, always possess a $\mathbb{Z}_{N}$ grading (called $N$
-ality).  For instance, in the case of the usual $A_{n}$ Dynkin diagrams,
vertices are either "even" or "odd".  All these graphs ``of type
$\mathcal{A}$'' have self - fusion (an associative multiplication law
with positive integral structure constants), but they are not the only
ones enjoying this property.  The obtained graph algebras are
associative and commutative algebras with a particular basis, they are
denoted by the same symbol as the graph itself.  For a given $N$ (the
choice of $SU(N)$), the first task is to determine all those diagrams
which simultaneously admit $N$-ality, generate a module (with integral
structure constants) over some associative algebras of type
$\mathcal{A}$ and also admit  self-fusion.  The next task is to
identify all those diagrams (with $N$-ality) which do not necessarily
enjoy self-fusion, but which nevertheless generate a module (with
integral structure constants) over one of the algebras defined by the
previous family.  A list of requirements\footnote{For instance, 
when looking for modules over commutative algebras associated with 
${\mathcal {A}}$ diagrams, one should impose that they have the same generalized 
Coxeter numbers.} that a given diagram should
obey in order to be a member of some ``generalized Coxeter - Dynkin
system'' was given in reference \cite{Zuber:generaldynkin}, but as
mentioned by A. Ocneanu in \cite{Ocneanu:Bariloche} (see also \cite{Ocneanu:MSRI}), 
this list was not complete, in the sense that a local condition of 
cohomological nature should also be imposed on its set of ``cells''; 
this is not discussed here.

\subsection{The classical and quantum systems of diagrams for $SU(2)$ and $
SU(3)$}

\paragraph{The classical {$\mathbf{SU(2)}$} system.}

Choose a finite subgroup of $SU(2)$, \textrm{i.e.,\/}\ one of the
so-called  binary polyhedral groups. The fundamental representation is again
$2$ dimensional  and the multiplication of any of its irreps by the
fundamental is encoded  by the corresponding diagram of tensorisation,
which, for the binary groups of symmetries of  platonic bodies
co\"{\i}ncides with the affine exceptional Dynkin diagrams $E_{6}^{(1)}$, $
E_{7}^{(1)}$, $E_{8}^{(1)}$ (McKay correspondence, \cite{McKay}).
The vector space generated by the set of irreducible representations
of such a subgroup is a module over the algebra generated by the set
of irreps of $SU(2)$ (reduce irreps from the group $SU(2)$ to its
subgroup and use tensor multiplication of representations).  In
diagrammatic parlance, we may say that affine $ADE$ diagrams are
modules over the $A_{\infty}$ diagram.
Irreps of a binary polyhedral group can also be tensor  multiplied and
decomposed into irreps (with positive integral  structure constants). In
other words : affine $ADE$ diagrams  have self fusion. In particular one of
the vertices $\sigma_{j}$ acts as the unit, we call it $\sigma_{0}$.
For each of these diagrams, call $G_{1}$ the adjacency matrix; its  highest
eigenvalue $\beta$ (called  the Perron - Frobenius norm of the diagram) is
equal to $2$ in all  cases
and it co\"\i ncides with the dimension of the
fundamental representation. For a given  diagram, dimensions of the irreps
are given by components of  the (unique) normalized eigenvector corresponding
to $\beta$ (it is normalized  to $1$ at the unit point $\sigma_{0}$).
The table of characters $S$ happens to be equal to the  matrix of
eigenvectors (properly normalized)  of $G_{1}$. This is a way to express the
general McKay correspondence  in the case of $SU(2)$.

\paragraph{The quantum {$\mathbf{SU(2)}$} system.}

Now we move to the quantum case and replace the $A_{\infty}$ diagram by $
A_{n}$ diagrams seen as truncated $A_{\infty}$ diagrams. These diagrams $
A_{n}$ have self - fusion. The next task is to determine those diagrams
(with bi-ality) that generate modules over the $A_{n}$ : we get the $A$, $D$
and $E$ diagrams. For example, $E_{6}$ is an $A_{11}$ module, $E_{7}$ an $
A_{17}$ module, and $E_{8}$ an $A_{29}$ module. Some of them have
self-fusion ($A_{n}$, $D_{eve}$, $E_{6}$, $E_{8}$), others don't ($D_{odd}$, $
E_{7}$). A  diagram $D_{eve}$ actually determines a two-parameters
family of associative structures, but only two of them have structure constants
which are positive integers (self fusion); these two structures can be
identified when we permute the two end points of the $D_{even}$ fork;
when such a phenomenon appears, the algebra of quantum symmetries
$Oc(G)$, to which we shall return later, appears to be non commutative.

The norm of a diagram $G$ is found to be $\beta = 2 \cos \frac{2 \pi
}{\kappa}$, where $\kappa = n+1$ if $G = A_{n}$ or when $G$ is a
module over $A_{n}$.  Note that $1 < \beta < 2$ (see also
\cite{Jones:book}).  The quantum dimensions $qdim_{a}$ of the vertices
of $G$ are obtained or defined as the components of the normalized
Perron Frobenius eigenvector (which corresponds to the eigenvalue
$\beta$).  For every $ADE$ diagram, \ie for every member $G$ of the
system that we may call ``the $SU(2)$ Coxeter-Dynkin system'', the
integer $ \kappa$ is called the \textsl{Coxeter number of the
diagram}.  All these diagrams (with or without self - fusion) can also
be labelled by an integer $k$, called the \textsl{level of the
diagram} and defined by $k = \kappa-2$. A description of the ADE diagrams 
in terms of representations of 
quantum subgroups (a quantum analogue of
the  McKay correspondance) was discussed by \cite{KirillovOstrik} in 
the framework  of modular categories.

\paragraph{The classical {$\mathbf{SU(3)}$} system.}

Representation theory for finite subgroups of $SU(3)$ is fully characterized
by a family of diagrams that have self -- fusion and generate modules over
the graph algebra of the generalized $\mathcal{A}_{\infty}$ diagram of $SU(3)
$. All of these diagrams have a norm equal to $3$.

\paragraph{The quantum {$\mathbf{SU(3)}$} system.}

Now we move to the quantum and replace $\mathcal{A}_{\infty}$ by
$\mathcal{A}_{k}$ (truncated $A_{\infty}$ diagrams).  These
$\mathcal{A}_{k}$ have self-fusion.  The next task is to determine
those diagrams (with tri-ality) that are modules over the
$\mathcal{A}_{k}$ : we get the Di Francesco -- Zuber diagrams.  Some
of them have self-fusion and others don't.  The system contains in
particular the $\mathcal{A}$ series and a finite number of ``genuine
exceptional'' cases ($\mathcal{E}_5$, $\mathcal{E}_9$ and
$\mathcal{E}_{21}$).  The other diagrams of the system are obtained as
orbifolds of the genuine diagrams (exceptional or not) and as twists
or conjugates (sometimes both) of the genuine diagrams and of their
orbifolds.  See references \cite{DiFrancescoZuber},
\cite{Zuber:generaldynkin} \cite{Ocneanu:Bariloche},
\cite{Zuber:Bariloche}.  All of them have a norm $\beta$ equal to $1+
2 \cos \left( \frac{2 \pi}{ \kappa} \right)$.  Note that $2 < \beta <
3$.  This again defines an integer $ \kappa$ called the
``generalized Coxeter number'' or ``altitude'' (like in
\cite{DiFrancescoZuber}).  The level $k$ of a \textsl{diagram}
belonging to this family is defined by the relation $k \doteq \kappa -
3$.  The truncated $\mathcal{A}_{\infty}$ diagrams that we call
$\mathcal{A}_{k}$ are of level $k$ (see the footnote in the next
subsection).  Even when it exists, the determination of the graph
algebra of a given diagram is not always unique; a phenomenon similar
to what happens for the $D_{even}$ diagrams (see a previous remark)
occurs for instance in the case of the ${\mathcal E}_{9}$ diagram of
the $SU(3)$ system.

\subsection{General notations and characteristic numbers for generalized
Coxeter-Dynkin diagrams}

The classical representation theory of $SU(N)$ can be encoded by a set of $
N-1$ diagrams (with oriented edges and infinitely many vertices)
generalizing the $A_{\infty}$ diagram of $SU(2)$; there is one such oriented
diagram for each fundamental representation. For definiteness, we choose the
basic representation of $SU(N)$; its Young tableau is given by a single box.
A given system of diagrams is then labelled by an integer $N$, it has the
same value for all diagrams of a system. For $ADE$ Dynkin diagrams, $N=2$,
the (dual) Coxeter number of $SU(2)$. For Di Francesco - Zuber diagrams,
$N = 3$, the (dual) Coxeter number of $SU(3)$.
The generalized Coxeter number (altitude) of a diagram $G$ is called
$\kappa$ in our paper, it can be defined directly from the norm
$\beta$ of $G$; for usual Dynkin diagrams, altitude is the usual dual
Coxeter number.
It is useful to define the root of unity $\mathbf{q} \doteq exp{\frac{i \pi}{
\kappa}}$, so that $\mathbf{q}^{2 \kappa} = 1$.
The level $k$ of a given \textsl{diagram} belonging to a given system
of type $SU(N)$ is defined by the relation $k \doteq \kappa - N$.
More generally, one could probably define generalized Coxeter-Dynkin
systems for any Lie group (the case $SU(2)$ corresponding
to the $ADE$ system), but such a theory remains to be investigated.

As we know, for a given system, members of the $\mathcal{A}$ family
(call them $\mathcal{A}_{k}$, with $k$ standing for the
level\footnote{ Another favorite notation is $\mathcal{A}^{(k+N)}$,
the upper index referring now to the altitude.  We shall stick to the
notation using level as a subscript.}) are obtained as
truncated\footnote{ Truncation is made by removing the parts of the
diagram with level higher than $k$; what we obtain is a truncated Weyl
chamber (``a Weyl alcove'').} $ \mathcal{A}_{\infty}$ diagrams.  They
can be related to a particular category of representations of quantum
groups at roots of unity, but we shall not discuss this aspect here.

A diagram $G$ of level $k$ belonging to such a generalized system is 
always such
that the vector space spanned by the set of its $m$ vertices is a module
over the member $\mathcal{A}_{k}$ of the $\mathcal{A}$ family with the same
level\footnote{
Warning, in the $SU(2)$ case, we have two notations for the same objects
since the subindex of $A_{n}$ refers usually to the number of vertices (the
rank), but in this particular case, $k = n-1$, so that $\mathcal{A}_{k=n-1}
= A_{n}$.} (the number of vertices of this corresponding diagram of type $
\mathcal{A}$ will be called $n$,
so $m=n$ when $G$ is of type $A$.
Notice that $n = k+1$ for usual $A_{n} = \mathcal{
A}_{k}$ Dynkin diagrams, but $n = (k+1)(k+2)/2$ for Di Francesco --
Zuber diagrams of type $\mathcal{A}_{k}$.

The list of exponents $\{r\}$ of a graph $G$ of type $ADE$ can be
defined directly from the table of eigenvalues of the adjacency matrix
$G_{1}$ of $G$:  these eigenvalues are of the form $2 Cos(r \pi
/\kappa)$.  For instance, in the case of $E_{6}$, from the list of
eigenvalues, $$\{2 Cos( \pi /12), 2 Cos(4 \pi /12), 2 Cos( 5\pi/12), 2
Cos(7\pi /12), 2 Cos(8\pi /12), 2 Cos(11\pi /12)\}$$ we read the
exponents $\{1, 4, 5, 7, 8, 11\}$.  Notice that exponents also refer
to particular vertices of the corresponding diagram of type
$\mathcal{A}$ with the same Coxeter number (for $E_{6}$, see the
circled vertices on figure \ref{fig: E6/A11induction}, and remember
that our indices for  labelling vertices are shifted by $1$).  The list of
exponents $\{r=(r_1,r_2)\}$ of a graph $G$ belonging to a generalized
system can also be defined directly from the adjacency matrix $G_{1}$ of
$G$.  For Di Francesco - Zuber diagrams (ie the $SU(3)$ system), they
can be read from the following general formula giving the eigenvalues
of $G_{1}$ (\cite{DiFrancescoZuber}) ${(1 + e^{\frac{2 i \pi
r_1}{\kappa}} + e^{\frac{2 i \pi (r_1 + r_2)}{\kappa}})}/{e^{\frac{2
i\pi (2 r_1 + r_2)}{3 \kappa}}}$.  For instance, the exponents of
${\mathcal E}_{5}$ are
$$\{(1,1),(3,3),(1,3),(4,3),(3,1),(3,4),(3,2),(1,6),(4,1),(1,4),(2,3),(6,1)\}$$
Here again, exponents refer to particular vertices of the
corresponding diagram of type $\mathcal{A}$ with the same Coxeter
number and remember that indices labelling vertices are usually shifted by
$(1,1)$.  Exponents appear in the expression giving the corresponding
modular invariant partition function (see the examples of ${E}_{6}$ or
of ${\mathcal E}_{5}$ in sections \ref{sec:E6partfun} and 
\ref{sec:E5partfun}) and in the usual
(or generalized) Rocha-Cariddi formulae.

\subsection{Paths, essential paths, the bi-algebra $\mathcal{B}G$ and
the algebra $Oc(G)$ of quantum symmetries}

We then move from the geometry of the ``space'' $G$ to the geometry of the
paths on $G$, a procedure  quite common in quantum physics! Paths on
$G$ generate a vector space $Paths$ which comes with a grading: paths of
homogeneous grade are associated with Young diagrams of SU(N). In the case of $
SU(2)$ this grading is just an integer (to be thought of as a length, a
Young frame with a single row, or as a point of $A_{n} \equiv
\mathcal{A}_{k=n-1}$).

What turns out to be most interesting is a particular vector subspace
of $Paths$ whose elements are called ``essential paths'' (we refer to
\cite{Ocneanu:paths}, \cite{Coque:Qtetra}, see also
\cite{Coquereaux:ClassicalTetra} for a definition).  The space of
essential paths $EssPaths$ is itself graded in the same way as $
Paths$ and one may consider the graded algebra of endomorphisms of
essential paths $\mathcal{B}G \doteq End_{\sharp}(EssPaths) =
\bigoplus_{j=0,r-1} End(EssPaths^{j})$.

By using the fact that paths on the chosen
diagram can be concatenated , one may define {\sl another}
multiplicative (associative) structure on
the vector space $\mathcal{B}G$ (see \cite{Ocneanu:paths} for a definition).
This leads to a {\sl di}-algebra $\mathcal{B}G$ which turns out to be
semi-simple for both structures, but existence of a scalar product
allows one to transmute one of the multiplications into a co-multiplication
compatible with the other structure and one obtains in this way a {\sl
bi}-algebra. This bi-algebra is sometimes called, by A. Ocneanu ``Algebra of
double triangles" (DTA), a terminology coming from the graphical representation
of the corresponding elementary matrices by diffusion graphs or,
dually, as double triangles.

For these two associative laws on the same space, that we may call
``composition law'' and ``convolution law'' (or ``vertical law'' and
``horizontal law''), there are two --- usually distinct --- block
decompositions for $\mathcal{\ B}G$ (ideals corresponding to simple
blocks).  The first type of blocks, labelled by $j$, corresponds to
the grading associated with points $\sigma_{j}$ of $\mathcal{\
A}_{k}$, \ie in the case of $SU(2)$, to the lengths of the paths,
and, more generally, to Young diagrams of $SU(N)$); interpretation of
this first block structure is therefore clear from the definition of
$\mathcal{B}G$ as sum of algebras of endomorphisms.  The second block
decomposition can be interpreted as follows: $ADE$ diagrams (or their
$SU(N)$ generalizations) may have classical symmetries, for instance,
all $A_{n}$ diagrams have an obvious $\ZZ_{2}$ symmetry; these
classical symmetries (action of a finite group on vertices) can be
promoted to the level of paths in an obvious way and therefore lead to
particular endomorphisms of $EssPaths$; but
there are more ``quantum symmetries'' acting on the space of essential
paths than classical symmetries: irreducible quantum symmetries
(call them $x$) are precisely associated with the blocks of
$\mathcal{B}G$ for the second multiplication.  We call $Oc(G)$ the
algebra spanned by the minimal central projectors associated with the
later blocks, using the first multiplicative structure.  $Oc(G)$ is
called the ``algebra of quantum symmetries''.  In all cases it is
  an associative algebra with two generators (called ``left'' and
``right'' generators) and the Cayley graph of multiplication by these
two generators, called the ``Ocneanu graph of $G$''  is also
denoted by $Oc(G)$.  The linear span of these generators are called
left and right chiral parts, and their intersection is called
``ambichiral''.

The number\footnote{This number is infinite in the classical situation
(finite subgroups of Lie groups).} of (simple) blocks of ${\cal B}G$
for its first multiplication, is $n$ (the number of points of the
corresponding ${\cal A}$ diagram); dimension of these blocks will be
called $d_{j}, j=1\ldots n$.  The number of (simple) blocks of ${\cal
B}G$ for its second multiplication, will be called $s$ (the number of
points of the corresponding Ocneanu diagram); dimension of these
blocks will be called $d_{x}, x=1\ldots s$.  Existence of two block
decompositions for the same underlying vector space $\mathcal{B}G$
leads obviously to the number-theoretical identity (quadratic sum
rule): $\sum_{i=1,n} d_{i}^{2} = \sum_{x=1,s} d_{x}^{2}$.  In all
cases explicitly studied so far, an unexpected linear sum rule also
holds (in some cases one has to introduce a natural correction
factor).

The direct determination of the algebra $Oc(G)$, using the definition
provided by A. Ocneanu, is not an easy task, and the corresponding
graphs were first known (published) for the $SU(2)$ Coxeter-Dynkin
system \cite {Ocneanu:paths}.  This algebra is not always commutative.
One of the purposes of \cite{Coque:Qtetra} and \cite{CoqueGil:ADE},
besides the calculation of the toric matrices, was actually to give an
algebraic construction providing a realization of the \textsl{algebra}
$Oc(G)$ in terms of graph algebras associated with appropriate Dynkin
diagrams.  In many relatively easy cases where $G$ admits self-fusion
and is also such that $Oc(G)$ is commutative, the algebra of quantum
symmetries is isomorphic with $G\otimes_{J} G$, where $J$ is a
particular subalgebra of the graph algebra of $G$; the tensor product
sign, taken ``above $J$'', means that we identify $au \otimes b$ and
$a \otimes ub$ whenever $u \in J \subset G$.  In those easy cases, and
as shown in \cite{CoqueGil:Tmodular}, the subalgebra $J$ can be
determined from the modular properties of the graph $G$; we shall
remind the reader how this is done in a later section.  Paradoxically,
for Dynkin diagrams, and besides the $A_{n}$ themselves, the
``simple'' cases happen to be those where $G$ is an exceptional
diagram equal to $E_{6}$ or $E_{8}$.  We refer to \cite{CoqueGil:ADE}
for a discussion of all $ADE$ cases and \cite{CoqueGil:Tmodular} for a
discussion of a number of cases belonging to the Di Francesco - Zuber
system.

\subsection{The matrices $N_{i}$, $F_{i}$, $G_{a}$, $E_{a}$, $S_{x}$ and $
W_{x,y}$}
\label{sec:diversesmatrices}
\subsubsection{Fusion matrices: the $N_{i}$'s}

Fusion matrices are defined for $\mathcal{A}_{k}$ diagrams.  They are
square matrices of dimension $n \times n$ called $N_{i}$.  They are
associated with the vertices $\tau_i$, with $i \in \{0, \ldots,
n-1\}$, and provide a faithful representation of the graph algebra.
Here $i$ is actually a multi-index referring to a Young frame of
$SU(N)$ and the cardinality of the indexing set is $n$.  When the
Young frame refers to a fundamental representation (only one column),
this fusion matrix is the adjacency matrix of the corresponding
oriented diagram.  Other matrices $N_{i}$ are obtained from the
fundamental ones by applying the particular recurrence relation
specific to $SU(N)$.  Example: in the case of $SU(2)$, each Young
diagram is an horizontal string of boxes and is characterized by its
length; the matrix $N_{1}$ is the adjacency matrix of $\mathcal{A}_{k}
= A_{n=k+1}$ and $N_{0}$ is the unit; the recurrence relation
(coupling of spins) is $$N_{i+1} = N_{1} N_{i} - N_{i-1}$$
Matrices $N_{i}$ have indices $(j,k)$ referring to vertices of
$\mathcal{A}_{k}$.  These matrices generate a (commutative)
associative algebra isomorphic with the algebra of the given $
\mathcal{A}$ diagram.  The indices $i,j$ runs from $0$ to $n-1$ but we
shall sometimes use indices $r=i+1, s=j+1$ running from $1$ to $n$.
In the case of $SU(3)$, the index $j$ labelling vertices $\tau_{j}$ of
the ${\mathcal A}_\kappa$ diagram is a pair $(j_1, j_2)$, with $j_1,
j_2 \geq 0$ and $j_1 + j_2 \leq k$.  The identity is $N_{0,0}$ and the
matrix $N_{1,0}$ denotes the adjacency matrix of the (oriented)
diagram $G$.  The recurrence formula reads
$$
\begin{array}{lcll}
N_{j_{1}, j_{2}} &=& 0 & {\text{if} \, j_{1} < 0 \;\, \text{or} \,
j_{2} <0 } \\
N_{j_{1}, 0} &=& N_{1,0} N_{j_{1}-1 , 0} - N_{j_{1}-2 , 1} & {}
\\
N_{j_{1}, j_{2}} &=& N_{1,0} N_{j_{1}-1 , j_{2}} - N_{j_{1}-1 , j_{2}
- 1} - N_{j_{1}-2 ,j_{2} + 1} & {\text{if} \, j_{2} \neq 0} \\
N_{0, j_{1}} &=& N_{j_{1}, 0}^T & {}
\end{array}
$$

\subsubsection{Fused adjacency matrices: the $F_{i}$'s}

The module property (external multiplication) of the vector space associated
with a diagram $G$, of level $k$ and possessing $m$ vertices, with respect
to the action of the  algebra $\mathcal{A}_{k}$ is encoded by a set of $n$
matrices $F_{i}, {i=0 \ldots n-1}$, of dimension $m \times m$, sometimes
called ``fused graph matrices'' (a somehow misleading terminology!): $
\tau_{i} \sigma_{a} = \sum_{b} (F_{i})_{ab} \sigma_{b}$.

If $G$ is of type $\mathcal{A}$, we shave $n=m$, $F_{i}=N_{i}$ and we are
done. More generally, call $F_{0}$ the unit matrix of dimension $m \times m$,
and $F_{1}$ the adjacency matrix of $G$. For usual $ADE$ diagrams, each
edge carries both orientations and $F_{1}$ is symmetric; for generalized
diagrams, this is not so. Other matrices $F_{i}$ are then obtained by
imposing the same recurrence relation as for the fusion matrices. Matrices $
F_{i}$ have indices $(a,b)$ referring to vertices of $G$; they characterize $
G$ as a module over the corresponding $\mathcal{A}$ graph. They are also in
one to one correspondence with the minimal central projectors diagonalizing
one of the two associative structures of the di-algebra $\mathcal{B}G$, in
other words they characterize the corresponding blocks and give their
dimensions $d_{i} = \sum_{a,b} (F_{i})_{a,b}$.

In the case of $SU(3)$ diagrams, remember that indices $j$ are pairs
$(j_1, j_2)$ and that fused adjacency matrices $F_{j}$,
associated with any graph $G$ of a given level, are determined by the
same recurrence relations as for matrices $N_i = N_{j_1,
j_2 }$ associated with the graph ${\mathcal A}$ of the same level;
only the seed is different: $F_{1} \doteq G_{1}$, the adjacency
matrix of $G$.

\subsubsection{Graph matrices: the $G_{a}$'s}

The diagram $G$ sometimes admits self-fusion. In those cases, the $m$ linear
generators $\sigma_a$ of $G$ ($a$ runs from $0$ to $m-1$) are represented by $m
$ commuting matrices $G_a$ of dimension $m\times m$ spanning a faithful
representation of the graph algebra. We call $G_{0} \doteq F_{0}$, $G_{1}
\doteq F_{1}$ and more generally $G_{a}$ the set of matrices (one for each
vertex of $G$) representing faithfully the multiplication of vertices.
Warning: with the exception of $F_{0}$ and $F_{1}$, the matrices $F_{i}$ and
$G_{a}$ are distinct (in the case of $\mathcal{A}$ diagrams, of course, they
are identical).

\subsubsection{Essential matrices: the $E_{a}$'s}
\label{sec:EssMat}

By definition, the $m$ essential matrices $E_{a}$ are rectangular
matrices of dimension $n \times m$ defined by setting\footnote{The
reader should be cautious about the meaning of indices: our indices
$i$ or $a$ refer to actual vertices of the graphs but the numbers
chosen for labelling matrix rows and columns depend on some arbitrary
ordering on these sets of vertices.  Moreover our labels $i$ and $a$
start from $0$, not from $1$.}, for every vertex $a$ of $G$,
$(E_{a})_{i,b} \doteq (F_{i})_{a,b}$.  These are rectangular matrices
of dimension $(n,m)$.  Matrices $E_{a}$ display ``visually'' the
structure of essential paths emanating from a vertex $a$ on the
diagram $G$.  One can check that, for graphs with self fusion,
$E_{a}=E_{0} G_{a}$.  The particular matrix $E_{0}$ is usually called
``intertwiner'', in the statistical physics literature.

As we know, vertices of the diagram $G$ should be thought of as an analogue
of irreducible representations for a subgroup of a group; the irreducible
representations of the bigger group are themselves represented by vertices
of the corresponding $A$ graph. In this analogy, the first column of each
matrix $F_{i}$ would describe the branching rule of $\tau_{i}$ with respect to
the chosen subgroup (restriction mechanism). In the same way, the columns of
the particular essential matrix $E_{0}$ would describe the induction mechanism:
the non-zero matrix elements of the column labelled by $\sigma_{b}$ tell us
what are those representations $\tau_{i}$ that contain $\sigma_{b}$ in their
decomposition (for the branching $A \rightarrow G$).

\subsubsection{Matrices for $Oc(G)$}

Since we have a di-algebra $\mathcal{B}G$ we have also
a set of matrices $S_{x}$ which characterize the blocks of the other
associative structure (one for each point of the Ocneanu graph). In
``simple cases'', like $E_{6}$ or $E_{8}$, the matrix $S_{x}$ associated
with the vertex $x = a \otimes_{J} b$ of the Ocneanu graph is simply equal
to the product $G_{a} G_{b}$.  The dimension $d_{x}$ of the block $x$ is
obtained by summing the matrix elements of $S_{x}$.

\subsubsection{Toric matrices and generalized toric matrices: the $W_{x}$
and $W_{x,y}$}
\label{sec:toricmat}

We know that $\mathcal{A}_{k}$ acts on $G$, but
$\mathcal{A}_{k}$ also acts (from both sides) on $Oc(G)$.
   In general $Oc(G)$ is an
$\mathcal{A}_{k}$ bimodule and the action is encoded as follows: $\tau_{i} \,
x \, \tau_{j} = \sum_{y \in Oc(G)} (W_{xy})_{j}^{i} \; y $, with $x,y \in
Oc(G)$ and $\tau_{i},\tau_{j} \in \mathcal{A}_{k}$.  In general, one
obtains $s \times s = s^{2} $ matrices $W_{xy}$ of dimension $ s
\times s$, (many of them may happen to be equal). In particular one
obtains the $s$ matrices
$W_{x} \doteq W_{x0}$ and the matrix $W_{0}= W_{00}$
associated with the origin of the Ocneanu graph.
Practically, once we have the $m$ rectangular matrices $E_{a}$, of dimension
$n \times m$, we first replace by $0$ all the matrix elements of the columns
labelled by vertices $b$ that \textsl{do not} belong to the subset $J$ of
the graph $G$, call $E_{a}^{red}$ these ``reduced'' matrices and obtain, for
each point\footnote{
In some cases, $x$ may be a linear combination of such elements.} $x = a
\otimesdot b$ of the Ocneanu graph $Oc(G)$, a ``toric matrix'' $W_{x} =
E_{a}\, (E_{b}^{red})^T$, of dimension $n \times n$.

We will explain in section \ref{sec:algo1} how to generalize the
previous method to obtain all the toric matrices $W_{x,y}$ (``first
algorithm'').  Actually, the $W_{x,y}$ can also be obtained from the
$W_{x}$, determined as above, by working out the multiplication table of
$Oc(G)$ (this is our ``second algorithm'').  All we have to do is to
decompose the product $x\times y$ on the basis generators: if
$x.y=\Sigma_{z} C_{x,y}^{z} z$ with $x,y,z\in Oc(G)$ then
$W_{x,y}=\Sigma _{z}C_{x,y}^{z} W_{0,z}$.  This can be seen as a
compatibility equation; indeed, the action of ${\cal A}_{k}$ is
central, so $\tau _{i}.x.\tau _{j} = x.  \tau _{i}.\underline{0}.\tau
_{j} $ implies
\begin{eqnarray*}
  \Sigma _{y}(W_{xy})^{i}_{j}y &=&  x . (\Sigma _{z}(W_{0z})^{i}_{j} z) =
  \Sigma _{z}(W_{0z})^{i}_{j} x. z \\
{} &=& \Sigma _{z}(W_{0z})^{i}_{j} \Sigma_{y} C_{x,z}^{y} y
  = \Sigma_{y} (\Sigma _{z}  C_{x,z}^{y} (W_{0z})^{i}_{j}) y
\end{eqnarray*}
Notice that linearity of this relation implies in particular $W_{xy,0}
= W_{x,y}$.Moreover, when $Oc(G)$ is commutative, \ie $xy=yx$, we
have $W_{x,y}= W_{y,x}$ (but the later equality does not imply the
former).

 From the toric matrices $W_{xy}$ describing the
bimodule structure of $Oc(G)$, one obtains the corresponding twisted
partition functions as sesquilinear forms in the complex vector space 
$\mathbb{C}^s$.
Introducing a basis $\chi$ of vectors $(\chi_{j})$, usually interpreted
as characters, we write
$$Z_{x,y} = \overline{\chi} W_{x,y} \chi$$ and $Z_x = Z_{x,\underline{0}}$.
The modular invariant
partition function is $Z_{\underline{0}}$ with $\underline{0} = 0
\otimesdot 0$.  The example of $E_{6}$ is discussed in section
\ref{section: example E6}.

\subsection{Modular aspects: $S$, $T$ and $SL(2,\mathbb{Z})$}

\subsubsection{The $S$ operator}
\label{sec:SfromMul}

Any finite subgroup of $SU(2)$ can be associated with an affine ADE
graph, in such a way that the normalized Perron Frobenius vector of
the graph gives the list of dimensions for irreducible representations
of the finite subgroup.  This observation, known as McKay
correspondence, was later generalized by observing that the whole
table of characters of a finite subgroup of $SU(2)$ can be identified
with the list of eigenvectors (properly normalized) of the adjacency
matrix of the corresponding affine Dynkin diagram (generalized McKay
correspondence).  For any finite group, not necessarily a subgroup of
$SU(2)$, the commutative and associative algebra generated by
irreducible characters (multiplication of representations) can be
realized by a set of commuting matrices (the analogue of our matrices
$G_a$ ) and the table of characters can be reconstructed, without
using the notion of conjugacy classes, by diagonalizing simultaneously
this set of (commuting) matrices: the character table $S$ is a
properly normalized diagonalizing matrix. The following ``quantum
construction'' is analogous.

In the quantum case (\ie diagrams ADE), there is no group, there are
no conjugacy classes and no table of characters.  Nevertheless, there
is an adjacency matrix for the chosen diagram.  The matrix $S$ that we
are looking for is precisely the quantum analogue of the table of
characters, and is obtained, for each level $k$ as the (properly
normalized) table of eigenvectors for the adjacency matrix of the
diagram ${\mathcal A}_{k}$.  The bonus in the quantum situation is
that one can interpret $S$ as one of the generators of the modular
group in a particular representation; this representation of $SL(2,\ZZ)$ appeared in
a work by Hurwitz \cite{Hurwitz} about a century ago.  $S$, interpreted as
a quantum table of characters (or a ``quantum Fourier transform'')
implements therefore a quantum analogue of the McKay correspondence.
For illustration, the modular matrix $S$ for the $A_{11}$ diagram is
determined in this way in section \ref{sec:SforA11}.  The general expression
for $S=s$, in the case of the $SU(2)$ system, with $\kappa =
k+2$, is
$$S_{ij} = \sqrt{\frac{2}{\kappa}} \sin(\pi
\frac{(i+1)(j+1)}{\kappa}) \, \textstyle{for} \,  0 \leq i,j \leq \kappa-2 $$

\subsubsection{$SL(2,\ZZ)$}

A projective representation of $SL(2,\ZZ)$ can be defined with two
matrices $s$ and $t$ and a phase $\zeta$ which are such that
$(st)^{3}= \zeta^{3} s^{2}$, $s^{2}= C$, $Ct=tC$ and $C^{2}=1$.  The
matrix $C$ is called ``conjugation matrix'' and $t$ is the ``modular
twist''.  Such representations of the modular group can be obtained on
the space generated by the simple objects in any braided modular
category \cite{KiriOsBook}.  The general formula for the modular
phase is $\zeta = e^{2 \mathsf{i} \pi c/24}$ with $c = (\kappa - N) d/\kappa$.
In the present context, \ie generalized Coxeter - Dynkin diagrams of
type $SU(N)$, $\kappa$ is the altitude (generalized Coxeter), $\kappa
- N = k$ is the level and $d = dim\, SU(N)$.  Therefore $c =
\frac{3k}{k+2}$ for $SU(2)$ and $c = \frac{8k}{k+3}$ for $SU(3)$.  The
modular phase $\zeta$ is then equal to $e^{\frac{i\pi}{4}} e^{\frac{
-i \pi} {2\kappa}}$ for an $ADE$ diagram and to $e^\frac{2 i
\pi}{3} e^{-\frac{2 i \pi}{\kappa}}$ for a Di Francesco - Zuber
diagram.  We use modular generators $S,T$ normalized as follows: $S =
s$ and $T = t / \zeta $.  The $SL(2,Z)$ relations then read
$(S T )^{3}=  S^{2}$, $S^{2}= 1$.

\subsubsection{The $T$ operator}
\label{sec:TgeneratorAff}
In the framework of modular categories, and for a Lie algebra
${\mathcal G}$, a general expression for the modular twist is $ t_{ij}
= \delta_{ij} \mathbf{q}^{<<j, j + 2 \rho>>}$ where $\mathbf{q} =
e^{\mathsf{i}\pi/\kappa}$, $\rho$ is half the sum of positive roots,
$i$, $j$ are elements of the weight lattice characterizing the
representation $\tau_{i}$ and $\tau_{j}$; moreover $<<.,.>>$ is an
invariant bilinear form on $\mathcal G$ normalized by
$<<\alpha,\alpha>>=2$ for a short root $\alpha$.  For $SU(2)$, with
$i,j = 0 \ldots \kappa -2$, the modular twist is $t_{ij} =
e^{\frac{\mathsf{i} \pi}{2\kappa} j(j+2)} \delta_{ij}$.  Its logarithm
is proportional to the Casimir operator: $j$ is related with the
(would be) spin $\ell$ by $j+1 = 2\ell +1$, therefore
${\frac{\mathsf{i} \pi}{2\kappa} j(j+2)} = {\frac{2 \mathsf{i}
\pi}{\kappa} \ell(\ell+1)}$.  With our normalization, 
the modular generator $T$ is therefore
$$T_{ij} = exp[2 \mathsf{i}\pi (\frac{(j+1)^2}{4
\kappa} - \frac{1}{8})] \delta_{ij} $$
The expression $(\frac{(j+1)^2}{4 \kappa} - \frac{1}{8})$ is the
``modular anomaly'', and it is convenient to call ``modular exponent''
the quantity $\hat T = (j+1)^{2}$ mod $4\kappa$ (we could as well use
$j(j+2)$ mod $4\kappa$ or any other expression differing by a constant
shift).

In the case of $SU(3)$, the action of the modular matrix $T$ on
vertices $\tau_{j} \equiv \tau_{(j_{1}, j_{2})}$ of ${\mathcal
A}_{k}$ is also diagonal and given by:
$$
\left( T \right)_{i j} = e_{\kappa} \left[ -(i_1+1)^2 -
(i_1+1).(i_2+1) - (i_2+1)^2 + \kappa \right]
\delta_{i j},
$$
where $i \doteq (i_1,i_2)$, $j \doteq
(j_1,j_2)$, $e_{\kappa}[x] \doteq \exp \left( \frac{-2 \mathsf{i} \pi
x}{3\kappa} \right)$, and $\kappa=k +3$.  We call ``modular exponent''
the quantity $\hat{T} = [-(i_1+1)^2 - (i_1+1).(i_2+1)
- (i_2+1)^2 + \kappa ] \, \textstyle{mod} \, 3 \kappa$.

\subsubsection{Modular invariance}

Modular invariance of the partition function $Z_{00}$ can be proven
either by checking that it is invariant when we replace the modular
parameter $\tau$ by $\tau + 1$ or $-1/\tau$ in the characters
$\chi_{r}$ (these functions are generalized Jacobi's theta functions)
or, much more simply, by showing that the matrix $W_{0}$ commutes with
the generators $S$ and $T$ of the modular group in this
representation.

It can be checked, from the explicit expressions of $S$ and $T$ in the
$SU(2)$ case, that, $T^{8 \kappa} = 1$ when $\kappa$ is odd and $T^{4
\kappa} = 1$ when $\kappa$ is even.  This, by itself, is not enough to
imply the following property, which is nevertheless true, and was
proven more than hundred years ago \cite{Hurwitz}: the Hurwitz --
Verlinde representation of $SL(2,\ZZ)$ factorizes over the finite
group $SL(2,\ZZ/8 \kappa \ZZ)$ when $\kappa$ is odd, and
over $SL(2,\ZZ/4 \kappa \ZZ)$ when $\kappa$ is even.  For
instance, $T^{40}=1$ for  $A_{4}$  ($40 = 8 \times 5$), but
$T^{48}=1$ for $A_{11}$ ($48 = 4 \times 12$).

\subsubsection{Determination of $Oc(G)$ from the modular properties of
the diagram $G$}
\label{sec:Ocstruct}
In general an Ocneanu cell system is defined by four graphs -- two
horizontal and two vertical -- satisfying a number of matching
properties (see \cite{Ocneanu:paragroups}, \cite{EvansKawa:book}).
Particularly interesting cell systems are obtained when one chooses
the two horizontal graphs as given by two Dynkin diagrams with the
same Coxeter number.  In the present situation, these two are given by
the same Dynkin diagram $G$ (we write ``Dynkin'' but this graph can be
a member of an higher system).  A priori, the determination of $Oc(G)$
results from the study of the the block structure of ${\mathcal B}G$
for its convolution law.  This, in turns, requires the determination
of the values of all Ocneanu cells for the graph system of type
$(G,G)$, a task that may involve rather long calculations\ldots but if
our only purpose is to determine $Oc(G)$, it is simpler to find a
short cut.  One possibility is to use the fact that we already know,
in many cases, the expression of the modular invariant (as calculated
by \cite{CIZ} for $SU(2)$ and \cite{Gannon} for $SU(3)$); such a
technique was apparently followed by A. Ocneanu himself in his
determination of the irreducible quantum symmetries $x$, also called
``irreducible connections'', associated with a given diagram.
However, if we do not want to use this {\it \/ a priori} knowledge,
there is another technique, which uses modular properties of the
diagram; this was one of the purposes of the article
\cite{CoqueGil:Tmodular}.

The  ${\mathcal A}$ series is always modular: one can define a
representation\footnote{Actually this representation
factors to a finite group.} of $SL(2,\ZZ)$ on the vector space of every diagram
of this class and the operator $T$ is diagonal on the vertices.
Take now $G$ some member of a generalized Dynkin-Coxeter system,
and call $A = {\cal A}(G)$ the corresponding member of the ${\mathcal
A}$ series (same Coxeter number or altitude).
Being a module over the algebra of $A$, there are
induction-restriction maps between $G$ and $A$.
These maps are described by the essential matrices $E_{a}$ or by
matrices $F_{i}$ (see section \ref{sec:EssMat} and 
\cite{Coque:Qtetra}, \cite{CoqueGil:ADE}).
One can try to define
an action of $SL(2,\ZZ)$ on the vector space of $G$ in a way that
should be compatible with those maps, but
this is not necessarily possible. In plain terms: suppose that the vertex
$\sigma$ of $G$ appears both in the branching
rules (restriction map from $A$ to $G$)
of vertices $\tau_p$ and $\tau_q$ of $A$; one could think of defining the
value of the modular generator $T$ on $\sigma$
       either as $T(\tau_p)$ or as $T(\tau_q)$, but
this is ambiguous, unless these two values are equal.
In general, there is only a subset $J$ of the vertices of $G$ for
which $T$ can be defined:
a vertex $\sigma$ will belong to this subset whenever $T$ is constant
along the vertices of $A$ whose restriction
to $G$  contains $\sigma$. The knowledge of this set $J$ allows one,
in the ``simple cases'', to determine
$Oc(G)$, the algebra of quantum symmetries of $G$: the set $J$
generates a particular subalgebra of $G$ and one finds $Oc(G) =
G \otimes_{J} G$.

     \paragraph{Results for the $ADE$ systems}: for diagrams of type
     ${\mathcal A}$, the subalgebra $J$ coincides with the algebra of
     the diagram itself, so that $Oc({\mathcal A})$ is isomorphic with
     ${\mathcal A}$.  For $E_{6}$, the subalgebra $J$, isomorphic with
     $A_{3}$ is generated by the three extremal points, and $Oc(E_{6})
     = E_{6} \otimes_{A_{3}} E_{6}$ has dimension $12$ (notice that
     $\kappa = 12$, as well, but this is an accident).  For $E_{8}$,
     the subalgebra $J$, isomorphic with $A_{2}$ is generated by the
     two extremal points of the long branches , and $Oc(E_{8}) = E_{8}
     \otimes_{A_{2}} E_{8}$ has dimension $32$ (notice that $\kappa =
     30$).  The other cases are more difficult to analyze: $Oc(E_{7}) =
     D_{10} \otimes_{\rho} D_{10}$, where the exceptional twist $\rho$
     can be determined from the modular properties (with respect to
     $T$) of the $A_{17}$ diagram; its dimension is $17$.  The algebra
     of quantum symmetries for a $D_{odd}$ diagram can be written as a
     quotient (using an identification map $\rho$) of the tensor square
     of the associated algebra of type ${\mathcal A}$ (for instance,
     $Oc(D_5) = A_7 \otimes_{\rho(A_7)} A_7$); the Ocneanu graph of
     $D_{2n+1}$ has $4n-1$ vertices.  In some respect, the
     determination of Ocneanu graphs for $D_{even}$ diagrams is more
     difficult; indeed, the algebra of quantum symmetries, in this
     case, is not commutative.  We sketch its construction because the
     result will be used later in our study of the twisted partition
     functions for the Potts model.  Starting from $D_{2n}$, one first
     obtains the induction-restriction rules with respect to the
     corresponding $A$ diagram with the same norm ($A_{4n-3}$) by
     calculating the essential matrices; from these rules and from the
     expression of the modular operator $T$ on $A_{4n-3}$, one
     determines the set $J$.  One finds that $Oc(D_{2n})$ consists of
     two separate components.  The first is given by
     $D_{2n}^{trunc}\otimes_{J^{'}} D_{2n}^{trunc}$, where
     $D_{2n}^{trunc}$ is the vector space corresponding to the
     subdiagram spanned by $\{\sigma_{0}, \sigma_{1}, \sigma_{2},
     \ldots \sigma_{2n-3} \}$, obtained by removing the fork, and $J' =
     \{\sigma_{0}, \sigma_{2}, \ldots \sigma_{2n-4} \}$ is the
     corresponding truncated subset\footnote{We choose the natural
     order to label vertices $\sigma_{a}$ of $D_{2n}$.} of $J$.  The
     second component is a non-commutative $2\times 2$ matrix algebra
     reflecting the indistinguishability of $\sigma_{2n-2}$ and
     $\sigma_{2n-2}^{'}$.  Ambichiral points are associated with the
     $n+1$ vertices of $J$ (\ie $n-1$ for the linear branch and $2$ for
     the fork); the Ocneanu graph of $D_{2n}$ has
     $\frac{(2n-2)(2n-2)}{n-1} + 4 = 4n$ vertices.

Results for the $SU(3)$ system: there is no complete treatment in the
available literature, but several examples have been worked out in
\cite{CoqueGil:Tmodular}.  Because we shall use it later (see section
\ref{sec:w3minimal}) in our study of twisted minimal models of type ${\mathcal
W}_{3}$, we just mention that the Ocneanu graph of the exceptional
${\mathcal E}_{5}$ diagram has $24$ points; both left and right chiral
subgraphs have $12$ points; the ambichiral subalgebra is of dimension
$6$ and the supplementary subspace has also dimension $6$.

\subsection{Characters for affine models}
\label{sec:CharacAff}

Strictly speaking, we do not need to use characters in this paper
since modular properties of the partition functions are to be
discussed in terms of commutation relations between the toric matrices
and the $S,T$ generators of $SL(2,\ZZ)$.  However, for completeness
sake, and for the reader who wants to check explicitly the results in
terms of invariance, or non invariance, with respect to transformations
$\tau \rightarrow -1/\tau $ and $\tau \rightarrow \tau +1$ (or $\tau
\rightarrow \tau +N$, for $T^{N}$), we remind the definitions of the
characters as functions of $\tau$, for affine models.  Here $\tau$ is
a point in the upper -- half plane and we set $q \doteq e^{2 i \pi
\tau}$.  These characters provide a basis of the vector space
$\CC^{n}$, for the defining representation (matrices $N_{i}$) of the
graph algebra of diagrams of type ${\mathcal A}$.  In the case of the
$SU(2)$ system, $k = \kappa - 2$ denotes the level, and for each
vertex $j = 0 \ldots k$ of a diagram $A_{k+1}={\mathcal A}_{k}$, we
set $r = j + 1 \equiv 2 \ell + 1$ and define

\[
\xi_{j}^{(k)} (q)=\frac{\sum\limits_{t=-\infty }^{\infty }\left(
2 \kappa t + r \right) q^{(2  \kappa t + r )^{2}/(4 \kappa)}}{\eta (\tau )}
\]

A closed form, for this expression, is
$$
\xi_{j}^{(k)} (q) = q^{\frac{{\left( 1 + j  \right) }^2}
           {4\,\left( 2 + k \right) }}\,\frac{
        \left( \left( 1 + j  \right) \,
           \left( 1 + {\theta}(3,
              \left( 1 + j  \right) \,\tau ,q^{2 + k})
             \right)  - \mathsf{i} \,\left( 2 + k \right) \,
           {\theta'}(3,
            \left( 1 + j  \right) \,\tau ,q^{2 + k})
          \right) }{{{\eta}(\tau )}^3}
          $$
          where  $\eta(\tau )$ is the Dedekind eta
          function,
         $\theta[3, u, v]$ is the third elliptic Jacobi theta function, and
              $\theta'[3, u, v]$ is its first derivative with respect to $u$.
More explicitly, these characters read

$$
\xi_{j}^{(k)} (q) = q^
          {-\left( \frac{1}{8} \right)  +
            \frac{{\left( 1 + j  \right) }^2}
             {4\,\left( 2 + k \right) }}\, \frac{
         \left( \sum_{t = -\infty }^{{+}\infty }
            \left( {j } + 1 +
               2\,t\,\left( k + 2 \right)  \right) \,
             {{q}}^
              {t\,\left( j  + 1 +
                  t\,\left( k + 2 \right)  \right) } \right) }
         {\sum_{t = -\infty }^{{+}\infty }
          \left( 1 + 4\,t \right) \,
           {{q}}^{t\,\left( 1 + 2\,t \right) }}
$$

When $\tau \rightarrow i \infty$,
then $\xi_{j}^{(k)} (q) \simeq (j + 1) q^{-\frac{1}{8} +
h}$, with
$h = \frac{(j + 1)^{2}}{4 \kappa}$. The power of $q$ is negative when 
$r = j + 1 <
\sqrt{\kappa /2}$. It is  often convenient to use expressions that are
valid in a neighborhood of infinity, for instance:

Graph $A_{1}$

$$\xi_{0}^{(0)} (q) = 1$$

Graph $A_{2}$
\begin{eqnarray*}
\xi_{0}^{(1)} (q)& = &{q^{-\frac{1}{24}}}(1 + 3\,q + 4\,q^2 + 7\,q^3
+ 13\,q^4 + 19\,q^5 +
        29\,q^6 +\ldots)
        \\
\xi_{1}^{(1)} (q)& = &{q^{\frac{5}{24}}}\,\left( 2 + 2\,q + 6\,q^2 + 8\,q^3 +
        14\,q^4 + 20\,q^5 + 34\,q^6 + \ldots \right)
         \end{eqnarray*}

Graph $A_{3}$
\begin{eqnarray*}
	 \xi_{0}^{(2)} (q)&=&{q^{\frac{- 1}{16}}}({1 + 3\,q + 9\,q^2 + 15\,q^3
+ 30\,q^4 + 54\,q^5 +
        94\,q^6 + \ldots})\\
\xi_{1}^{(2)} (q)&=&q^{\frac{1}{8}}\,\left( 2 + 6\,q + 12\,q^2 + 26\,q^3 +
        48\,q^4 + 84\,q^5 + 146\,q^6 +\ldots \right)\\
\xi_{2}^{(2)} (q)&=&q^{\frac{7}{16}}\,\left( 3 + 4\,q + 12\,q^2 + 21\,q^3 +
        43\,q^4 + 69\,q^5 + 123\,q^6 +\ldots \right)
    \end{eqnarray*}

$SU(3)$ characters have similar expressions, but indices $j$
refer then to a Young frame with two rows.
\noindent
Because of the two existing conventions, $(i,j)$ or $(r,s)$,
for the label of the origin ($0$ or $1$)
it is convenient to set $\chi_{1} \equiv \xi_{0}$, $\chi_{2} \equiv
\xi_{1}$ \, \etc :
$$\chi_{j + 1}^{(k)} \doteq \xi_{j}^{(k)}$$

\section{Torus structures for affine models}

\subsection{Example of an affine model: the $E_{6}$ case}
\label{section: example E6}

Toric matrices $W_{x0}$ have been determined for all $ADE$ cases and a
few others.  Since we shall need them later, we summarize the
situation for $E_{6}$.  We also present, in this case, several results
that were not available before: the full multiplication table of
$Oc(E6)$, the determination of the frustrated partition functions with
two twists, and a discussion of modular properties of these functions.
We also show how to display the expression of these partition
functions in a compact way, by using induction -- restriction rules for
the pair $(A_{11},E_{6})$.

\subsubsection{The $E_{6}$ diagram and its Ocneanu graph (summary)}

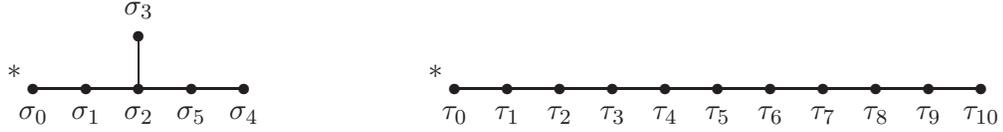
\begin{figure}[h]
\unitlength 0.7mm
\begin{center}
\begin{picture}(170,20)
\multiput(0,5)(10,0){5}{\circle*{2}}
\put(20,15){\circle*{2}}
\put(0,5){\line(1,0){40}}
\put(20,5){\line(0,1){10}}
\put(0,0){\makebox(0,0){$\sigma_0$}}
\put(10,0){\makebox(0,0){$\sigma_1$}}
\put(20,0){\makebox(0,0){$\sigma_2$}}
\put(30,0){\makebox(0,0){$\sigma_5$}}
\put(40,0){\makebox(0,0){$\sigma_4$}}
\put(20,20){\makebox(0,0){$\sigma_3$}}

\multiput(80,5)(10,0){11}{\circle*{2}}
\put(80,5){\line(1,0){100}}
\put(80,0){\makebox(0,0){$\tau_0$}}
\put(90,0){\makebox(0,0){$\tau_1$}}
\put(100,0){\makebox(0,0){$\tau_2$}}
\put(110,0){\makebox(0,0){$\tau_3$}}
\put(120,0){\makebox(0,0){$\tau_4$}}
\put(130,0){\makebox(0,0){$\tau_5$}}
\put(140,0){\makebox(0,0){$\tau_6$}}
\put(150,0){\makebox(0,0){$\tau_7$}}
\put(160,0){\makebox(0,0){$\tau_8$}}
\put(170,0){\makebox(0,0){$\tau_9$}}
\put(180,0){\makebox(0,0){$\tau_{10}$}}

\put(-5,7){$\ast$}
\put(75,7){$\ast$}

\end{picture}
\end{center}
\caption{The $E_6$ and $A_{11}$ Dynkin diagrams}
\label{Fig: E6labels}
\end{figure}

Figure \ref{Fig: E6labels} displays $E_{6}$ and the related diagram
$A_{11}$. Vertices $\sigma_{a}$ of $E_{6}$ are
labelled $0,1,2,5,4,3$ as shown on the picture.
$A_{11}$ acts on $E_{6}$, hence $ A_{11}$ also acts from the left and
from the right on the Ocneanu algebra\footnote{This tensor product is
taken above the subalgebra $A_{3}$ generated by vertices $0,4,3$, so
that $a \otimesdot ub = a u\otimesdot b$ when $u \in A_{3}$.} of
quantum symmetries which can be shown to be equal (\cite
{Coque:Qtetra}, \cite{CoqueGil:ADE}) to $Oc(E_{6}) =
E_{6} \otimes_{A_{3}} E_{6}$. It has dimension $12$.

\begin{figure}[hhh]
\unitlength 0.8mm
\par
\begin{center}
{\tiny {\ $W_{0} = \left(
\begin{array}{ccccccccccc}
1 & . & . & . & . & . & 1 & . & . & . & . \cr . & . & . & . & . & . & . & .
& . & . & . \cr . & . & . & . & . & . & . & . & . & . & . \cr . & . & . & 1
& . & . & . & 1 & . & . & . \cr . & . & . & . & 1 & . & . & . & . & . & 1
\cr . & . & . & . & . & . & . & . & . & . & . \cr 1 & . & . & . & . & . & 1
& . & . & . & . \cr . & . & . & 1 & . & . & . & 1 & . & . & . \cr . & . & .
& . & . & . & . & . & . & . & . \cr . & . & . & . & . & . & . & . & . & . &
.. \cr . & . & . & . & 1 & . & . & . & . & . & 1 \cr
\end{array}
\right) $} \qquad 
\begin{picture}(50,70)
\multiput(25,5)(0,10){3}{\circle*{2}}
\multiput(25,45)(0,10){3}{\circle{2}}
\multiput(5,25)(0,10){3}{\circle*{2}}
\multiput(45,25)(0,10){3}{\circle*{2}}

\thicklines
\put(5,45){\line(1,1){20}}
\put(5,35){\line(1,1){20}}
\put(5,25){\line(1,1){20}}
\put(5,25){\line(0,1){20}}

\thinlines
\put(45,45){\line(-1,-1){20}}
\put(45,35){\line(-1,-1){20}}
\put(45,25){\line(-1,-1){20}}
\put(25.3,5){\line(0,1){20}}

\thicklines
\dashline[50]{1}(45,45)(25,65)
\dashline[50]{1}(45,35)(25,55)
\dashline[50]{1}(45,25)(25,45)
\dashline[50]{1}(45,25)(45,45)

\thinlines
\dashline[50]{1}(5,45)(25,25)
\dashline[50]{1}(5,35)(25,15)
\dashline[50]{1}(5,25)(25,5)
\dashline[50]{1}(24.7,5)(24.7,25)

\small
\put(25,68){\makebox(0,0){\ud0}}
\put(25,58){\makebox(0,0){\ud3}}
\put(25,48){\makebox(0,0){\ud4}}
\put(1,45){\makebox(0,0){\ud1}}
\put(1,35){\makebox(0,0){\ud2}}
\put(1,25){\makebox(0,0){\ud5}}
\put(49,45){\makebox(0,0){$\ud{1^{'}}$}}
\put(49,35){\makebox(0,0){$\ud{31^{'}}$}}
\put(49,25){\makebox(0,0){$\ud{41^{'}}$}}
\put(21,25){\makebox(0,0){$\ud{11^{'}}$}}
\put(21,15){\makebox(0,0){$\ud{21^{'}}$}}
\put(21,5){\makebox(0,0){$\ud{51^{'}}$}}
\normalsize

\end{picture}
}
\end{center}
\caption{The $E_6$ Ocneanu graph and its modular invariant}
\label{Fig: Oc(E6)}
\end{figure}
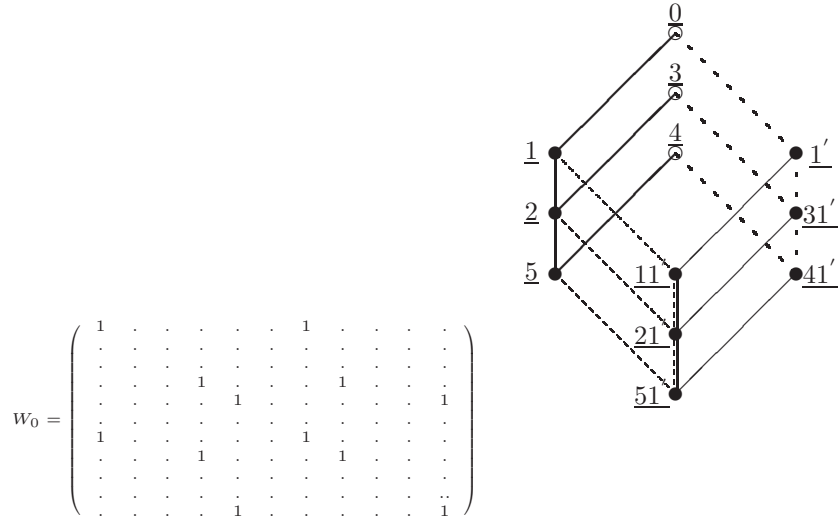

The bimodule structure of $Oc(E_{6})$ over $A_{11}$ is encoded by $12
\times 12 = 144 $ matrices $W_{xy}$ of dimension $ 11 \times 11$, (as
we shall see, many of them are equal).  In particular one obtains the
$12$ matrices $W_{x} \doteq W_{x0}$, one for each point of the Ocneanu
graph, and the matrix $W_{0} \doteq 0 \otimesdot_{A_{3}} 0$ associated
with the origin $\underline 0$.  Figure \ref{Fig: Oc(E6)} displays the
Ocneanu graph and the matrix $W_{0}$.  Continuous and dashed lines on
this graph describe respectively the multiplications by the left and
right chiral generators $\underline 1 = 1 \otimesdot 0$, $\underline
1' = 0 \otimesdot 1$.  We use the notations $\underline a = a
\otimesdot 0$, $\underline a' = 0 \otimesdot a$ and $\underline{ab'}
\equiv {\underline a} \, {\underline b'} \equiv a \otimesdot b$.
There are many identities hidden in this graph, like for instance
$\underline{31'}=\underline{2'}$; to see them, the reader should work
out for himself the multiplication table of the graph algebra of
$E_{6}$ or refer to references \cite{Coque:Qtetra} or
\cite{CoqueGil:ADE}.

\subsubsection{Induction - restriction mechanism and $Oc(G)$}

\begin{figure}[hhh]
\unitlength 0.7mm
\begin{center}
\begin{picture}(170,35)
\multiput(0,20)(10,0){5}{\circle*{2}}
\put(20,30){\circle*{2}}
\put(0,20){\line(1,0){40}}
\put(20,20){\line(0,1){10}}
\put(0,25){\makebox(0,0){$\star $}}
\put(0,15){\makebox(0,0){$\tau_{0}$}}
\put(0,10){\makebox(0,0){$\tau_{6}$}}
\put(10,15){\makebox(0,0){$\tau_{1}$}}
\put(10,10){\makebox(0,0){$\tau_{5}$}}
\put(10,5){\makebox(0,0){$\tau_{7}$}}
\put(20,15){\makebox(0,0){$\tau_{2}$}}
\put(20,10){\makebox(0,0){$\tau_{4}$}}
\put(20,5){\makebox(0,0){$\tau_{6}$}}
\put(20,0){\makebox(0,0){$\tau_{8}$}}
\put(30,15){\makebox(0,0){$\tau_{3}$}}
\put(30,10){\makebox(0,0){$\tau_{5}$}}
\put(30,5){\makebox(0,0){$\tau_{9}$}}
\put(40,15){\makebox(0,0){$\tau_{4}$}}
\put(40,10){\makebox(0,0){$\tau_{10}$}}
\put(20,35){\makebox(0,0){$\tau_{3},\tau_{7}$}}

\put(0,20){\circle{4}}
\put(20,30){\circle{4}}
\put(40,20){\circle{4}}

\multiput(80,20)(10,0){11}{\circle*{2}}
\put(80,20){\line(1,0){100}}
\put(80,15){\makebox(0,0){$\tau_0$}}
\put(90,15){\makebox(0,0){$\tau_1$}}
\put(100,15){\makebox(0,0){$\tau_2$}}
\put(110,15){\makebox(0,0){$\tau_3$}}
\put(120,15){\makebox(0,0){$\tau_4$}}
\put(130,15){\makebox(0,0){$\tau_5$}}
\put(140,15){\makebox(0,0){$\tau_6$}}
\put(150,15){\makebox(0,0){$\tau_7$}}
\put(160,15){\makebox(0,0){$\tau_8$}}
\put(170,15){\makebox(0,0){$\tau_9$}}
\put(180,15){\makebox(0,0){$\tau_{10}$}}

\put(70,6){\makebox(0,0){$\hat{T}:$}}

\put(80,5){\makebox(0,0){ 1}}
\put(90,5){\makebox(0,0){4}}
\put(100,5){\makebox(0,0){9}}
\put(110,5){\makebox(0,0){ 16}}
\put(120,5){\makebox(0,0){25}}
\put(130,5){\makebox(0,0){36}}
\put(140,5){\makebox(0,0){ 1}}
\put(150,5){\makebox(0,0){ 16}}
\put(160,5){\makebox(0,0){33}}
\put(170,5){\makebox(0,0){4}}
\put(180,5){\makebox(0,0){ 25}}

\end{picture}
\end{center}
\caption{The $E_6 \hookleftarrow A_{11}$ induction graph relative to
vertex $\sigma_{0}$, and the values
of $\hat T$ on irreps of $A_{11}$}
\label{fig: E6/A11induction}
\end{figure}
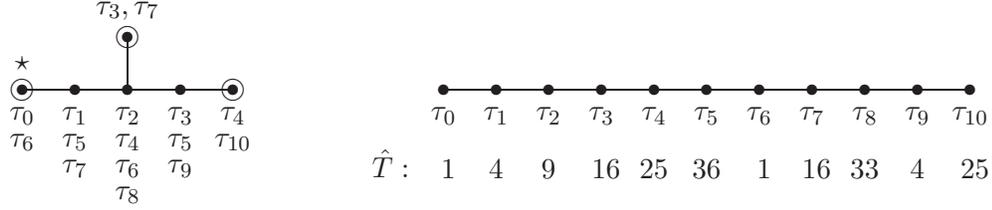

   From the diagram $E_{6}$ alone, we can determine the six essential
matrices $E_{a}$ of dimension $(11,6)$, as explained before.  Rows
of $E_{0}$ give the restriction (branching) rules $A_{11} \rightarrow
E_{6}$ and columns give the induction rules.  Induction rules are
displayed on Fig \ref{fig: E6/A11induction}.  We also give
the values of the modular exponent $\hat T$ for the vertices
$\tau_i$'s of $A_{11}$.

We notice that the value of the modular matrix $T$ on $\tau_{0}$ and
$\tau_{6}$ is the same (also for $\tau_{3}$ and $\tau_{7}$, and for
$\tau_{4}$ and $\tau_{10}$).  This allows one to assign a fixed value
of $T$ to three particular vertices of $E_6$.  For every other point
of the $E_6$ graph, the value of $T$ that would be inherited from the
$A_n$ graph by this induction mechanism is not uniquely determined.
These elements $\{\sigma_{0}, \sigma_{3}, \sigma_{4} \}$ span the
subalgebra $J$ isomorphic with the graph algebra of $A_3$; it admits
an invariant supplement in the graph algebra of $E_6$.  Using this
determination of $J$, as explained in section \ref{sec:Ocstruct} (or
\cite{CoqueGil:Tmodular}), the algebra $Oc(E_{6})$ is found to be
equal to $E_{6} \otimes_{A_{3}} E_{6}$.
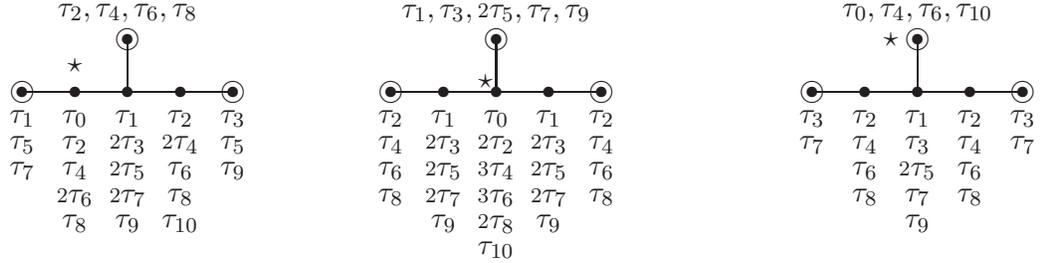
\begin{figure}[hhh]
\unitlength 0.7mm
\begin{center}
\begin{picture}(170,35)
\multiput(0,20)(10,0){5}{\circle*{2}}
\put(20,30){\circle*{2}}
\put(0,20){\line(1,0){40}}
\put(20,20){\line(0,1){10}}

\put(0,15){\makebox(0,0){$\tau_{1}$}}
\put(0,10){\makebox(0,0){$\tau_{5}$}}
\put(0,5){\makebox(0,0){$\tau_{7}$}}

\put(10,25){\makebox(0,0){$\star $}}
\put(10,15){\makebox(0,0){$\tau_{0}$}}
\put(10,10){\makebox(0,0){$\tau_{2}$}}
\put(10,5){\makebox(0,0){$\tau_{4}$}}
\put(10,0){\makebox(0,0){${\scriptstyle{2}} \tau_{6}$}}
\put(10,-5){\makebox(0,0){$\tau_{8}$}}

\put(20,15){\makebox(0,0){$\tau_{1}$}}
\put(20,10){\makebox(0,0){${\scriptstyle{2}}{\tau_{3}}$}}
\put(20,5){\makebox(0,0){${\scriptstyle{2}}{\tau_{5}}$}}
\put(20,0){\makebox(0,0){${\scriptstyle{2}}{\tau_{7}}$}}
\put(20,-5){\makebox(0,0){$\tau_{9}$}}

\put(30,15){\makebox(0,0){$\tau_{2}$}}
\put(30,10){\makebox(0,0){${\scriptstyle{2}}{\tau_{4}}$}}
\put(30,5){\makebox(0,0){$\tau_{6}$}}
\put(30,0){\makebox(0,0){$\tau_{8}$}}
\put(30,-5){\makebox(0,0){$\tau_{10}$}}

\put(40,15){\makebox(0,0){$\tau_{3}$}}
\put(40,10){\makebox(0,0){$\tau_{5}$}}
\put(40,5){\makebox(0,0){$\tau_{9}$}}

\put(20,35){\makebox(0,0){$\tau_{2},\tau_{4}, \tau_{6}, \tau_{8}$}}

\put(0,20){\circle{4}}
\put(20,30){\circle{4}}
\put(40,20){\circle{4}}
\multiput(70,20)(10,0){5}{\circle*{2}}
\put(90,30){\circle*{2}}
\put(70,20){\line(1,0){40}}
\put(90,20){\line(0,1){10}}

\put(70,15){\makebox(0,0){$\tau_{2}$}}
\put(70,10){\makebox(0,0){$\tau_{4}$}}
\put(70,5){\makebox(0,0){$\tau_{6}$}}
\put(70,0){\makebox(0,0){$\tau_{8}$}}

\put(80,15){\makebox(0,0){$\tau_{1}$}}
\put(80,10){\makebox(0,0){${\scriptstyle{2}}\tau_{3}$}}
\put(80,5){\makebox(0,0){${\scriptstyle{2}}\tau_{5}$}}
\put(80,0){\makebox(0,0){${\scriptstyle{2}} \tau_{7}$}}
\put(80,-5){\makebox(0,0){$\tau_{9}$}}

\put(88,22){\makebox(0,0){$\star $}}

\put(90,15){\makebox(0,0){$\tau_{0}$}}
\put(90,10){\makebox(0,0){${\scriptstyle{2}}{\tau_{2}}$}}
\put(90,5){\makebox(0,0){${\scriptstyle{3}}{\tau_{4}}$}}
\put(90,0){\makebox(0,0){${\scriptstyle{3}}{\tau_{6}}$}}
\put(90,-5){\makebox(0,0){${\scriptstyle{2}}\tau_{8}$}}
\put(90,-10){\makebox(0,0){$\tau_{10}$}}

\put(100,15){\makebox(0,0){$\tau_{1}$}}
\put(100,10){\makebox(0,0){${\scriptstyle{2}}\tau_{3}$}}
\put(100,5){\makebox(0,0){${\scriptstyle{2}}\tau_{5}$}}
\put(100,0){\makebox(0,0){${\scriptstyle{2}} \tau_{7}$}}
\put(100,-5){\makebox(0,0){$\tau_{9}$}}

\put(110,15){\makebox(0,0){$\tau_{2}$}}
\put(110,10){\makebox(0,0){$\tau_{4}$}}
\put(110,5){\makebox(0,0){$\tau_{6}$}}
\put(110,0){\makebox(0,0){$\tau_{8}$}}

\put(90,35){\makebox(0,0){$\tau_{1},\tau_{3}, {\scriptstyle{2}} 
\tau_{5}, \tau_{7},
\tau_{9}$}}

\put(70,20){\circle{4}}
\put(90,30){\circle{4}}
\put(110,20){\circle{4}}
\multiput(150,20)(10,0){5}{\circle*{2}}
\put(170,30){\circle*{2}}
\put(150,20){\line(1,0){40}}
\put(170,20){\line(0,1){10}}

\put(150,15){\makebox(0,0){$\tau_{3}$}}
\put(150,10){\makebox(0,0){$\tau_{7}$}}

\put(160,15){\makebox(0,0){$\tau_{2}$}}
\put(160,10){\makebox(0,0){$\tau_{4}$}}
\put(160, 5){\makebox(0,0){$\tau_{6}$}}
\put(160, 0){\makebox(0,0){$\tau_{8}$}}

\put(170,15){\makebox(0,0){$\tau_{1}$}}
\put(170,10){\makebox(0,0){$\tau_{3}$}}
\put(170,5){\makebox(0,0){${\scriptstyle{2}} \tau_{5}$}}
\put(170,0){\makebox(0,0){$\tau_{7}$}}
\put(170,-5){\makebox(0,0){$\tau_{9}$}}

\put(180,15){\makebox(0,0){$\tau_{2}$}}
\put(180,10){\makebox(0,0){$\tau_{4}$}}
\put(180,5){\makebox(0,0){$\tau_{6}$}}
\put(180,0){\makebox(0,0){$\tau_{8}$}}

\put(190,15){\makebox(0,0){$\tau_{3}$}}
\put(190,10){\makebox(0,0){$\tau_{7}$}}

\put(165,30){\makebox(0,0){$\star $}}

\put(170,35){\makebox(0,0)
{$\tau_{0},\tau_{4}, \tau_{6}, \tau_{10}$}}

\put(150,20){\circle{4}}
\put(170,30){\circle{4}}
\put(190,20){\circle{4}}

\end{picture}
\end{center}
\caption{The $E_6 \hookleftarrow A_{11}$ induction graphs relative to
vertices $\sigma_{1}$, $\sigma_{2}$ and $\sigma_{3}$ }
\label{fig:E6/A11 inductionfromsigma1}
\end{figure}

\subsubsection{Linear and quadratic sum rules}

Dimensions of the $11$ blocks $d_{i}$ are equal to
$(6,10,14,18,20,20,20,18,14,10,6)$.
Dimension of the $12$ blocks $d_{x}$ are equal to
$(6,8,6,10,14,10,10,14,10,20,28,20)$.
The quadratic sum rule reads: $\sum_{i} d_{i}^{2} = \sum_{x} d_{x}^{2} = 2512.$
The linear sum rule also holds: $\sum_{i} d_{i} = \sum_{x} d_{x} = 156.$

There are also quantum sum rules (mass relations): define $o(G)
\doteq \sum_{a \in G} qdim_{a}^{2}$, where $qdim_{a}$ are the quantum
dimensions of the vertices $a$ of $G$ (for example $o(E_{6})=4(3+\sqrt
3)$, $o(A_{11})=24 (2 + \sqrt 3)$, $o(A_{3}) = 1 + (\sqrt 2))^{2} =
4)$; then, if $G$ is a module
over ${\cal A}_{k}$ (for some $k$) and when $Oc(G) = G \otimes_{J}
G$, one can check that $o(Oc(G))$ defined as $\frac{o(G) \times
o(G)}{o(J)}$ is equal to $o({\cal A}_{k})$, for instance $\frac{o(E_{6})
\times o(E_{6})}{ o(A_{3})} = o(A_{11})$; we do not know any general formal
proof of these quantum relations.

\subsubsection{Toric matrices $W_{x \underline 0}$ and frustrated
functions with one twist (results)}

The toric matrices $W_{x \underline 0}$ calculated as explained in
section \ref{sec:toricmat} were explicitly listed in
\cite{Coque:Qtetra} and the corresponding partition functions $Z_{x
\underline 0}$ also appear in \cite{CoqueGil:ADE}.  We recall the
results\footnote{For $x = a \otimesdot b$, we simply call $W_{ab}
\doteq W_{ a \otimesdot b, \underline 0}$.} :

{\small
$$
\begin{array}{cc}

W_{00} & W_{11}  \\
{} & {} \\


\left( \begin{array}{ccccccccccc}
      1 & . & . & . & . & . & 1 & . &
      . & . & . \cr . & . & . & . &
      . & . & . & . & . & . & . \cr
      . & . & . & . & . & . & . & . &
      . & . & . \cr . & . & . & 1 &
      . & . & . & 1 & . & . & . \cr
      . & . & . & . & 1 & . & . & . &
      . & . & 1 \cr . & . & . & . &
      . & . & . & . & . & . & . \cr
      1 & . & . & . & . & . & 1 & . &
      . & . & . \cr . & . & . & 1 &
      . & . & . & 1 & . & . & . \cr
      . & . & . & . & . & . & . & . &
      . & . & . \cr . & . & . & . &
      . & . & . & . & . & . & . \cr
      . & . & . & . & 1 & . & . & . &
      . & . & 1 \cr
    \end{array} \right)

&

\left( \begin{array}{ccccccccccc}
    . & . & . & . & . & . & . & . &
      . & . & . \cr . & 1 & . & . &
      . & 1 & . & 1 & . & . & . \cr
      . & . & 1 & . & 1 & . & 1 & . &
      1 & . & . \cr . & . & . & 1 &
      . & 1 & . & . & . & 1 & . \cr
      . & . & 1 & . & 1 & . & 1 & . &
      1 & . & . \cr . & 1 & . & 1 &
      . & 2 & . & 1 & . & 1 & . \cr
      . & . & 1 & . & 1 & . & 1 & . &
      1 & . & . \cr . & 1 & . & . &
      . & 1 & . & 1 & . & . & . \cr
      . & . & 1 & . & 1 & . & 1 & . &
      1 & . & . \cr . & . & . & 1 &
      . & 1 & . & . & . & 1 & . \cr
      . & . & . & . & . & . & . & . &
      . & . & . \cr
    \end{array} \right)

\\

{} & {} \\

\end{array}
$$
}

{\small

$$
\begin{array}{cc}

{} & {} \\
W_{30} & W_{21}  \\
{} & {} \\

    \left( \begin{array}{ccccccccccc}
      . & . & . & 1 & . & . & . & 1 &
      . & . & . \cr . & . & . & . &
      . & . & . & . & . & . & . \cr
      . & . & . & . & . & . & . & . &
      . & . & . \cr 1 & . & . & . &
      1 & . & 1 & . & . & . & 1 \cr
      . & . & . & 1 & . & . & . & 1 &
      . & . & . \cr . & . & . & . &
      . & . & . & . & . & . & . \cr
      . & . & . & 1 & . & . & . & 1 &
      . & . & . \cr 1 & . & . & . &
      1 & . & 1 & . & . & . & 1 \cr
      . & . & . & . & . & . & . & . &
      . & . & . \cr . & . & . & . &
      . & . & . & . & . & . & . \cr
      . & . & . & 1 & . & . & . & 1 &
      . & . & . \cr
    \end{array} \right)

&

\left( \begin{array}{ccccccccccc}
    . & . & . & . & . & . & . & . &
      . & . & . \cr . & . & 1 & . &
      1 & . & 1 & . & 1 & . & . \cr
      . & 1 & . & 1 & . & 2 & . & 1 &
      . & 1 & . \cr . & . & 1 & . &
      1 & . & 1 & . & 1 & . & . \cr
      . & 1 & . & 1 & . & 2 & . & 1 &
      . & 1 & . \cr . & . & 2 & . &
      2 & . & 2 & . & 2 & . & . \cr
      . & 1 & . & 1 & . & 2 & . & 1 &
      . & 1 & . \cr . & . & 1 & . &
      1 & . & 1 & . & 1 & . & . \cr
      . & 1 & . & 1 & . & 2 & . & 1 &
      . & 1 & . \cr . & . & 1 & . &
      1 & . & 1 & . & 1 & . & . \cr
      . & . & . & . & . & . & . & . &
      . & . & . \cr
    \end{array} \right)

\\
{} & {} \\
{} & {} \\
W_{40} & W_{51}  \\
{} & {} \\

\left( \begin{array}{ccccccccccc}
      . & . & . & . & 1 & . & . & . &
      . & . & 1 \cr . & . & . & . &
      . & . & . & . & . & . & . \cr
      . & . & . & . & . & . & . & . &
      . & . & . \cr . & . & . & 1 &
      . & . & . & 1 & . & . & . \cr
      1 & . & . & . & . & . & 1 & . &
      . & . & . \cr . & . & . & . &
      . & . & . & . & . & . & . \cr
      . & . & . & . & 1 & . & . & . &
      . & . & 1 \cr . & . & . & 1 &
      . & . & . & 1 & . & . & . \cr
      . & . & . & . & . & . & . & . &
      . & . & . \cr . & . & . & . &
      . & . & . & . & . & . & . \cr
      1 & . & . & . & . & . & 1 & . &
      . & . & . \cr
    \end{array} \right)

&

\left( \begin{array}{ccccccccccc}
      . & . & . & . & . & . & . & . &
      . & . & . \cr . & . & . & 1 &
      . & 1 & . & . & . & 1 & . \cr
      . & . & 1 & . & 1 & . & 1 & . &
      1 & . & . \cr . & 1 & . & . &
      . & 1 & . & 1 & . & . & . \cr
      . & . & 1 & . & 1 & . & 1 & . &
      1 & . & . \cr . & 1 & . & 1 &
      . & 2 & . & 1 & . & 1 & . \cr
      . & . & 1 & . & 1 & . & 1 & . &
      1 & . & . \cr . & . & . & 1 &
      . & 1 & . & . & . & 1 & . \cr
      . & . & 1 & . & 1 & . & 1 & . &
      1 & . & . \cr . & 1 & . & . &
      . & 1 & . & 1 & . & . & . \cr
      . & . & . & . & . & . & . & . &
      . & . & . \cr
    \end{array} \right)

\end{array}
$$
}

{\small

$$
\begin{array}{cc}

W_{10} & W_{01}  \\
{} & {} \\

\left( \begin{array}{ccccccccccc}
      . & . & . & . & . & . & . & . &
      . & . & . \cr 1 & . & . & . &
      . & . & 1 & . & . & . & . \cr
      . & . & . & 1 & . & . & . & 1 &
      . & . & . \cr . & . & . & . &
      1 & . & . & . & . & . & 1 \cr
      . & . & . & 1 & . & . & . & 1 &
      . & . & . \cr 1 & . & . & . &
      1 & . & 1 & . & . & . & 1 \cr
      . & . & . & 1 & . & . & . & 1 &
      . & . & . \cr 1 & . & . & . &
      . & . & 1 & . & . & . & . \cr
      . & . & . & 1 & . & . & . & 1 &
      . & . & . \cr . & . & . & . &
      1 & . & . & . & . & . & 1 \cr
      . & . & . & . & . & . & . & . &
      . & . & . \cr
    \end{array} \right)

&

\left( \begin{array}{ccccccccccc}
    . & 1 & . & . & . & 1 & . & 1 &
      . & . & . \cr . & . & . & . &
      . & . & . & . & . & . & . \cr
      . & . & . & . & . & . & . & . &
      . & . & . \cr . & . & 1 & . &
      1 & . & 1 & . & 1 & . & . \cr
      . & . & . & 1 & . & 1 & . & . &
      . & 1 & . \cr . & . & . & . &
      . & . & . & . & . & . & . \cr
      . & 1 & . & . & . & 1 & . & 1 &
      . & . & . \cr . & . & 1 & . &
      1 & . & 1 & . & 1 & . & . \cr
      . & . & . & . & . & . & . & . &
      . & . & . \cr . & . & . & . &
      . & . & . & . & . & . & . \cr
      . & . & . & 1 & . & 1 & . & . &
      . & 1 & . \cr
    \end{array} \right)

\\
{} & {} \\

\end{array}
$$
}

{\small

$$
\begin{array}{cc}

W_{20} & W_{02} = W_{31}  \\
{} & {} \\

\left( \begin{array}{ccccccccccc}
     . & . & . & . & . & . & . & . &
      . & . & . \cr . & . & . & 1 &
      . & . & . & 1 & . & . & . \cr
      1 & . & . & . & 1 & . & 1 & . &
      . & . & 1 \cr . & . & . & 1 &
      . & . & . & 1 & . & . & . \cr
      1 & . & . & . & 1 & . & 1 & . &
      . & . & 1 \cr . & . & . & 2 &
      . & . & . & 2 & . & . & . \cr
      1 & . & . & . & 1 & . & 1 & . &
      . & . & 1 \cr . & . & . & 1 &
      . & . & . & 1 & . & . & . \cr
      1 & . & . & . & 1 & . & 1 & . &
      . & . & 1 \cr . & . & . & 1 &
      . & . & . & 1 & . & . & . \cr
      . & . & . & . & . & . & . & . &
      . & . & . \cr
    \end{array} \right)

&

\left( \begin{array}{ccccccccccc}
    . & . & 1 & . & 1 & . & 1 & . &
      1 & . & . \cr . & . & . & . &
      . & . & . & . & . & . & . \cr
      . & . & . & . & . & . & . & . &
      . & . & . \cr . & 1 & . & 1 &
      . & 2 & . & 1 & . & 1 & . \cr
      . & . & 1 & . & 1 & . & 1 & . &
      1 & . & . \cr . & . & . & . &
      . & . & . & . & . & . & . \cr
      . & . & 1 & . & 1 & . & 1 & . &
      1 & . & . \cr . & 1 & . & 1 &
      . & 2 & . & 1 & . & 1 & . \cr
      . & . & . & . & . & . & . & . &
      . & . & . \cr . & . & . & . &
      . & . & . & . & . & . & . \cr
      . & . & 1 & . & 1 & . & 1 & . &
      1 & . & . \cr
    \end{array} \right)

\\

{} & {} \\
{} & {} \\
W_{50} & W_{05} = W_{41}  \\
{} & {} \\

\left( \begin{array}{ccccccccccc}
      . & . & . & . & . & . & . & . &
      . & . & . \cr . & . & . & . &
      1 & . & . & . & . & . & 1 \cr
      . & . & . & 1 & . & . & . & 1 &
      . & . & . \cr 1 & . & . & . &
      . & . & 1 & . & . & . & . \cr
      . & . & . & 1 & . & . & . & 1 &
      . & . & . \cr 1 & . & . & . &
      1 & . & 1 & . & . & . & 1 \cr
      . & . & . & 1 & . & . & . & 1 &
      . & . & . \cr . & . & . & . &
      1 & . & . & . & . & . & 1 \cr
      . & . & . & 1 & . & . & . & 1 &
      . & . & . \cr 1 & . & . & . &
      . & . & 1 & . & . & . & . \cr
      . & . & . & . & . & . & . & . &
      . & . & . \cr
    \end{array} \right)

&

\left( \begin{array}{ccccccccccc}
      . & . & . & 1 & . & 1 & . & . &
      . & 1 & . \cr . & . & . & . &
      . & . & . & . & . & . & . \cr
      . & . & . & . & . & . & . & . &
      . & . & . \cr . & . & 1 & . &
      1 & . & 1 & . & 1 & . & . \cr
      . & 1 & . & . & . & 1 & . & 1 &
      . & . & . \cr . & . & . & . &
      . & . & . & . & . & . & . \cr
      . & . & . & 1 & . & 1 & . & . &
      . & 1 & . \cr . & . & 1 & . &
      1 & . & 1 & . & 1 & . & . \cr
      . & . & . & . & . & . & . & . &
      . & . & . \cr . & . & . & . &
      . & . & . & . & . & . & . \cr
      . & 1 & . & . & . & 1 & . & 1 &
      . & . & . \cr
    \end{array} \right)

\end{array}
$$
}

We shall come back to these toric matrices with a single twist at the
end of the next section and write the corresponding partition
functions in a compact way.

\subsubsection{Toric matrices $W_{xy}$ from induction graphs}
\label{sec:algo1}
Here we give a first algorithm allowing a simple determination of all
the $W_{xy}$.  For simplicity, we choose to carry this discussion in
the case of the $E_{6}$ graph.  In the case of toric matrices with a
single twist, this algorithm was described in \cite{Coque:Qtetra} and
\cite{CoqueGil:ADE}, it is described here in the case of arbitrary
toric matrices (two twists).  It therefore generalizes the method of
the previous section and uses the data given by essential matrices
($E_{6}/A_{11}$ induction rules). Another algorithm for the
determination of the $W_{xy}$, using the multiplication table of the
algebra of quantum symmetries, will be described later.

Call ${\mathcal V}_{034}[a]$ the $(11,3)$ rectangular matrix
describing the $E_{6} \leftarrow A_{11}$ induction graph relative to
the vertex $\sigma_{a}$ and restricted to vertices $\sigma_{0},
\sigma_{3}, \sigma_{4}$ of $E_{6}$ (spanning the subalgebra $J$
isomorphic with $A_{3}$).  Call ${\mathcal V}_{125}[a]$ the analogous
$(11,3)$ matrix relative to the same vertex $\sigma_{a}$ but obtained
by restriction to vertices $\sigma_{1}, \sigma_{2}, \sigma_{5}$
(spanning a supplement of $J$).  Both matrices (and induction graphs)
can be obtained from the $(11,6)$ essential matrix $E_{a}$ by keeping
only the columns labelled by $0,3,4$ (respectively those labelled by
$1,2,5$).  The induction graph relative to vertex $\sigma_{0}$ was
given on figure \ref{fig: E6/A11induction}; we also give the induction
graph relative to vertices $\sigma_{1}$, $\sigma_{2}$ and $\sigma_{3}$
on figure \ref{fig:E6/A11 inductionfromsigma1}; graphs relative
to $\sigma_{5}$ and $\sigma_{4}$ are obtained from those relative to
$\sigma_{1}$ and $\sigma_{0}$ by $\ZZ_{2}$ symmetry.

We need to use the three graph matrices of $A_{3}$, obviously
given by
$$
\begin{array}{ccc}
W_{0}({A_{3}})=\left(
\begin{array}{lll}
1 & 0 & 0 \\
0 & 1 & 0 \\
0 & 0 & 1
\end{array}
\right) \;, & \;W_{1}({A_{3}})=\left(
\begin{array}{lll}
0 & 1 & 0 \\
1 & 0 & 1 \\
0 & 1 & 0
\end{array}
\right) \;, & \;W_{2}({A_{3}})=\left(
\begin{array}{lll}
0 & 0 & 1 \\
0 & 1 & 0 \\
1 & 0 & 0
\end{array}
\right)
\end{array}$$
Since $A_{3}$ is a member of the $A$
series, graph matrices, essential matrices and toric matrices of type
$W_{x0}$ are equal.
Remember that, in the isomorphism $J \simeq A_{3}$, indices $0,1,2$ 
of $A_{3}$ are associated with indices
$0,3,4$ of $E_{6}$. Toric matrices of $E_{6}$ are of dimension 
$(11,11)$; they can be written
      as products of matrices of dimension
      $(11,3)(3,3)(3,11)$ where the $(3,3)$ matrices are the
      toric matrices of $A_{3}$ and where the rectangular matrices of
      dimensions $(11,3)$ or $(3,11)$ give the induction/restriction
      rules from $A_{11}$ to $A_{3}$.

      \smallskip

The set of toric matrices $W_{xy}$ with twists $x = a \otimesdot
b$ and $y = c \otimesdot d$, written $W_{ab,cd}$ is
$$
\{W_{x,cd}\}_{x \in Oc(E_{6})} =
\left\{
\begin{array}{ccc}

\begin{array}{ccc}
       {} & {\mathcal V}_{034}[c] W_{i}(A_{3})
{\mathcal V}_{034}^T [d] & {}
\end{array}
\\
\begin{array}{ccc}
       {\mathcal V}_{125}[c] W_{i}(A_{3})
{\mathcal V}_{034}^T [d]  & , & {\mathcal V}_{034}[c] W_{i}(A_{3})
{\mathcal V}_{125}^T [d]
\end{array}
\\
\begin{array}{ccc}
       {} & {\mathcal V}_{125}[c] W_{i}(A_{3})
{\mathcal V}_{125}^T [d] & {}
\end{array}
\end{array}
\right\}_{i=1,2,3}
$$

The above table exhibits, on purpose, a one to one correspondence with
the drawing of the Ocneanu graph (figure \ref{Fig: Oc(E6)}), with ambichiral
generators on the first line, left and right chiral generators on the
second line and supplementary generators on the third line.  Moreover
the correspondence with $i$ indices of $A_{3}$ runs from top to bottom
on each vertical of figure \ref{Fig: Oc(E6)}.  For instance, $x = \underline {2
1'} = 2 \otimesdot 1$, is the second supplementary generator, so that
a matrix $W_{21,cd}$ is equal to ${\mathcal V}_{125}[c] W_{1}(A_{3})
{\mathcal V}_{125}^T [d]$.

Introducing the ``adapted vectors'' $ w[a] \doteq {\mathcal
V}_{034}[a] .  \chi$ and $v[a] \doteq {\mathcal V}_{125}[a] .  \chi$ ,
where $ \chi$ is the vector of characters\footnote{We denote the
eleven characters of $A_{11}$ by $\chi_{j + 1} \equiv \xi_{j}$, with
$j + 1 = 1 \ldots 11$, dropping the upper index $k$ which is always
equal to $10$ in this case.}, we write the partition functions
$Z_{xy}$ associated with matrices $W_{xy}$ as
$$
\{Z_{x,cd}\}_{x \in Oc(E_{6})} =
\left\{
\begin{array}{ccc}

\begin{array}{ccc}
       {} & (\overline \chi . w[c]) W_{i}(A_{3})
(w[d]^{T} . \chi) & {}
\end{array}
\\
\begin{array}{ccc}
      (\overline \chi . v[c]) W_{i}(A_{3})
(w[d]^{T} .  \chi) & , & (\overline \chi . w[c]) W_{i}(A_{3})
(v[d]^{T} . \chi)
\end{array}
\\
\begin{array}{ccc}
       {} &  (\overline \chi . v[c]) W_{i}(A_{3})
(v[d]^{T} . \chi) & {}
\end{array}
\end{array}
\right\}_{i=1,2,3}
$$
For instance,
$Z_{21,cd} = (\overline \chi . v[c]) W_{1}(A_{3}) (v[d]^{T} . \chi) $,
and,
$Z_{21} \doteq Z_{21,00} = (\overline \chi . v[0]) W_{1}(A_{3}) 
(v[0]^{T} . \chi) $.

Altogether, we have six adapted vectors $v[a]$ and six adapted vectors
$w[a]$, all of them have three components\footnote{When studying the $E_{8}$
graph and its induction pattern relative to $A_{29}$, $J$ happens to
be two - dimensional, so the $w[a]$ will have two components and the
$v[a]$ will have six.}.  The use of these two adapted vectors $w[a]$
and $v[a]$ associated with induction rules for the vertex $a$ allows
one to write all the results for $W_{xy}$ (or $Z_{xy}$) in a very
compact way.  Let us now rewrite the partition functions with one
twist, already obtained in the last section (matrices $W_{x}$) in
terms of these adapted vectors.

\textbf{Adapted vectors for the vertex $0$} (see figure \ref{fig: 
E6/A11induction}):

$$
\begin{array}{ccc}
w_{1}\doteq w_{1}[0] =\chi_{1}+\chi_{7}& , & v_{1}\doteq v_{1}[0] 
=\chi_{2}+\chi_{6}+\chi_{8} \\
w_{2}\doteq w_{2}[0] =\chi_{4}+\chi_{8}& , & v_{2}\doteq v_{2}[0] 
=\chi_{3}+\chi_{5}+\chi_{7}+\chi_{9} \\
w_{3}\doteq w_{3}[0] =\chi_{11}+\chi_{5}& , & v_{3}\doteq v_{3}[0] 
=\chi_{4}+\chi_{6}+\chi_{10}
\end{array}
$$

With $v = v[0] = \{v_{1},v_{2}, v_{3}\}$ and  $w = w[0] =
\{w_{1},w_{2}, w_{3}\}$, the twisted partition functions of type $Z_{x0}$ read
$$
\begin{array}{ccc}

\begin{array}{ccc}
       {} & Z_{00} = \overline{{w}} W_{0}(A_{3})
{w} & {} \\
       {} & Z_{30} = \overline{{w}} W_{1}(A_{3})
{w} {} \\
       {} & Z_{40} = \overline{{w}} W_{2}(A_{3})
{w} & {}
\end{array}
\\
\begin{array}{ccc}
       Z_{10} = \overline{{w}} W_{0}(A_{3})  {
       v} & {}  &  Z_{01} = \overline{{v}} ({
W}_{0}(A_{3})  {w}  \\
       Z_{20} = \overline{{w}} W_{1}(A_{3})  {
       v} & {} & Z_{31} = \overline{{v}} ({
W}_{1}(A_{3})  {w} \\
       Z_{50} = \overline{{w}} W_{2}(A_{3})  {
      v} & {}  & Z_{41} = \overline{{v}} ({
W}_{2}(A_{3})  {w}
\end{array}
\\
\begin{array}{ccc}
       {} & Z_{11} = \overline{{v}} W_{0}(A_{3})  {
       v} & {} \\
       {} & Z_{21} = \overline{{v}} W_{1}(A_{3})  {
       v} {} \\
       {} & Z_{51} = \overline{{v}} W_{2}(A_{3})  {
      v} & {}
\end{array}
\end{array}
      $$

      The first entry, $Z_{00}$, is the usual (modular invariant) partition
      function.      \label{sec:E6partfun}
      Explicitly, we rewrite the 12 partition functions $Z_{ab} \doteq
Z_{ab,00}$ (ambichiral, left and right chiral, and  supplementary)
in terms of these six linear combinations of characters $v$ and $w$ as follows:
$$
\begin{array}{ccc}
\begin{array}{ccc}
       {} & Z_{00}(q) =\left| w_{1}\right| ^{2}+\left| w_{2}\right| ^{2}+\left|
w_{3}\right| ^{2}
  & {} \\
       {} & Z_{30}(q) =\left( \overline{w_{1}}+\overline{w_{3}}\right) w_{2}+
\overline{w_{2}} \left(
w_{1}+w_{3}\right)
  & {} \\
       {} & Z_{40}(q) 
=\overline{w_{1}}w_{3}+\overline{w_{3}}w_{1}+\left| w_{2}\right|
^{2}  & {} \\
{} & {} & {}
\end{array}
\\
\begin{array}{ccc}
      Z_{10}(q) =\overline{v_{3}}w_{3}+\overline{v_{1}}w_{1}+\overline{v_{2}}
w_{2} & {} &  Z_{01} = \overline Z_{10} \\
  Z_{20}(q) =\left( \overline{v_{1}}+\overline{v_{3}}\right) w_{2}+
\overline{v_{2}} \left(
w_{1}+w_{3}\right) & {} & Z_{02} =  \overline Z_{20}\\
Z_{50}(q) =\overline{v_{3}}w_{1}+\overline{v_{1}}w_{3}+\overline{v_{2}}
w_{2} & {}  & Z_{05} =  \overline Z_{50}
\end{array}
\\
\begin{array}{ccc}
     {} & {} & {} \\
       {} &Z_{11}(q) =\left| v_{1}\right| ^{2}+\left| v_{2}\right| ^{2}+\left|
v_{3}\right| ^{2} & {} \\
       {} &Z_{21}(q) =\left( \overline{v_{1}}+\overline{v_{3}}\right) v_{2}+
\overline{v_{2}} \left(
v_{1}+v_{3}\right) & {} \\
       {} &Z_{51}(q) 
=\overline{v_{1}}v_{3}+\overline{v_{3}}v_{1}+\left| v_{2}\right|
^{2} & {}
\end{array}
\end{array}
      $$

\smallskip

The table of twisted partition functions $Z_{xy}$ for all ADE
models appearing at
the end of reference \cite{CoqueGil:Tmodular} could be greatly simplified by
using this compact reformulation.

\subsubsection{The multiplication table for $Oc(E_{6})$}

As we know, the Ocneanu graph encodes the result of multiplication of
basis elements of $Oc(G)$ by the two chiral left and right
generators.  Determination of the full table of multiplication of
$Oc(G)$ can then be obtained in a straightforward manner.  It is given
below\footnote{Part of this multiplication table was obtained by G.
Schieber.} for the algebra $Oc(E_{6})$.
\landscape
\begin{table}[hhh]
\tiny
$$
\begin{array}{|c||c|c|c|c|c|c|c|c|c|c|c|c|}
\hline
{}& & & & & & & & & & & & \\
{} & \ud0 & \ud1 & \ud2 & \ud3 & \ud4 & \ud5 & \ud{1^{'}} &
\ud{11^{'}} & \ud{21^{'}} & \ud{31^{'}} & \ud{41^{'}} & \ud{51^{'}} \\
{}& & & & & & & & & & & & \\
\hline
\hline
{}& & & & & & & & & & & & \\
\ud0 & \ud0 & \ud1 & \ud2 & \ud3 & \ud4 & \ud5 & \ud{1^{'}} &
\ud{11^{'}} &
\ud{21^{'}} & \ud{31^{'}} & \ud{41^{'}} & \ud{51^{'}} \\
{}& & & & & & & & & & & & \\
\hline
{}& & & & & & & & & & & & \\
\ud1 & \ud1 & \ud0+\ud2 & \ud1+\ud3+\ud5 & \ud2 & \ud5 & \ud2+\ud4 &
\ud{11^{'}} & \ud{1^{'}} + \ud{21^{'}} &
\ud{11^{'}}+\ud{31^{'}}+\ud{51^{'}} & \ud{21^{'}} & \ud{51^{'}} &
\ud{21^{'}}+\ud{41^{'}}\\
{}& & & & & & & & & & & & \\
\hline
{}& & & & & & & & & & & & \\
\ud2 & \ud2
&\ud1+\ud3+\ud5&\ud0+\ud2+\ud2+\ud4&\ud1+\ud5&\ud2&\ud1+\ud3+\ud5&\ud{21^{'}}&\ud{11^{'}}+\ud{31^{'}}+\ud{51^{'}}&\ud{1^{'}}+\ud{21^{'}}+\ud{21^{'}}+\ud{41^{'}}&\ud{11^{'}}+\ud{51^{'}}&\ud{21^{'}}&\ud{11^{'}}+

\ud{31^{'}}+\ud{51^{'}} \\
{}& & & & & & & & & & & & \\
\hline
{}& & & & & & & & & & & & \\
\ud3 & \ud3
&\ud2&\ud1+\ud5&\ud0+\ud4&\ud3&\ud2&\ud{31^{'}}&\ud{21^{'}}&\ud{11^{'}}+\ud{51^{'}}&\ud{1^{'}}+\ud{41^{'}}&\ud{31^{'}}&\ud{21^{'}}

\\
{}& & & & & & & & & & & & \\
\hline
{}& & & & & & & & & & & & \\
\ud4 & \ud4
&\ud5&\ud2&\ud3&\ud0&\ud1&\ud{41^{'}}&\ud{51^{'}}&\ud{21^{'}}&\ud{31^{'}}&\ud{1^{'}}&\ud{11^{'}}

\\
{}& & & & & & & & & & & & \\
\hline
{}& & & & & & & & & & & & \\
\ud5 & \ud5 &\ud2+\ud4&\ud1+\ud3+\ud5&\ud2&\ud1&\ud0+\ud2&\ud{51^{'}}&
\ud{21^{'}}+\ud{41^{'}}&\ud{11^{'}}+\ud{31^{'}}+\ud{51^{'}}&\ud{21^{'}}&
\ud{11^{'}} &\ud{1^{'}}+\ud{21^{'}} \\
{}& & & & & & & & & & & & \\
\hline
{}& & & & & & & & & & & & \\
\ud{1^{'}} & \ud{1^{'}} &
\ud{11^{'}}&\ud{21^{'}}&\ud{31^{'}}&\ud{41^{'}}&
\ud{51^{'}}&\ud0+\ud{31^{'}}&\ud1+\ud{21^{'}}&\ud2+\ud{11^{'}}+\ud{51^{'}}&
\ud3+\ud{1^{'}}+\ud{41^{'}}&\ud4+\ud{31^{'}}&\ud5+\ud{21^{'}} \\
{}& & & & & & & & & & & & \\
\hline
{}& & & & & & & & & & & & \\
\ud{11^{'}} & \ud{11^{'}} & \ud{1^{'}}+\ud{21^{'}}&
\ud{11^{'}}+\ud{31^{'}}+\ud{51^{'}}&\ud{21^{'}}&\ud{51^{'}}&
\ud{21^{'}}+\ud{41^{'}}&\ud1+\ud{21^{'}}&
\ud0+\ud2+\ud{11^{'}}+\ud{31^{'}}+\ud{51^{'}}&
\ud1+\ud3+\ud5+\ud{1^{'}} + &
\ud2+\ud{11^{'}}+\ud{51^{'}}&\ud5+\ud{21^{'}}&
\ud2+\ud4+\ud{11^{'}}+\ud{31^{'}}+\ud{51^{'}}
\\
{}& & & & & & & & & + \ud{21^{'}}+\ud{21^{'}}+\ud{41^{'}}+\ud{41^{'}} & & & \\
{}& & & & & & & & & & & & \\
\hline
{}& & & & & & & & & & & & \\
\ud{21^{'}} & \ud{21^{'}} & \ud{11^{'}}+ \ud{31^{'}}+
\ud{51^{'}}&\ud{1^{'}}+\ud{21^{'}}+\ud{21^{'}}+\ud{41^{'}}&
\ud{11^{'}}+\ud{51^{'}}&\ud{21^{'}}&
\ud{11^{'}}+\ud{31^{'}}+\ud{51^{'}}&
\ud2+\ud{11^{'}}+\ud{51^{'}}&
\ud1+\ud3+\ud5+\ud{1^{'}}+&
\ud0+\ud2+\ud2+\ud4+ &
\ud1+\ud5+\ud{21^{'}}+\ud{21^{'}} &\ud2+\ud{11^{'}}+\ud{51^{'}}&
\ud1+\ud3+\ud5+\ud{1^{'}}+\\
{}& & & & & & & & \ud{21^{'}}+\ud{21^{'}}+\ud{41^{'}} &
+ \ud{11^{'}}+ \ud{11^{'}}+\ud{31^{'}}+\ud{31^{'}}+  &  & &
+ \ud{21^{'}}+\ud{21^{'}}+\ud{41^{'}} \\
{}& & & & & & & & & + \ud{51^{'}}+\ud{51^{'}} & & & \\
{}& & & & & & & & & & & & \\
\hline
{}& & & & & & & & & & & & \\
\ud{31^{'}} & \ud{31^{'}} &  \ud{21^{'}} &\ud{11^{'}}+
\ud{51^{'}}& \ud{1^{'}}+\ud{41^{'}}&\ud{31^{'}}&
\ud{21^{'}}&\ud3+\ud{1^{'}}+\ud{41^{'}}&\ud2+\ud{11^{'}}+\ud{51^{'}}&
\ud1+\ud5+\ud{21^{'}}+\ud{21^{'}}&\ud0+\ud4+\ud{31^{'}}+\ud{31^{'}}&
\ud3+\ud{1^{'}}+\ud{41^{'}}&
\ud2+\ud{11^{'}}+\ud{51^{'}} \\
{}& & & & & & & & & & & & \\
\hline
{}& & & & & & & & & & & & \\
\ud{41^{'}} & \ud{41^{'}} &  \ud{51^{'}}&\ud{21^{'}}&
\ud{31^{'}}&\ud{1^{'}}&\ud{11^{'}}&\ud4+\ud{31^{'}}&
\ud5+\ud{21^{'}}&\ud2+\ud{11^{'}}+\ud{51^{'}}&
\ud3+\ud{1^{'}}+\ud{41^{'}}&\ud0+\ud{31^{'}}& \ud1+\ud{21^{'}}\\
{}& & & & & & & & & & & & \\
\hline
{}& & & & & & & & & & & & \\
\ud{51^{'}} & \ud{51^{'}} &
\ud{21^{'}} +\ud{41^{'}} &\ud{11^{'}}+\ud{31^{'}}+\ud{51^{'}}& \ud{21^{'}}&
\ud{11^{'}}&\ud{1^{'}}+\ud{21^{'}}&\ud5+\ud{21^{'}}&
\ud2+\ud4+\ud{11^{'}}+\ud{31^{'}}+\ud{51^{'}}&
\ud1+\ud3+\ud5+\ud{1^{'}}+&
\ud2+\ud{11^{'}}+\ud{51^{'}}& \ud1+\ud{21^{'}}&
\ud0+\ud2+\ud{11^{'}}+\ud{31^{'}}+\ud{51^{'}}\\
{}& & & & & & & & & + \ud{21^{'}}+\ud{21^{'}}+\ud{41^{'}} & & & \\
{}& & & & & & & & & & & & \\
\hline
\end{array}
$$
\caption{Multiplication table for the Ocneanu algebra of $E_6$}
\end{table}
\endlandscape
Once toric matrices with one twist are determined, the knowledge of
this multiplication table allows one to determine all toric matrices
with two twists (``second algorithm'').  Besides the general property
$W_{x,y}=W_{y,x}$, which holds in the present case since $Oc(E_{6})$
is commutative, this table also allows one
to obtain many other identities between toric matrices; for instance,
from the fact that $(1 \otimesdot 1) .  (0 \otimesdot 4) = (5
\otimesdot 1) = (4 \otimesdot 1) .  (1 \otimesdot 0) = (5 \otimesdot
1) .  (0 \otimesdot 0)$ we deduce the identities $W_{11,04} =
W_{51,00} = W_{41,10}$.

\subsubsection{Toric matrices $W_{xy}$ and frustrated functions
with two twists (results)}

Since $dim(Oc(E_{6})=12$, we have, \textsl{a priori}, $12^{2}$
generalized toric structures $W_{x,y}$ for the graph $E_{6}$.  However
taking into account the symmetry $W_{x,y}=W_{yx}$ and other identities
encoded by the previous table, it happens that only $36$, among the
expected $144$ toric structures, are distinct.  It is interesting to
restrict our attention to those that are symmetric, but we already
know that six among the twelve toric matrices with one twist are
symmetric (the three ambichiral ones $W_{00,00}, W_{30,00}, W_{40,00}$
and the three which are neither ambichiral nor chiral, $W_{11,00},
W_{21,00}, W_{51,00}$).  Therefore we are left with only six new
matrices that are given below:

{\small
$$
\begin{array}{cc}

W_{30,21} =
\left(
\begin{array}{ccccccccccc}
0 & 0 & 0 & 0 & 0 & 0 & 0 & 0 & 0 & 0 & 0 \\
0 & 1 & 0 & 1 & 0 & 2 & 0 & 1 & 0 & 1 & 0 \\
0 & 0 & 2 & 0 & 2 & 0 & 2 & 0 & 2 & 0 & 0 \\
0 & 1 & 0 & 1 & 0 & 2 & 0 & 1 & 0 & 1 & 0 \\
0 & 0 & 2 & 0 & 2 & 0 & 2 & 0 & 2 & 0 & 0 \\
0 & 2 & 0 & 2 & 0 & 4 & 0 & 2 & 0 & 2 & 0 \\
0 & 0 & 2 & 0 & 2 & 0 & 2 & 0 & 2 & 0 & 0 \\
0 & 1 & 0 & 1 & 0 & 2 & 0 & 1 & 0 & 1 & 0 \\
0 & 0 & 2 & 0 & 2 & 0 & 2 & 0 & 2 & 0 & 0 \\
0 & 1 & 0 & 1 & 0 & 2 & 0 & 1 & 0 & 1 & 0 \\
0 & 0 & 0 & 0 & 0 & 0 & 0 & 0 & 0 & 0 & 0
\end{array}
\right)

\bigskip

&

W_{11,51} =
\left(
\begin{array}{ccccccccccc}
0 & 0 & 1 & 0 & 2 & 0 & 1 & 0 & 1 & 0 & 1 \\
0 & 1 & 0 & 2 & 0 & 2 & 0 & 2 & 0 & 1 & 0 \\
1 & 0 & 2 & 0 & 3 & 0 & 3 & 0 & 2 & 0 & 1 \\
0 & 2 & 0 & 4 & 0 & 4 & 0 & 4 & 0 & 2 & 0 \\
2 & 0 & 3 & 0 & 4 & 0 & 5 & 0 & 3 & 0 & 1 \\
0 & 2 & 0 & 4 & 0 & 4 & 0 & 4 & 0 & 2 & 0 \\
1 & 0 & 3 & 0 & 5 & 0 & 4 & 0 & 3 & 0 & 2 \\
0 & 2 & 0 & 4 & 0 & 4 & 0 & 4 & 0 & 2 & 0 \\
1 & 0 & 2 & 0 & 3 & 0 & 3 & 0 & 2 & 0 & 1 \\
0 & 1 & 0 & 2 & 0 & 2 & 0 & 2 & 0 & 1 & 0 \\
1 & 0 & 1 & 0 & 1 & 0 & 2 & 0 & 1 & 0 & 0
\end{array}
\right)

\\

W_{11,21} =
\left(
\begin{array}{ccccccccccc}
0 & 1 & 0 & 2 & 0 & 2 & 0 & 2 & 0 & 1 & 0 \\
1 & 0 & 2 & 0 & 3 & 0 & 3 & 0 & 2 & 0 & 1 \\
0 & 2 & 0 & 4 & 0 & 4 & 0 & 4 & 0 & 2 & 0 \\
2 & 0 & 4 & 0 & 6 & 0 & 6 & 0 & 4 & 0 & 2 \\
0 & 3 & 0 & 6 & 0 & 6 & 0 & 6 & 0 & 3 & 0 \\
2 & 0 & 4 & 0 & 6 & 0 & 6 & 0 & 4 & 0 & 2 \\
0 & 3 & 0 & 6 & 0 & 6 & 0 & 6 & 0 & 3 & 0 \\
2 & 0 & 4 & 0 & 6 & 0 & 6 & 0 & 4 & 0 & 2 \\
0 & 2 & 0 & 4 & 0 & 4 & 0 & 4 & 0 & 2 & 0 \\
1 & 0 & 2 & 0 & 3 & 0 & 3 & 0 & 2 & 0 & 1 \\
0 & 1 & 0 & 2 & 0 & 2 & 0 & 2 & 0 & 1 & 0
\end{array}
\right)

&

W_{03,03} =
\left(
\begin{array}{ccccccccccc}
1 & 0 & 0 & 0 & 1 & 0 & 1 & 0 & 0 & 0 & 1 \\
0 & 0 & 0 & 0 & 0 & 0 & 0 & 0 & 0 & 0 & 0 \\
0 & 0 & 0 & 0 & 0 & 0 & 0 & 0 & 0 & 0 & 0 \\
0 & 0 & 0 & 2 & 0 & 0 & 0 & 2 & 0 & 0 & 0 \\
1 & 0 & 0 & 0 & 1 & 0 & 1 & 0 & 0 & 0 & 1 \\
0 & 0 & 0 & 0 & 0 & 0 & 0 & 0 & 0 & 0 & 0 \\
1 & 0 & 0 & 0 & 1 & 0 & 1 & 0 & 0 & 0 & 1 \\
0 & 0 & 0 & 2 & 0 & 0 & 0 & 2 & 0 & 0 & 0 \\
0 & 0 & 0 & 0 & 0 & 0 & 0 & 0 & 0 & 0 & 0 \\
0 & 0 & 0 & 0 & 0 & 0 & 0 & 0 & 0 & 0 & 0 \\
1 & 0 & 0 & 0 & 1 & 0 & 1 & 0 & 0 & 0 & 1
\end{array}
\right)

\\

{} & {}

\\

W_{51,51} =
\left(
\begin{array}{ccccccccccc}
1 & 0 & 1 & 0 & 1 & 0 & 2 & 0 & 1 & 0 & 0 \\
0 & 1 & 0 & 2 & 0 & 2 & 0 & 2 & 0 & 1 & 0 \\
1 & 0 & 2 & 0 & 3 & 0 & 3 & 0 & 2 & 0 & 1 \\
0 & 2 & 0 & 4 & 0 & 4 & 0 & 4 & 0 & 2 & 0 \\
1 & 0 & 3 & 0 & 5 & 0 & 4 & 0 & 3 & 0 & 2 \\
0 & 2 & 0 & 4 & 0 & 4 & 0 & 4 & 0 & 2 & 0 \\
2 & 0 & 3 & 0 & 4 & 0 & 5 & 0 & 3 & 0 & 1 \\
0 & 2 & 0 & 4 & 0 & 4 & 0 & 4 & 0 & 2 & 0 \\
1 & 0 & 2 & 0 & 3 & 0 & 3 & 0 & 2 & 0 & 1 \\
0 & 1 & 0 & 2 & 0 & 2 & 0 & 2 & 0 & 1 & 0 \\
0 & 0 & 1 & 0 & 2 & 0 & 1 & 0 & 1 & 0 & 1
\end{array}
\right)

&

W_{21,21} =
\left(
\begin{array}{ccccccccccc}
1 & 0 & 2 & 0 & 3 & 0 & 3 & 0 & 2 & 0 & 1 \\
0 & 2 & 0 & 4 & 0 & 4 & 0 & 4 & 0 & 2 & 0 \\
2 & 0 & 4 & 0 & 6 & 0 & 6 & 0 & 4 & 0 & 2 \\
0 & 4 & 0 & 8 & 0 & 8 & 0 & 8 & 0 & 4 & 0 \\
3 & 0 & 6 & 0 & 9 & 0 & 9 & 0 & 6 & 0 & 3 \\
0 & 4 & 0 & 8 & 0 & 8 & 0 & 8 & 0 & 4 & 0 \\
3 & 0 & 6 & 0 & 9 & 0 & 9 & 0 & 6 & 0 & 3 \\
0 & 4 & 0 & 8 & 0 & 8 & 0 & 8 & 0 & 4 & 0 \\
2 & 0 & 4 & 0 & 6 & 0 & 6 & 0 & 4 & 0 & 2 \\
0 & 2 & 0 & 4 & 0 & 4 & 0 & 4 & 0 & 2 & 0 \\
1 & 0 & 2 & 0 & 3 & 0 & 3 & 0 & 2 & 0 & 1
\end{array}
\right)

\end{array}
$$
}
Using our ``first algorithm'',
the last toric matrix $W_{21,21}$, for instance, is
$$
W_{21,21} = {\mathcal V}_{125}[2] . W_{1}(A_{3}) . {\mathcal V}_{125}[1]
$$
The structure of the corresponding partition function
$Z_{21,21} = \overline \chi . W_{21,21} . \chi$ is better
understood if we use this last expression, leading to
$$
Z_{21,21} = (\overline v[2]_{1} + \overline v[2]_{3}) v[1]_{2} +
\overline v[2]_{2} (v[1]_{1} + v[1]_{3})
$$
with
$$
\begin{array}{cc}
  {v[2]_{1} = \chi_{2} + 2 \chi_{4} + 2 \chi_{6} + 2 \chi_{8} +
  \chi_{10}, } &
  {v[1]_{1} = \chi_{1} + \chi_{3} + \chi_{5} + 2 \chi_{7} +  \chi_{9}}
  \\
  {v[2]_{2} = \chi_{1} + 2 \chi_{3} + 3 \chi_{5} + 3 \chi_{7} + 2
  \chi_{9} + \chi_{11},} &
  {v[1]_{2} = \chi_{2} + 2 \chi_{4} + 2 \chi_{6} + 2 \chi_{8} +
  \chi_{10} }\\
  {v[2]_{3} = \chi_{2} + 2 \chi_{4} + 2 \chi_{6} + 2 \chi_{8} +
\chi_{10},  } &
{v[1]_{3} = \chi_{3} + 2 \chi_{5} + \chi_{7} + \chi_{9} + \chi_{11}} \\
     \end{array}
     $$
than if we just consider its fully developed form obtained by
using the explicit expression for matrix $W_{21,21}$.
The same toric matrix can also be obtained, using our ``second
algorithm'', as a linear combination of toric matrices with a single twist;
indeed, the $Oc(E_{6})$ multiplication table tells us that
$$
(2 \otimesdot 1)(2 \otimesdot 1) = 0 \otimesdot 0 + 2 \otimesdot 0 + 
2 \otimesdot 0 + 4 \otimesdot 0 + 1 \otimesdot
1 + 1 \otimesdot 1 + 3 \otimesdot 1 + 3 \otimesdot 1 + 5 \otimesdot 1 
+ 5 \otimesdot 1
$$
therefore, once we have determined the toric matrices with a
  single twist, we get $$W_{21,21} = W_{00,00} + 2 W_{20,00} +
  W_{40,00} + 2 W_{11,00} + 2 W_{31,00} + 2 W_{51,00}$$

\subsubsection{Modular properties of $E_{6}$ }

\paragraph{The modular matrix $S$ for $A_{11}$}
\label{sec:SforA11}

As discussed in section \ref{sec:SfromMul}, rather than using a general
formula, we determine directly the modular matrix $S$ from the
properties of the diagram $A_{11}$.   The following
table gives, for each eigenvalue $\delta $ of the adjacency matrix of
$A_{11}$ (so $k=10, \kappa =12$), the components of the associated
eigenvector $\psi$, chosen in such a way that it takes the value $1$
at the vertex $\tau_{0}$.  The table also gives the norm $\overline
\psi \psi$.  Define $\phi$ as the normalized eigenvector corresponding
to $\psi$ (\ie $\phi = \psi / \sqrt{\overline \psi \psi}$).  The
modular matrix $S$ is then obtained from the table $s$ of the $11$
eigenvectors $\phi$ (with our conventions, $S = s$).

$$
\tiny{
\begin{array}{c|ccccccccccc|c|}
        \delta & \tau_{0} & \tau_{1} &\tau_{2} &\tau_{3} &\tau_{4}
        &\tau_{5} &\tau_{6} &\tau_{7} &\tau_{8} &\tau_{9} & \tau_{10}
        &\overline \psi \psi \\ \hline \\
        \beta =\sqrt{2 + \sqrt{3}} & [1] & [2]=\beta & [3] & [4] & [5] & [6] &
        [7] & [8] & [9] & [10] & [11] & 24(2 + \sqrt{3}) \\  \sqrt{3} &
        1 & \sqrt{3} & 2 &  \sqrt{3} & 1 & 0 & -1 & - \sqrt{3} & -2 & -
        \sqrt{3} & -1 & 24 \\
         \sqrt{2} & 1 &  \sqrt{2} & 1 & 0 & -1 & -  \sqrt{2} & -1 & 0 &
         1 &  \sqrt{2} & 1 & 12 \\
         1 & 1 & 1 &0 &-1 & -1 &0& 1& 1& 0& -1& -1& 8 \\
         \sqrt{2 - \sqrt{3}} & [1'] & [2'] & [3'] & [4'] & [5'] & [6'] &
        [7'] & [8'] & [9'] & [10'] & [11'] & 24(2 - \sqrt{3}) \\
        0 & 1 &0 &-1& 0& 1& 0& -1& 0& 1& 0& -1& 6 \\
         -\sqrt{2 - \sqrt{3}} & [1'] & -[2'] & [3'] & -[4'] & [5'] & -[6'] &
        [7'] & -[8'] & [9'] & -[10'] & [11'] & 24(2 - \sqrt{3}) \\
         -1 & 1 & -1 &0 &1 & -1 &0& 1& -1& 0& 1& -1& 8 \\
         - \sqrt{2} & 1 &  -\sqrt{2} & 1 & 0 & -1 &  \sqrt{2} & -1 & 0 &
         1 &  -\sqrt{2} & 1 & 12 \\
          -\sqrt{3} &
        1 & -\sqrt{3} & 2 & - \sqrt{3} & 1 & 0 & -1 & \sqrt{3} & -2 &
        \sqrt{3} & -1 & 24 \\
       - \beta = -\sqrt{2 + \sqrt{3}} & [1] & -[2] & [3] & -[4] & [5] & -[6] &
        [7] & -[8] & [9] & -[10] & [11] & 24(2 + \sqrt{3})
\end{array}
}
$$
With $[n]\doteq [n]_{+}$, $[n'] \doteq [n]_{-}$ and
$$
\begin{array}{ccc}
        [1]_{\pm} = [11]_{\pm} = 1, & [2]_{\pm}=[10]_{\pm} = \sqrt{2
{\pm} \sqrt{3}}, & [3]_{\pm}=[9]_{\pm} =
        1 {\pm} \sqrt{3}, \cr [4]_{\pm}=[8]_{\pm} = {\pm} \sqrt{3(2 {\pm}
        \sqrt{3})} ,& [5]_{\pm}=[7]_{\pm}=  2 {\pm} \sqrt{3}, &
        [6]_{\pm}= 2 \sqrt{2 {\pm} \sqrt{3}}
\end{array}
$$

\paragraph{The modular matrix $T$ for $A_{11}$}
$$T = e^{\mathsf{i} \pi/24} \, 
diag[7,10,15,22,-17,-6,7,22,-9,10,-17)]$$

\paragraph{Characters of $A_{11}.$}
In a neighborhood of $i \infty$, the characters $\chi_{r}^{(10)}
\doteq
\xi_{r-1}^{(10)}$ of $A_{11}$
read\footnote{Of course\ldots the
       coefficients of $\xi_{j}^{(k)}$ are not simply $j +
       1$ times bigger than those of $\xi_{0}^{(k)}$\; !}
        $$
\begin{array}{cc}
\chi_{1}^{(10)} (q)   =
q^{- \frac{5}{48}  }
\,(1 + 3\,q + 9\,q^2  +\ldots), &
\chi_{2}^{(10)} (q)  =  q^{- \frac{1}{24}  }
\,(2 + 6\,q + 18\,q^2 + \ldots)\\
\chi_{3}^{(10)} (q)  =  q^{\frac{1}{16}}
\,(3 + 9\,q + 27\,q^2 + \ldots), &
\chi_{4}^{(10)} (q)  =  q^{\frac{5}{24}}
\,(4 + 12\ q + 36\, q^2  + \ldots)\\
\chi_{5}^{(10)} (q)  =  q^{\frac{19}{48}}
\,(5 + 15\,q + 45\,q^2 + \ldots), &
\chi_{6}^{(10)} (q) =  q^{\frac{5}{8}}
\,(6 + 18\,q + 54\,q^2 + \ldots)\\
\chi_{7}^{(10)} (q)   =  q^{\frac{43}{48}}
\,(7 + 21\,q + 63\,q^2 + \ldots), &
\chi_{8}^{(10)} (q)   =  q^{\frac{29}{24}}
\,(8 + 24\,q + 72\,q^2  + \ldots)\\
\chi_{9}^{(10)} (q) =  q^{\frac{25}{16}}
\,(9 + 27\,q + 81\,q^2 + \ldots ), &
\chi_{10}^{(10)} (q)   =  q^{\frac{47}{24}}
\,(10 + 30\,q + 76\,q^2  + \ldots)\\
\chi_{11}^{(10)} (q) =  q^{\frac{115}{48}}
\,(11 + 20\,q + 60\,q^2  + \ldots).
\end{array}
$$

\paragraph{Modular properties for the twisted partition
functions of $E_{6}$}

Since $\kappa = 12$ is even, the representation of the modular group
factorizes over the finite group $SL(2,\ZZ/4\kappa \ZZ) =
SL(2,\ZZ/48\ZZ)$.  A presentation of $SL(2,\ZZ/N \ZZ)$, by generators
and relations, for $N > 4$, can be found in \cite{CosteGannon:modular}); all
necessary relations can be checked here (in particular $T^{48}=1$).
Notice that, since $48 = 16 \times 3$ and since integers $16$
and $3$ are relatively prime, this finite group is isomorphic with
$SL(2,\ZZ/3\ZZ) \times SL(2,\ZZ/2^{4}\ZZ)$, of order $24 \times 3072$.

The eleven dimensional vector space spanned by the characters of
$A_{11}$ carry a representation of $SL(2,\ZZ/48\ZZ)$ which is not
irreducible since the three dimensional vector subspace spanned by
vectors $w_{1}=w[0]_{1}$, $w_{2}=w[0]_{2}$ and $w_{3}=w[0]_{2}$ is invariant.
Indeed, under $S: \tau \mapsto -1/\tau$, $w_{1} \mapsto
\frac{1}{2}(w_{1}+w_{2}) - \frac{1}{\sqrt 2} w_{2}$,
$w_{2} \mapsto \frac{1}{\sqrt 2} (w_{3}-w_{1})$,
$w_{3} \mapsto
\frac{1}{2}(w_{1}+w_{2}) + \frac{1}{\sqrt 2} w_{2}$ and, under $T: \tau
\mapsto \tau + 1$, $w_{1} \mapsto e^{\frac{19 i \pi}{24}}  w_{1}$,
$w_{2} \mapsto e^{\frac{5 i \pi}{12}} w_{2}$, $w_{3} \mapsto
e^{\frac{- 5 i \pi}{24}} w_{3}$.
Bilinear forms on this three dimensional irreducible subspace build a
vector space
$\CC^{3} \otimes \CC^{3} \simeq \CC^{9}$, which itself contains an irreducible
subspace of dimension one, spanned by the $W_{0,0}$ matrix.
The above transformation properties of characters allow one to check that
$Z_{0,0}$ is indeed invariant, but it is much easier to check that
the toric matrix $W_{0,0}$ commutes with both $S$ and $T$.

Twisted partition functions are not, a priori, invariant under the
modular group.  By inspection, we found the following remarkable
property: besides $W_{00,00}$ itself, none\footnote{ Actually $ (0
\otimesdot 4)(0 \otimesdot 4) = 0\otimesdot 0$ in the multiplication
table of $Oc(E_{6})$, so that $W_{04,04} = W_{00,00}$ and the
corresponding entry in the table (it commutes with $T$ !)  is just the
usual modular invariant.} of the toric matrices $W_{x,y}$ commutes
with $S$, but they all commute with the operator $S T^{-1} S$;
moreover, toric matrices also commute with particular powers of the
operator $T$.  The results are summarized in the following table:
columns and rows are labelled by vertices $x, y$ of the Ocneanu graph,
the corresponding entry gives the smallest power $p$, such that
$W_{x,y}$ commutes with $T^p$; dots stand for $p = 48$ (but this
commutation property is trivial since $T^{48}=1$ anyway).

$$
\begin{array}{c||ccc|ccc|ccc|ccc}
      W_{x,y} & 00 & 03 & 04 & 10 & 20 & 50 & 01 & 02 & 05 & 11 & 21 &
      51 \\
\hline
\hline
00 & 1 & 16 & 2 & . & 12 & . & . & 12 & . & 12 & . & 12 \\
03 & 16 & 2 & 16 & 12 & . & 12 & 12 & . & 12 & . & 12 & . \\
04 &  2 & 16 & 1 & . & 12 & . & . & 12 & . & 12 & . & 12 \\
\hline
10 & . & 12 & .  & 12 & . & 12 & 12 & . & 12 & . & 12 & . \\
20 & 12 & . & 12 & . & 12 & . & . & 12 & . & 12 & . & 12 \\
50 & . & 12 & .  & 12 & . & 12 & 12 & . & 12 & . & 12 & . \\
\hline
01 & . & 12 & .  & 12 & . & 12 & 12 & . & 12 & . & 12 & . \\
02 & 12 & . & 12 & . & 12 & . & . & 12 & . & 12 & . & 12 \\
05 & . & 12 & .  & 12 & . & 12 & 12 & . & 12 & . & 12 & . \\
\hline
11 &  12 & . & 12 & . & 12 & . & . & 12 & . & 12 & . & 12 \\
21 &  . & 12 & .  & 12 & . & 12 & 12 & . & 12 & . & 12 & . \\
51 &  12 & . & 12 & . & 12 & . & . & 12 & . & 12 & . & 12 \\
\end{array}
$$

The operator $T^{N}$ represents the shift $\tau \mapsto \tau + N$ and
$S T^{-1} S$ represents the transformation $\tau \mapsto
\frac{\tau}{\tau + 1}$.  Together, these two elements generate
$\Gamma_{0}(N)$, a congruence subgroup of level $N$.  The usual
partition function is invariant with respect to the
modular\footnote{We write ``modular'' but the relevant group is
$SL(2,\ZZ)$, not $PSL(2,\ZZ)$.} group, but twisted partition functions
are invariant only with respect to appropriate congruence subgroups.
For instance, $Z_{03,03}$ is invariant under the subgroup
$\Gamma_{0}(2)$.  Actually, we should remember that, in this case, the
whole representation factorizes through the principal congruence
subgroup $\Gamma(48)$.

\subsection{Example of an affine model of type $SU(3)$: the 
${\mathcal E}_{5}$ case}
\label{sec:E5partfun}
The Di Francesco -- Zuber $G = \mathcal{E}_{5}$ diagram is displayed
below, it is a module over the $\mathcal{A}_{5}$ diagram (the generator
corresponding to the given orientation is the vertex $(1,0)$).

\bigskip

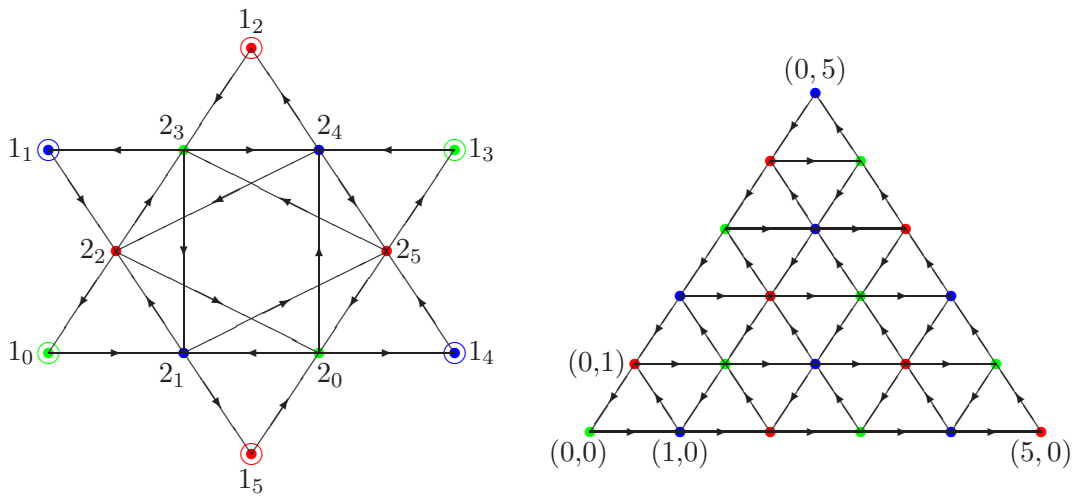
\begin{figure}[hhh]
\begin{center}
\unitlength 0.30mm
\par
\begin{picture}(440,190)(0,-15)

\put(0,45){\begin{picture}(60,45)
\put(0,0){\color{green} \circle*{5}}
\put(0,0){\color{green} \circle{9}}
\put(60,0){\color{blue} \circle*{5}}
\put(30,45){\color{red} \circle*{5}}
\put(0,0){\vector(1,0){32.5}}
\put(30,0){\line(1,0){30}}
\put(60,0){\vector(-2,3){16.5}}
\put(45,22.5){\line(-2,3){15}}
\put(30,45){\vector(-2,-3){16.5}}
\put(15,22.5){\line(-2,-3){15}}
\end{picture}}

\put(120,45){\begin{picture}(60,45)
\put(0,0){\color{green} \circle*{5}}
\put(60,0){\color{blue} \circle{9}}
\put(60,0){\color{blue} \circle*{5}}
\put(30,45){\color{red} \circle*{5}}
\put(0,0){\vector(1,0){32.5}}
\put(30,0){\line(1,0){30}}
\put(60,0){\vector(-2,3){16.5}}
\put(45,22.5){\line(-2,3){15}}
\put(30,45){\vector(-2,-3){16.5}}
\put(15,22.5){\line(-2,-3){15}}
\end{picture}}

\put(60,135){\begin{picture}(60,45)
\put(0,0){\color{green} \circle*{5}}
\put(60,0){\color{blue} \circle*{5}}
\put(30,45){\color{red} \circle*{5}}
\put(30,45){\color{red} \circle{9}}
\put(0,0){\vector(1,0){32.5}}
\put(30,0){\line(1,0){30}}
\put(60,0){\vector(-2,3){16.5}}
\put(45,22.5){\line(-2,3){15}}
\put(30,45){\vector(-2,-3){16.5}}
\put(15,22.5){\line(-2,-3){15}}
\end{picture}}

\put(0,90){\begin{picture}(60,45)
\put(0,45){\color{blue} \circle*{5}}
\put(0,45){\color{blue} \circle{9}}
\put(60,45){\vector(-1,0){32.5}}
\put(30,45){\line(-1,0){30}}
\put(30,0){\vector(2,3){16.5}}
\put(45,22.5){\line(2,3){15}}
\put(0,45){\vector(2,-3){16.5}}
\put(15,22.5){\line(2,-3){15}}
\end{picture}}

\put(120,90){\begin{picture}(60,45)
\put(60,45){\color{green} \circle*{5}}
\put(60,45){\color{green} \circle{9}}
\put(60,45){\vector(-1,0){32.5}}
\put(30,45){\line(-1,0){30}}
\put(30,0){\vector(2,3){16.5}}
\put(45,22.5){\line(2,3){15}}
\put(0,45){\vector(2,-3){16.5}}
\put(15,22.5){\line(2,-3){15}}
\end{picture}}

\put(60,0){\begin{picture}(60,45)
\put(30,0){\color{red} \circle*{5}}
\put(30,0){\color{red} \circle{9}}
\put(60,45){\vector(-1,0){32.5}}
\put(30,45){\line(-1,0){30}}
\put(30,0){\vector(2,3){16.5}}
\put(45,22.5){\line(2,3){15}}
\put(0,45){\vector(2,-3){16.5}}
\put(15,22.5){\line(2,-3){15}}
\end{picture}}

\put(60,135){\vector(0,-1){47.5}}
\put(60,90){\line(0,-1){45}}
\put(60,45){\vector(2,1){47.2}}
\put(105,67.5){\line(2,1){45}}
\put(150,90){\vector(-2,1){47.2}}
\put(105,112.5){\line(-2,1){45}}

\put(120,45){\vector(0,1){47.5}}
\put(120,90){\line(0,1){45}}
\put(120,135){\vector(-2,-1){47.2}}
\put(75,112.5){\line(-2,-1){45}}
\put(30,90){\vector(2,-1){47.2}}
\put(75,67.5){\line(2,-1){45}}

\put(240,10){\begin{picture}(200,180)

\put(0,0){\begin{picture}(40,40)
\put(0,0){\color{green} \circle*{5}}
\put(40,0){\color{blue} \circle*{5}}
\put(20,30){\color{red} \circle*{5}}
\put(0,0){\vector(1,0){21}}
\put(20,0){\line(1,0){20}}
\put(40,0){\vector(-2,3){11.5}}
\put(30,15){\line(-2,3){10}}
\put(20,30){\vector(-2,-3){11.5}}
\put(10,15){\line(-2,-3){10}}\end{picture}}

\put(40,0){\begin{picture}(40,40)
\put(40,0){\color{red} \circle*{5}}
\put(20,30){\color{green} \circle*{5}}
\put(0,0){\vector(1,0){21}}
\put(20,0){\line(1,0){20}}
\put(40,0){\vector(-2,3){11.5}}
\put(30,15){\line(-2,3){10}}
\put(20,30){\vector(-2,-3){11.5}}
\put(10,15){\line(-2,-3){10}}\end{picture}}

\put(80,0){\begin{picture}(40,40)
\put(40,0){\color{green} \circle*{5}}
\put(20,30){\color{blue} \circle*{5}}
\put(0,0){\vector(1,0){21}}
\put(20,0){\line(1,0){20}}
\put(40,0){\vector(-2,3){11.5}}
\put(30,15){\line(-2,3){10}}
\put(20,30){\vector(-2,-3){11.5}}
\put(10,15){\line(-2,-3){10}}\end{picture}}

\put(120,0){\begin{picture}(40,40)
\put(40,0){\color{blue} \circle*{5}}
\put(20,30){\color{red} \circle*{5}}
\put(0,0){\vector(1,0){21}}
\put(20,0){\line(1,0){20}}
\put(40,0){\vector(-2,3){11.5}}
\put(30,15){\line(-2,3){10}}
\put(20,30){\vector(-2,-3){11.5}}
\put(10,15){\line(-2,-3){10}}\end{picture}}

\put(160,0){\begin{picture}(40,40)
\put(40,0){\color{red} \circle*{5}}
\put(20,30){\color{green} \circle*{5}}
\put(0,0){\vector(1,0){21}}
\put(20,0){\line(1,0){20}}
\put(40,0){\vector(-2,3){11.5}}
\put(30,15){\line(-2,3){10}}
\put(20,30){\vector(-2,-3){11.5}}
\put(10,15){\line(-2,-3){10}}\end{picture}}

\put(20,30){\begin{picture}(40,40)
\put(20,30){\color{blue} \circle*{5}}
\put(0,0){\vector(1,0){21}}
\put(20,0){\line(1,0){20}}
\put(40,0){\vector(-2,3){11.5}}
\put(30,15){\line(-2,3){10}}
\put(20,30){\vector(-2,-3){11.5}}
\put(10,15){\line(-2,-3){10}}\end{picture}}

\put(60,30){\begin{picture}(40,40)
\put(20,30){\color{red} \circle*{5}}
\put(0,0){\vector(1,0){21}}
\put(20,0){\line(1,0){20}}
\put(40,0){\vector(-2,3){11.5}}
\put(30,15){\line(-2,3){10}}
\put(20,30){\vector(-2,-3){11.5}}
\put(10,15){\line(-2,-3){10}}\end{picture}}

\put(100,30){\begin{picture}(40,40)
\put(20,30){\color{green} \circle*{5}}
\put(0,0){\vector(1,0){21}}
\put(20,0){\line(1,0){20}}
\put(40,0){\vector(-2,3){11.5}}
\put(30,15){\line(-2,3){10}}
\put(20,30){\vector(-2,-3){11.5}}
\put(10,15){\line(-2,-3){10}}\end{picture}}

\put(140,30){\begin{picture}(40,40)
\put(20,30){\color{blue} \circle*{5}}
\put(0,0){\vector(1,0){21}}
\put(20,0){\line(1,0){20}}
\put(40,0){\vector(-2,3){11.5}}
\put(30,15){\line(-2,3){10}}
\put(20,30){\vector(-2,-3){11.5}}
\put(10,15){\line(-2,-3){10}}\end{picture}}

\put(40,60){\begin{picture}(40,40)
\put(20,30){\color{green} \circle*{5}}
\put(0,0){\vector(1,0){21}}
\put(20,0){\line(1,0){20}}
\put(40,0){\vector(-2,3){11.5}}
\put(30,15){\line(-2,3){10}}
\put(20,30){\vector(-2,-3){11.5}}
\put(10,15){\line(-2,-3){10}}\end{picture}}

\put(80,60){\begin{picture}(40,40)
\put(20,30){\color{blue} \circle*{5}}
\put(0,0){\vector(1,0){21}}
\put(20,0){\line(1,0){20}}
\put(40,0){\vector(-2,3){11.5}}
\put(30,15){\line(-2,3){10}}
\put(20,30){\vector(-2,-3){11.5}}
\put(10,15){\line(-2,-3){10}}\end{picture}}

\put(120,60){\begin{picture}(40,40)
\put(20,30){\color{red} \circle*{5}}
\put(0,0){\vector(1,0){21}}
\put(20,0){\line(1,0){20}}
\put(40,0){\vector(-2,3){11.5}}
\put(30,15){\line(-2,3){10}}
\put(20,30){\vector(-2,-3){11.5}}
\put(10,15){\line(-2,-3){10}}\end{picture}}

\put(60,90){\begin{picture}(40,40)
\put(20,30){\color{red} \circle*{5}}
\put(0,0){\vector(1,0){21}}
\put(20,0){\line(1,0){20}}
\put(40,0){\vector(-2,3){11.5}}
\put(30,15){\line(-2,3){10}}
\put(20,30){\vector(-2,-3){11.5}}
\put(10,15){\line(-2,-3){10}}\end{picture}}

\put(100,90){\begin{picture}(40,40)
\put(20,30){\color{green} \circle*{5}}
\put(0,0){\vector(1,0){21}}
\put(20,0){\line(1,0){20}}
\put(40,0){\vector(-2,3){11.5}}
\put(30,15){\line(-2,3){10}}
\put(20,30){\vector(-2,-3){11.5}}
\put(10,15){\line(-2,-3){10}}\end{picture}}

\put(80,120){\begin{picture}(40,40)
\put(20,30){\color{blue} \circle*{5}}
\put(0,0){\vector(1,0){21}}
\put(20,0){\line(1,0){20}}
\put(40,0){\vector(-2,3){11.5}}
\put(30,15){\line(-2,3){10}}
\put(20,30){\vector(-2,-3){11.5}}
\put(10,15){\line(-2,-3){10}}\end{picture}}

\put(-5,-10){\makebox(0,0){(0,0)}}
\put(40,-10){\makebox(0,0){(1,0)}}
\put(3,30){\makebox(0,0){(0,1)}}
\put(200,-10){\makebox(0,0){$(5,0)$}}
\put(100,160){\makebox(0,0){$(0,5)$}}

\end{picture}}

\put(-12,45){\makebox(0,0){$1_0$}}
\put(192,45){\makebox(0,0){$1_4$}}
\put(-12,135){\makebox(0,0){$1_1$}}
\put(192,135){\makebox(0,0){$1_3$}}
\put(90,-12){\makebox(0,0){$1_5$}}
\put(90,192){\makebox(0,0){$1_2$}}

\put(55,35){\makebox(0,0){$2_1$}}
\put(125,35){\makebox(0,0){$2_0$}}
\put(55,145){\makebox(0,0){$2_3$}}
\put(125,145){\makebox(0,0){$2_4$}}
\put(20,90){\makebox(0,0){$2_2$}}
\put(160,90){\makebox(0,0){$2_5$}}

\end{picture}
\par
\end{center}
\caption{The $\mathcal{E}_5$ and $\mathcal{A}_5$ generalized Dynkin diagrams}
\end{figure}

Induction-restriction rules between these two diagrams and
determination of the corresponding Ocneanu graph was analyzed in
\cite{CoqueGil:Tmodular}.
The dimension of the space of paths on $\mathcal{E}_{5}$ is infinite, but
when we restrict our attention to essential paths (one type of essential
path for every vertex of $\mathcal{A}_{5}$), one finds $21$ possibilities
\textrm{i.e.,\/}\ $21$ blocks of dimensions $(d_{j}, d_{j})$ for the first
algebra structure of $\mathcal{B}G$. The integers $d_{j}$ are given by the
list:

$(12)$, $(24,24)$, $(36,48,36)$, $(36, 60, 60, 36)$, $(24, 48,60,48,24)$, $
(12,24,36,36,24,12)$

For its other multiplicative structure, $\mathcal{B}G$ has $24$ blocks. Its
dimensions $d_{x}$ are as follows: six blocks with $d_{x}= 12$, twelve
blocks with $d_{x}= 24$ and six blocks with $d_{x}= 60$.

Notice that $\sum_{j} d_{j}^{2} = 29376$ and $\sum_{x} d_{x}^{2} = 29376$;
moreover $\sum_{p} d_{j} = 720$ and $\sum_{x} d_{x} = 720$. The indexing set
for $x$, \textrm{i.e.,\/}\ the Ocneanu graph of $\mathcal{E}_{5}$, has $24$
points; it was obtained in \cite{CoqueGil:Tmodular} and is displayed on
Figure \ref{fig:OcE5}, at the end of this article.

One obtains in this way $24$ toric matrices (and partition functions) of type
$W_{x,0}$, and $24^2$ matrices of type $W_{x,y}$. Many of them happen to
co\"\i ncide. The modular invariant partition function is associated with $
W_{0,0}$ and is given by
\[
\begin{array}{lcc}
\mathcal{Z}_{\mathcal{E}_5}\doteq \mathcal{Z}_{1_0 \otimesdot 1_0} & = & |
\chi_{(1,1)} + \chi_{(3,3)} |^2 + | \chi_{(1,3)} + \chi_{(4,3)} |^2 + |
\chi_{(3,1)} + \chi_{(3,4)} |^2 \\
{} & + & | \chi_{(3,2)} + \chi_{(1,6)} |^2 + | \chi_{(4,1)} +
\chi_{(1,4)}
|^2 + | \chi_{(2,3)} + \chi_{(6,1)} |^2
\end{array}
\]

It agrees with the expression first obtained by \cite{Gannon}, using
entirely different techniques.  One could then determine all toric
matrices with one or two twists and perform the same kind of analysis
as the one that was carried out for the $E_{6}$ diagram.

\section{From graphs to minimal models}
\subsection{Central charges}

\paragraph{Affine $SU(2)$ models.}
These are the models considered in the last section; they are
associated with an $ADE$ diagram $G$ of level $k$ (with Coxeter number,
or altitude, $\kappa = k+2$).  For an affine Lie algebra
$\widehat{\mathrm{g}}_{k}$ at level $k$, the central charge is obtained
from the modular phase (or from the expression of the modular $T$
operator, see sections \ref{sec:TgeneratorAff},
or from the principal part of characters near complex infinity, see
section \ref{sec:CharacAff}) equal to $\frac{dim({\mathrm{g}}) .  k}{k +
Coxeter({\mathrm{g}})}$.  $SU(2)$ models have therefore a central charge
$c(k) \doteq \frac{3 k}{k + 2}$, value that we may {\sl define} as the
central charge of the underlying $ADE$ diagram.  All these models are
unitary ($c \ge 1$).  The limiting case $c = 1$ is obtained for $k=1$,
\ie for the graph $A_{2}$.

Affine $SU(2)$ models can be identified with WZW $\widehat{su}(2)_{k}$
models or with coset models obtained from conformal embeddings (\ie same
central charge) of type $\widehat{su}(2)_{k} \subset 
\widehat{{\mathrm{g}}}_{1}$,
here $\widehat{{\mathrm{g}}}_{1}$ is some affine Lie algebra at level $1$.
For instance, both models associated with diagrams $A_{11}$ and
$E_{6}$ have the same central charge ($c = 5/2$) and the second model
can be obtained from a conformal embedding
$\widehat{su}(2)_{10} \subset \widehat{spin}(5)_{1}$; we can check that
$dim(Spin(5)) = 10$,  $Coxeter(Spin(5)) = 3$, and $\frac{3.10}{10+2} =
\frac{10.1}{1+3}$.

\paragraph{Minimal models.}

Minimal models of type ${\cal W}_{2}$ (or ``minimal models'', for
short\footnote{${\cal W}_{2}$ denotes the Virasoro algebra.})
are defined by a pair of diagrams $(G_{1},G_{2})$ belonging to the $SU(2)$
system, \ie two $ADE$ diagrams. We call $k_{1}$ and $k_{2}$ their
respective levels (so that Coxeter numbers $\kappa_{1}$ and
$\kappa_{2}$ are respectively equal to $k_{1}+2$ and $k_{2} + 2$).
Assuming that the Coxeter numbers of the diagrams are relatively
prime, the general formula for the central charge is

$$
c(k_{1},k_{2}) = 1 - (k_{1} - k_{2})(c(k_{1}) - c(k_{2})) =
  1 - 6\frac{(\kappa_{1}-\kappa_{2})^{2}}{\kappa_{1} \kappa_{2}} $$

Unitary minimal models are obtained when $\vert \kappa_{1} -
\kappa_{2} \vert = 1$, then $c = 1 - \frac{6}{\kappa_{1} \kappa_{2}}$
and ($0 < c < 1$).  Ordering $k_{1} = k-1 < k_{2} = k$, one gets
$c(k-1,k)= 1 - \frac{6}{(k+1)(k+2)}$, which is the value obtained in
particular for models of type $({\cal A}_{k-1}, {\cal A}_{k}) \equiv
(A_{k},A_{k+1})$.  The ordered set of values starts with
$\{0,1/2,7/10,4/5,6/7,25/28, \ldots\}$.  The limiting case $c = 0$ is
obtained for the pair $(A_{1},A_{2})$.  In particular, $c = 21/22$ for
the $(A_{10},A_{11})$ model; the same value of $c$ is obtained for the
$(A_{10},E_{6})$ model.
The previous formula giving $c$, for unitary
minimal models, can be written $c(k-1,k) = c(1) + c(k-1) - c(k)$,
indeed $c(1)=1$.  This expression is therefore compatible with a
coset model $\frac{\widehat{SU(2)}_{k-1}\otimes
\widehat{SU(2)}_{1}}{\widehat{SU(2)}_{k}}$, and it is a particular case
of a more general formula, valid for coset models
$\frac{\widehat{\mathrm{g}}_{k_{1}} \otimes
\widehat{\mathrm{g}}_{k_{2}}}{\widehat{\mathrm{g}}_{k_{1}+k_{2}}}$, namely
$c(k_{1},k_{2}) = dim(\mathrm{g}) (\frac{k_{1}}{k_{1} + h} +
\frac{k_{2}}{k_{2}+h} - \frac{k_{1} + k_{2}}{k_{1} + k_{2 } + h})$,
where $h$ is the dual Coxeter number of $g$.

\paragraph{Affine $SU(3)$ models.}
These are the models considered in the last section and associated
with a Di Francesco -- Zuber diagram $G$ of level $k$ (with
generalized Coxeter number, or altitude, $\kappa = k+3$).
 From the general formula for the modular phase, we see that all
affine $SU(3)$ models have a central charge $c(k) \doteq \frac{8
k}{k +  3}$.  All these models are
unitary ($c \ge 2$). The limiting case $c = 2$ is obtained for
$k=1$, \ie for the graph ${\cal A}_{1}$.

Affine $SU(3)$ models can be identified with WZW $\widehat{su}(3)_{k}$
models or with coset models obtained from conformal embeddings (\ie same
central charge) of type $\widehat{su}(3)_{k} \subset 
\widehat{{\mathrm{g}}}_{1}$,
here $\widehat{{\mathrm{g}}}_{1}$ is some affine Lie algebra at level $1$.
For instance, both models associated with diagrams ${\cal A}_{5}$ and
${\cal E}_{5}$ have the same central charge ($c = 5$) but the second model
can be obtained from a conformal embedding
$\widehat{su}(3)_{5} \subset \widehat{su}(6)_{1}$; we can check that
$dim(SU(6)) = 35$ and $Coxeter(SU(6)) = 6$, so that $\frac{8\times
5}{5 +3} = \frac{35\times 1}{6 + 1}$.


\paragraph{Minimal models of type ${\cal W}_{3}$.}

Minimal models of type ${\cal W}_{3}$
are defined by a pair of diagrams $(G_{1},G_{2})$ belonging to the $SU(3)$
system, \ie two Di Francesco -- Zuber diagrams. We call $k_{1}$ and 
$k_{2}$ their
respective levels, so that the generalized Coxeter numbers $\kappa_{1}$ and
$\kappa_{2}$ are respectively equal to $k_{1}+3$ and $k_{2} + 3$.
Again, assuming that the Coxeter numbers of the diagrams are relatively
prime, the general formula for the central charge is

$$
c(k_{1},k_{2}) = 2 - (k_{1} - k_{2})(c(k_{1}) - c(k_{2})) =
  2(1 - 12 \frac{(\kappa_{1}-\kappa_{2})^{2}}{\kappa_{1} \kappa_{2}}) $$

Unitary minimal models of type ${\cal W}_{3}$ are obtained when $\vert
\kappa_{1} - \kappa_{2} \vert = 1$, then $c = 2(1-12/\kappa_{1}
\kappa_{2})$ and ($4/5 \leq c < 2)$.  Ordering $k_{1} = k-1 < k_{2} =
k$, one gets $c(k-1,k)= c = 2 - \frac{24}{(k+2)(k+3)}$, and this holds
in particular for models of type $({\cal A}_{k-1}, {\cal A}_{k})$.
The ordered set of values starts with $\{4/5, 6/5, 10/7, 11/7, 5/3,
\ldots\}$.  In particular, $c = 11/7$ for the $({\cal A}_{4},{\cal
A}_{5})$ model; the same value of $c$ is obtained for the $({\cal
A}_{4},{\cal E}_{5})$ model.  The limiting case (which is rather
special), $c = 4/5 < 1$ is obtained for the pair (${\cal A}_{1},{\cal
A}_{2}$).  For unitary models, the central charge can also be written
$c(k-1,k) = c(1) + c(k-1) - c(k)$, indeed $c(1)=2$.  This expression
is therefore compatible with a coset model
$\frac{\widehat{SU(3)}_{k-1}\otimes
\widehat{SU(3)}_{1}}{\widehat{SU(3)}_{k}}$, and is a particular case
of an already mentioned more general formula, valid for all coset
models.

\paragraph{Remark.}
Minimal models of type ${\cal W}_{N}$ involve, by definition, a finite
number of irreducible representations of the algebra ${\cal W}_{N}$.
The Virasoro algebra ${\cal W}_{2}$ is subalgebra of ${\cal W}_{N}$,
for $N>2$ and, in particular, of ${\cal W}_{3}$.  Under the
restriction (``branching rules'') ${\cal W}_{3} \mapsto {\cal W}_{2}$,
an irreducible representation of ${\cal W}_{3}$ can be decomposed as a
sum of irreducible representations of ${\cal W}_{2}$, but this sum is
in general infinite.  For this reason, ${\cal W}_{3}$ minimal models
do not give rise, in general, to usual (${\cal W}_{2}$) minimal
models, although this may happens : it is the case for the smallest
member (${\cal A}_{1},{\cal A}_{2}$) of the diagonal ${\cal
W}_{3}$ series (its central charge $4/5$ is smaller than $1$) which
can be identified with the Potts model, \ie the non -- diagonal
  minimal model $(A_{4}, D_{4})$.

\paragraph{Affine $SU(N)$ models and minimal models of type ${\cal W}_{N}$.}
Let us just mention that for a diagram of level $k$ belonging to a 
generalized Coxeter -- Dynkin
system of type $SU(N)$, the altitude is $\kappa = k + N$, the central
charge is $c(k) = \frac{(N^{2}-1) k}{k + N}$. A minimal model of type
${\cal W}_{N}$ is defined by a pair of such diagrams and the central
charge is
$$
c(k_{1},k_{2}) = (N-1) - (k_{1} - k_{2})(c(k_{1}) - c(k_{2})) =
  (N-1) (1 - N (N+1) \frac{(\kappa_{1}-\kappa_{2})^{2}}{\kappa_{1} 
\kappa_{2}}) $$
  More generally, if we replace $SU(N)$ by a Lie group of rank $r$ and
  dual Coxeter number $N$, the last formula reads \cite{Bouwknegt}:
  $$
c(k_{1},k_{2}) =
r (1 - N (N+1) \frac{(\kappa_{1}-\kappa_{2})^{2}}{\kappa_{1} \kappa_{2}}) $$
In the later case the concept of ${\cal W}_{N}$ algebras has to
be generalized.

\subsection{Characters, symmetry of Kac tables and partition functions}

A generalized minimal model is defined by a pair $(G_{1},G_{2})$ of
diagrams which are members of some (generalized) Coxeter - Dynkin
system.  Characters are now labelled by a pair $(r,s)$ of vertices
belonging to ${\cal A}(G_{1}) \times {\cal A}(G_{2})$, where ${\cal
A}(G_{1})$ and ${\cal A}(G_{2})$ refer to the diagrams of the
$\cal{A}$ series which have respectively same Coxeter number (or
altitude) as the given two diagrams.  As it will be recalled below, in
the case of minimal models of type ${\cal W}_{N}$, what matters is a
quotient of this product of diagrams by the $\ZZ_{N}$ group.

\paragraph{Minimal models.}
Vertices are labelled by integers $r$ or $s$ and we have a $\ZZ_{2}$
action\footnote{$\ZZ_{2}$ acts separately on the two
diagrams but we take the diagonal action.} on $(A_{k_{1}+1},
A_{k_{2}+1})$: $(r,s) \mapsto (\sigma(r) \doteq k_{1}+2 - r, \sigma(s)
\doteq k_{2}+2 - s)$.  We take $1 \leq r \leq k_{1} + 1 $ and $1 \leq
s \leq k_{2} + 1$.

\paragraph{Minimal models of type ${\cal W}_{3}$.}
Vertices are labelled by $SU(3)$ Young diagrams or by the (integer)
components $(r_{1},r_{2})$ of the chosen vertex with respect to the
two fundamental weights of $SU(3)$, and we have a $\ZZ_{3}$
action\footnote{ $\ZZ_{3}$ acts separately
(counterclockwise) on the two diagrams but we take the diagonal
action.} on $({\cal A}_{k_{1}}, {\cal A}_{k_{2}})$, with
$(r=(r_{1},r_{2}),s=(s_{1},s_{2})) \mapsto \sigma(r) \doteq ((k_{1} + 3
- (r_{1} + r_{2}), r_{1}) , \sigma(s) \doteq(k_{2} + 3 - (s_{1} +
s_{2}), s_{1}))$.  Here we take $1 \leq r_{1},r_{2} \leq k_{1} +1 $
and $1 \leq s_{1}, s_{2} \leq k_{2} +1$.

\paragraph{The different types of frustrated partition functions.}
Partition functions for minimal models (twisted or not) can be thought
as sesquilinear forms $Z = \overline \phi .  W .  \phi$ and the matrix
$W$ is obtained as a tensor product of matrices $W = W(G_{1}) \otimes
W(G_{2})$ where $W(G_{1})$ and $W(G_{2})$ are respectively toric
matrices for the affine models associated with diagrams $G_{1}$ and
$G_{2}$.  Calling $k_{1}$ and $k_{2}$ the levels of these two
diagrams, we obtain in this way -- for minimal models of type Virasoro
-- a square matrix of dimension $((k_{1}+1)\times (k_{2}+1))^{2}$; for
minimal models of type ${\cal W}_{3}$, it is a square matrix of
dimension $((k_{1}+1)(k_{1}+2)(k_{2}+1)(k_{2}+2)/4)^{2}$.  Naively,
the elements of a vector space basis on which this $W$ matrix acts
could be labelled $\chi_{r} \otimes \chi_{s}$ in the first case, and
the same thing in the second, but with $r=(r_{1},r_{2})$ and
$s=(s_{1},s_{2})$.  However, at this point one has to take into
account the $\ZZ_{2}$ action (or the $\ZZ_{3}$ action, in the case of
${\cal W}_{3}$) that identifies basis vectors labelled by ${(r,s)}$
and by ${(\sigma(r),\sigma(s))}$.  A priori, for each pair
$(x_{1},y_{1})$ of vertices of the Ocneanu graph $Oc(G_{1})$ of the
diagram $G_{1}$, we have a toric matrix $W_{x_{1},y_{1}} (G_{1})$.
Same thing for the diagram $G_{2}$.  The most general twisted
(or frustrated ) partition function, for a minimal model defined by the
pair $(G_{1}, G_{2})$ is obtained as the $\ZZ(h)$ quotient of the
sesquilinear form associated with the tensor product of toric matrices
$W_{x_{1},y_{1}} (G_{1}) \otimes W_{x_{2},y_{2}} (G_{2})$.

Because of the $\ZZ_{N}$ identification ($N=2$ for Virasoro and $N=3$ 
for ${\cal
W}_{3}$), the formula for partition
functions reads:
$$
Z=\frac{1}{N}\overline{\chi }\left( W_{\underline x, \underline
y}^{\prime}\otimes W_{\underline z, \underline t}^{\prime\prime}\right) \chi
\label{ga}
$$
Since any of the indices $x_{i}$ or $y_{i}$ can be equal to
$\underline 0$, we obtain the six types of twisted partition functions
announced in the introduction; they are respectively obtained (up to a
trivial permutation of the diagrams $G_{1}, G_{2}$) by choosing $
((x,y),(z,t))$ to be of one of the following: $((0,0),(0,0)$,
$((x,0),(0,0))$ $((x,y),(0,0))$, $((x,0),(z,0))$, $((x,y),(z,t))$.
These six cases exhaust all possibilities for a conformal theory
specified by a pair of Dynkin diagrams; of course the last case is the
most general since it encompasses all the others and the usual
partition function is recovered when all four indices are equal to
$\underline 0$.  In principle, we should denote the most general
twisted partition functions of minimal models by the symbol
$W_{(x_{1},y_{1});(x_{2},y_{2})}$ and remember that $x_{i}$ themselves
are in general given by products of the type $a \otimesdot b$.  To
ease the reading we often drop these indices $x, y$ when they are
equal to $\underline 0 = 0 \otimesdot 0$ and hope that this will be
clear from the context.  We shall examine several examples in a later
section.  In the so called ``diagonal cases'', one takes two diagrams
of type ${\cal A}$; the different types of matrices (and indices)
introduced before co\" \i ncide: $i = a = x$,  $G_{a} = N_{i} = W_{x}$
and the undeformed toric matrix $W_{0}$ is just the unit matrix.

\paragraph{Torus structures versus twisted boundary conditions}
\label{sec:zubertwist}
The above partition functions, also called ``frustrated partition
functions'' are not, in general, modular invariant --- but they are
not arbitrary either!
Another definition of the same objects together with the
following interpretation was given in \cite{PetkovaZuber:Oc}
and we repeat it here.
A usual  partition function on a torus function can be written as
$Z=\sum\limits_{i,j}Z_{ij}\chi _{i}(q)\overline{\chi _{j}}(q)$, with
$q=
\exp (2 \mathsf{i} \pi \tau)
$,
where the calculation is made by identifying the states at the end of
the cylinder through the trace operation.  Then let us imagine that we
incorporate the action of an operator $X$ attached to the non --
trivial cycle of the cylinder before identifying the two ends.
The operator $X$ called \emph{twisting operator} should commute with
the Virasoro operators $L_{n}$ and it is invariant under a distortion of the
line to which it is attached.  $X$ is therefore attached to the
homotopy class of the contour $C$ and can be thought in general as a
linear combination of operators intertwining the different copies of
$\mathcal{V}_{i}\otimes \overline{\mathcal{V}_{j}}$ (Verma modules
corresponding to the holomorphic and antiholomorphic sectors of the
theory).  In other words, the effect of $X$ is basically to twist the
boundary conditions. The partition function reads
$$
Z_{X}=tr_{H}X \, T=e^{-2LH}
\;\;\;\;
{\textstyle{with}}
\;\;\;\;
\left[ L_{n},X\right] =\left[ \overline{L_{n}},X\right] =0
$$
An explicit expression, in the presence of two twists $X$ and $Y$,
  was written for $Z_{X}$ in \cite{PetkovaZuber:Oc}, in terms of the
  matrix elements of the modular operator $S$ but we do not use this
  formula in our approach since we prefer to use directly the induction
  - restriction rules associated with the diagrams (we do not use the
  Verlinde formula either since the expression of the $S$ operator
  itself -- and not the converse -- comes from the graph algebra of the
  ${\cal A}_{k}$ diagrams).


\subsection{Conformal weights and generalized Rocha - Cariddi formulae}
\label{RochaGene}
If our goal is only to give expressions for the (twisted) partition
functions, we do not need to use any explicit expressions for the
characters $\phi_{r,s}$ of minimal models but an expression of
conformal weights is needed when one wants to discuss the physical
contents of a given theory.  Let us recall briefly these standard
results.

\paragraph{Characters for minimal models}

Call $\kappa_{1} = k_{1} + 2$ and $\kappa_{2} = k_{2} + 2$ the Coxeter
numbers of the two diagrams $G_{1}$, $G_{2}$, the conformal weights
are given by the the Rocha - Cariddi formula
$$
h_{r,s} = \frac{( r\kappa_{2} - s \kappa_{1})^{2} -
(\kappa_{2}-\kappa_{1})^{2} }{4 \kappa_{1}\kappa_{2}}
\;  \textstyle{with} \; 1 \leq r \leq \kappa_{1} - 1 \;
\textstyle{and} \;  1 \leq s \leq \kappa_{2} - 1
$$
This expression is invariant under the $\ZZ_{2}$ diagonal action
defined previously.
In the unitary cases, $\vert \kappa_{2} - \kappa_{1}\vert =  1$ and
the above expression can be simplified.
For unitary minimal models and near complex infinity, the Virasoro
characters read
$$
\phi_{r,s} \simeq q^{-c(k_{1},k_{2})/24 + h_{r,s}} (1 + \ldots)
$$
where the expression of the central charge $c$ for a pair of $ADE$
diagrams was recalled before; notice that one can recover
the conformal weights $h_{r,s}$ from these asymptotic expressions
(without using the expression of the Kac determinant).

The general expression of characters $\phi_{r,s}$, for minimal models, is

$$
\frac{q^{\frac{1}{24} - \frac{c({\kappa_{1}},{\kappa_{2}})}{24}}}{\eta(q)}
\left( - \sum_{u = -\infty }^{{+}\infty }
{{q}}^{\frac{{\left( 2\,u\,{\kappa_{1}}\,{\kappa_{2}} +
r\,{\kappa_{1}} + s\,{\kappa_{2}}\right)}^2 - {\left( {\kappa_{1}} -
{\kappa_{2}}  \right) }^2}{4\,{\kappa_{1}}\,{\kappa_{2}}}}  +
\sum_{u = -\infty }^{{+}\infty
}{{q}}^{\frac{{\left( 2\,u\,{\kappa_{1}}\,{\kappa_{2}} +
r\,{\kappa_{1}} - s\,{\kappa_{2}} \right) }^2 -{\left( {\kappa_{1}} -
{\kappa_{2}} \right) }^2}{4\,{\kappa_{1}}\,{\kappa_{2}}}} \right)
$$

Both $\phi_{r,s}$ and $h_{r,s}$ are invariant under the $\ZZ_{2}$
action (a symmetry of the Kac table).

\paragraph{Relation between affine and Virasoro characters}
Call $\chi_{1}$ and $\chi_{2}$ the two (affine) characters associated
with the graph $A_{2}$. Call $\chi_{r} (G_{1})$ and $\chi_{s} 
(G_{2})$ the affine characters of
the graph $G_{1}$ and $G_{2}$ and $\phi_{r,s}(G_{1},G_{2})$ the
Virasoro characters. In the unitary case, \ie for consecutive graphs $G_{i}$ (
$\kappa_{2} = \kappa_{1} +1$), we have the useful relations:

\begin{eqnarray*}
\chi_{r} (G_{1})\, \chi_{1} &=& \sum_{s, odd}  \,
\phi_{r,s}(G_{1},G_{2}) \, \chi_{s} (G_{2}) \\
\chi_{r} (G_{1}) \, \chi_{2} &=& \sum_{s, even}  \,
\phi_{r,s}(G_{1},G_{2}) \, \chi_{s} (G_{2})
\end{eqnarray*}

\paragraph{Characters for ${\cal W}_{3}$ minimal models} \label{sec:caracw3}
In that case one has to consider two
conformal weights; the first one, called $h = h^{(2)}$) is usually 
defined as the
eigenvalue of the Virasoro generator $L_{0}$ for the highest weight
vector of the representation, and the other, called $h^{(3)}$, is defined
as the corresponding eigenvalue for the ``$W_{3}$ generator''
(\cite{Bouwknegt} and \cite{PilchEtAl}).
These values can also be obtained from the principal parts, near
complex infinity, of the ${\cal W}_{3}$ characters.
We just give the formula for $h$; here $r$ and $s$ are vectors with
two indices.
$$
h_{r,s}= \frac{(\kappa_{2} r - \kappa_{1}s) . (K) . (\kappa_{2} r - \kappa_{1}
s) - 2 (\kappa_{2}- \kappa_{1})^{2} }{2 \kappa_{1}\kappa_{2}}
$$
\noindent
where $K = (K)_{u,v} =  \frac{1}{3} \left(
\begin{array}{ll}
     2 & 1 \\ 1 & 2 \end{array} \right)$ is the
inverse Cartan matrix of $sl(3)$.  Both characters and conformal
weights are invariant under the $\ZZ_{3}$ action ($\ZZ_{3}$ symmetry
of the ${\cal W}_{3}$ Kac table).

\subsection{Modular operators for minimal models}

We just remind the reader that the general formula giving the modular
operator $T$, for minimal models, can for example be obtained from
the prefactor giving the asymptotic form
of Virasoro characters near complex infinity and that it
is \cite{yellowbook}:
$$
T_{(r,s); (t,u)} = \delta_{r,t} \, \delta_{s,u} \,  e^{2 i \pi(h_{rs} - c/24)}
$$

$$
S_{(m,n);(r,s)}=2\sqrt{\frac{2}{\kappa_{1}\kappa_{2}}}(-1)^{1+nr+ms}\sin [\pi
\frac{\kappa_{2}}{\kappa_{1}}mr]\sin [\pi \frac{\kappa_{1}}{\kappa_{2}}sn]
$$

\section{Torus structures for unitary minimal models}

\subsection{Examples from the $ADE$ series}

The most general twisted partition functions are of the type $$1/2 \,
W_{x_{1},y_{1}}(G_{1}) \widehat \otimes W_{x_{2},y_{2}}(G_{2})$$ where
the $\widehat \otimes$ symbol means that we first calculate the tensor
product of the toric matrices, and then identify pairwise basis
elements $\phi_{r,s}$ according to the $\ZZ_{2}$ symmetry of the Kac
table.  As we know, $W_{x_{1},y_{1}}(G_{1})$ can be gotten from the
knowledge of the toric matrices with only one twist
$W_{x_{1},0}(G_{1})$ and from the multiplication table of $Oc(G_{1})$
which, for members of the ${\cal A}$ series, coïncide with the
graph matrices (fusion matrices) themselves, their determination is
then relatively easy, see section \ref{sec:diversesmatrices}).  We
have an analogous comment for the graph $G_{2}$.  For minimal models,
it is therefore enough to study twisted partition functions of the
type $Z_{(x_{1},0);(x_{2},0)} = 1/2 \, W_{x_{1},0}(G_{1}) \widehat
\otimes W_{x_{2},0}(G_{2})$, that we shall call ``fundamental toric
structures'', for short.  Now, if we if we consider {\sl unitary}
cases and further restrict our attention to the so-called ``diagonal
cases'', \ie when both graphs are of type ${\cal A}$, unitarity
requirement tells us that $G_{1}=A_{n}$ and $G_{2}=A_{n+1}$.  In such
cases, we would expect $n (n+1)$ fundamental toric structures, but
$\ZZ_{2}$ symmetry brings this number down to $n(n+1)/2$.  More
general unitary models are obtained when the corresponding Coxeter
numbers are consecutive integers; for pairs of diagrams involving
members from the $D$ or $E$ series, a general general determination of
all fundamental toric structures has to take fully into account the
structure of the corresponding Ocneanu graphs.  For illustration, we
shall examine three unitary cases in this section: the Ising model ---
it is associated with $(A_{2},A_{3})$, the Potts model --- it is
associated with $(A_{4},D_{4})$), and the $(A_{10},E_{6})$ model.

\subsubsection{Ising model}
The first non trivial case of the minimal models series corresponds to
the Ising model with central charge $c=1/2$.  This model is associated
with a pair of Dynkin diagrams $(A_2,A_3)$ with Coxeter numbers
$(\kappa_{A_2}=3,\kappa_ {A_3}=4)$.  The modular invariant partition
function can be build from the tensor product of the corresponding
fundamental toric matrices $W_{0}^{(A_2)}\otimes W_{0}^{(A_3)}$ which
respectively describe the undeformed torus structures of the two
diagrams.  In what follows we present the partition functions
associated with the $3 = 2\times 3/2$ fundamental toric structures, as
discussed above, they are of the form $W_{x_{1}}^{(A_2)}\widehat \otimes
W_{x_{2}}^{(A_3)} \doteq W_{x_{1},0}^{(A_2)}\widehat \otimes
W_{x_{2},0}^{(A_3)} $.
Following \cite{PetkovaZuber:twist}, they can be interpreted as the 
result of the
insertion of twisted boundary conditions (defect lines).  The toric
matrices of the type $W_x$ of the pair $(A_2,A_3)$ are

$$ W_{0}({A_{2}})=\left(
\begin{array}{ll}
1 & 0 \\
0 & 1
\end{array}
\right) \;,
W_{1}({A_{2}})=\left(
\begin{array}{ll}
0 & 1 \\
1 & 0
\end{array}
\right) \;$$

$$ W_{0}({A_{3}})=\left(
\begin{array}{lll}
1 & 0 & 0 \\
0 & 1 & 0 \\
0 & 0 & 1
\end{array}
\right) \;,\;W_{1}({A_{3}})=\left(
\begin{array}{lll}
0 & 1 & 0 \\
1 & 0 & 1 \\
0 & 1 & 0
\end{array}
\right) \;,\;W_{2}({A_{3}})=\left(
\begin{array}{lll}
0 & 0 & 1 \\
0 & 1 & 0 \\
1 & 0 & 0
\end{array}
\right) $$

The fundamental twisted partitions functions are given by
\begin{eqnarray*}
Z_{xy} &=&\frac{1}{2}\overline{\phi }\left( W_{x}({A_{2}})\widehat 
\otimes W_{y}({\
A_{3}})\right) \phi
\end{eqnarray*}
where  $\phi =\left(\phi _{11} ,
\phi _{21} ,
\phi _{12} ,
\phi _{22} ,
\phi _{13} ,
\phi _{23}\right)$ are the characters identifying highest weight 
representations
with conformal dimensions given in the following table

\begin{center}
\begin{tabular}{|l|}
\hline
$\;h_{11}=h_{23}=0$ \\ \hline
$\;h_{12}=h_{22}=1/16$ \\ \hline
$h_{13}=h_{21}=1/2$ \\ \hline
\end{tabular}
\end{center}
Here and below, the Virasoro characters $\phi$ are labelled with
indices $r,s$ starting from $1$ (because this is standard), but
indices $i,j$ labelling partition functions or toric matrices in the $A$ cases,
start from $0$ (because this is our convention for
labelling vertices of $A_{n}$ diagrams).  Such a choice is admittedly
confusing but we hope that the reader, being warned, will not be
mistaken. The characters will also be sometimes labelled by
the corresponding conformal weights: $\phi_{h(r,s)}  \equiv
\phi_{r,s} $.

The six possible cases are listed and explicitly build as
follows
\begin{eqnarray*}
  Z_{00} &=&\frac{1}{2}\overline{\phi }\left( W_{0}({A_{2}})\otimes W_{0}({\
A_{3}})\right) \phi \\
&=&\frac{1}{2}\overline{\phi }\left(
\begin{tabular}{ll|ll|ll}
1 &  &  &  &  &  \\
& 1 &  &  &  &  \\ \hline
&  & 1 &  &  &  \\
&  &  & 1 &  &  \\ \hline
&  &  &  & 1 &  \\
&  &  &  &  & 1
\end{tabular}
\right) \phi \\
&=&\frac{1}{2}\left[ \left( \left| \phi _{1,1}\right| ^{2}+\left| \phi
_{2,3}\right| ^{2}\right) +\left( \left| \phi _{1,2}\right| ^{2}+\left| \phi
_{2,2}\right| ^{2}\right) +\left( \left| \phi _{1,3}\right| ^{2}+\left| \phi
_{2,1}\right| ^{2}\right) \right] \\
&=&\left| \phi _{1,1}\right| ^{2}+\left| \phi _{1,2}\right| ^{2}+\left| \phi
_{1,3}\right| ^{2}=\left| \phi _{0}\right| ^{2}+\left| \phi _{1/2}\right|
^{2}+\left| \phi _{1/16}\right| ^{2}
\end{eqnarray*}

Actually, we should have written $\widehat \otimes$ rather than
$\otimes$ in the above first line, but lines two and three
also involve a (hidden) $\ZZ_{2}$
identification, which is explicitly performed on line four
($\phi_{1,1}=\phi_{2,3}$, etc). From now on we shall not mention it
explicitly but it should always be understood that a $\ZZ_{2}$
identification of characters, corresponding to the symmetry of the
Kac table (conformal weights), should be performed at the end of
calculations.

\begin{eqnarray*}
  Z_{01} &=&\frac{1}{2}\overline{\phi }\left( W_{0}({A_{2}})\otimes W_{1}({\
A_{3}})\right) \phi \\
&=&\frac{1}{2}\overline{\phi }\left(
\begin{tabular}{ll|ll|ll}
&  & 1 &  &  &  \\
&  &  & 1 &  &  \\ \hline
1 &  &  &  & 1 &  \\
& 1 &  &  &  & 1 \\ \hline
&  & 1 &  &  &  \\
&  &  & 1 &  &
\end{tabular}
\right) \phi \\
&=&\phi _{1/16}\left( \overline{\phi _{1/2}}+\overline{\phi _{0}}\right) +
\overline{\phi _{1/16}}\left( \phi _{1/2}+\phi _{0}\right)
\end{eqnarray*}

\begin{eqnarray*}
  Z_{02} &=&\frac{1}{2}\overline{\phi }\left( W_{0}({A_{2}})\otimes W_{2}({\
A_{3}})\right) \phi \\
&=&\frac{1}{2}\overline{\phi }\left(
\begin{tabular}{ll|ll|ll}
&  &  &  & 1 &  \\
&  &  &  &  & 1 \\ \hline
&  & 1 &  &  &  \\
&  &  & 1 &  &  \\ \hline
1 &  &  &  &  &  \\
& 1 &  &  &  &
\end{tabular}
\right) \phi \\
&=&\overline{\phi _{0}}\phi _{1/2}+\overline{\phi _{1/2}}\phi _{0}+\left|
\phi _{1/16}\right| ^{2}
\end{eqnarray*}

\begin{eqnarray*}
  Z_{10} &=&\frac{1}{2}\overline{\phi }\left( W_{1}({A_{2}})\otimes W_{0}({\
A_{3}})\right) \phi \\
&=&\frac{1}{2}\overline{\phi }\left(
\begin{tabular}{ll|ll|ll}
& 1 &  &  &  &  \\
1 &  &  &  &  &  \\ \hline
&  &  & 1 &  &  \\
&  & 1 &  &  &  \\ \hline
&  &  &  &  & 1 \\
&  &  &  & 1 &
\end{tabular}
\right) \phi \\
&=&\overline{\phi _{0}}\phi _{1/2}+\overline{\phi _{1/2}}\phi _{0}+\left|
\phi _{1/16}\right| ^{2}=Z_{02}
\end{eqnarray*}

\begin{eqnarray*}
  Z_{11} &=&\frac{1}{2}\overline{\phi }\left( W_{1}({A_{2}})\otimes W_{1}({\
A_{3}})\right) \phi \\
&=&\frac{1}{2}\overline{\phi }\left(
\begin{tabular}{ll|ll|ll}
&  &  & 1 &  &  \\
&  & 1 &  &  &  \\ \hline
& 1 &  &  &  & 1 \\
1 &  &  &  & 1 &  \\ \hline
&  &  & 1 &  &  \\
&  & 1 &  &  &
\end{tabular}
\right) \phi \\
&=&\phi _{1/16}\left( \overline{\phi _{1/2}}+\overline{\phi _{0}}\right) +
\overline{\phi _{1/16}}\left( \phi _{1/2}+\phi _{0}\right) =Z_{01}
\end{eqnarray*}

\begin{eqnarray*}
  Z_{12} &=&\frac{1}{2}\overline{\phi }\left( W_{1}({A_{2}})\otimes W_{2}({\
A_{3}})\right) \phi \\
&=&\frac{1}{2}\overline{\phi }\left(
\begin{tabular}{ll|ll|ll}
&  &  &  &  & 1 \\
&  &  &  & 1 &  \\ \hline
&  &  & 1 &  &  \\
&  & 1 &  &  &  \\ \hline
& 1 &  &  &  &  \\
1 &  &  &  &  &
\end{tabular}
\right) \phi \\
&=&\left| \phi _{0}\right| ^{2}+\left| \phi _{1/2}\right| ^{2}+\left| \phi
_{1/16}\right| ^{2}=Z_{00}
\end{eqnarray*}

We therefore obtain only three distinct partition functions, as
expected: equalities $Z_{i,j}=Z_{2-i,3-j}$ are consequences of
the $\phi_{r,s}=\phi_{3-r,4-s}$ identifications (remember that $i,j$
indices are shifted by one unit, compared with $r,s$ indices).
In the graph algebra $A_{3}$, we have $\sigma_{1}^2 = \sigma_{0} +
\sigma_{2}$, and, since $Oc(A_3)=A_3$, a corresponding toric structure,
with two twists, described by the matrix $W_{1;1}(A_3) = W_{0}(A_3) +
W_{2}(A3)$; in the Ising model, we can therefore also consider the (non
fundamental) partition function associated with the toric matrix
$W_{0}(A_2)\otimes W_{1;1}(A_3)$, \ie $Z_{00}+Z_{02} = \vert \phi_0 +
\phi_{1/2} \vert^2 + 2 \vert \phi_{1/16}\vert^2$ \cite{zuber86}.
We summarize the results for the fundamental partition functions in the
following table:

\begin{center}
\begin{tabular}{|l|}
\hline
$ Z_{00}=Z_{12}=\left| \phi _{0}\right| ^{2}+\left| \phi
_{1/2}\right| ^{2}+\left| \phi _{1/16}\right| ^{2}$ \\ \hline
$ Z_{01}=Z_{11}=\phi _{1/16}\left( \overline{\phi _{1/2}}+
\overline{\phi _{0}}\right) +\overline{\phi _{1/16}}\left( \phi _{1/2}+\phi
_{0}\right) $ \\ \hline
$ Z_{02}=Z_{10}=\overline{\phi _{0}}\phi _{1/2}+\overline{\phi
_{1/2}}\phi _{0}+\left| \phi _{1/16}\right| ^{2}$ \\ \hline
\end{tabular}
\end{center}

The representations of the modular group appearing in these theories
are usually not faithful: it can be checked that $T^{24}=1$ for the
affine $A_{2}$ case, $T^{16}=1$ for the affine $A_{3}$ case and
$T^{48}=1$ for the minimal model $(A_{2},A_{3})$.  $Z_{00}$,
determined above, is the usual modular invariant partition function of
the Ising model, and the associated toric matrix commutes in
particular with $T$.  Toric matrices associated with $Z_{02}$ and
$Z_{01}$ respectively commute respectively with $T^{2}$ and $T^{16}$.
The twisted partition function $Z_{02}$ is invariant under the congruence
subgroup $\Gamma_{0}(2)$, which involves an additional $Z_{2}$ symmetry (in
general, $\Gamma_{0}(k)$ is not an invariant subgroup of the modular group
$\Gamma$ but it contains, as well as all its conjugates, the principal
congruence subgroup $\Gamma(k)$, which is invariant in $\Gamma$, moreover,
$\Gamma_{0}(k)$ / $\Gamma_(k) \simeq \ZZ_{k}$).  In the language of twisted
boundary conditions, one assumes that the fields corresponding to given
characters are invariant only up to a phase under translations of the
lattice, \ie one assumes that they transform
according to one -- dimensional representations of the cyclic group
$\ZZ_{k}$. In the case of $Z_{0,2}$, these are periodic boundary conditions imposed on
the spin in one direction, and antiperiodic ones in the other \cite{zuber86}.
The interpretation of $Z_{01}$, for which the partition function is invariant
under the congruence subgroup $\Gamma_{0}(16)$ would be interesting to
study further.

\subsubsection{Potts model}
\label{sec:PottsSU2}
As a second example we consider the (non diagonal) Potts model with
central charge $c=\frac{4}{5}$.  This example differs from the
previous one since the pair of Dynkin diagrams involved are not both
of the $A_n$ type but $(A_{4},D_{4})$ with dual Coxeter numbers
$(\kappa_{A_4}=5,\kappa_ {D_4}=6)$.  Following the same steps as
before we present the fundamental partition functions.

The ${\cal A}$ diagram corresponding to $D_{4}$ is $A_{5}$ (same
Coxeter number) so that the twisted partition functions in this case
take the form
$$
  Z_{xy} = \frac{1}{2}\overline{\phi }\left( W_{x}({A_{4}})\otimes W_{y}(D_{{
4 }})\right) \phi
$$
$$
\begin{array}{l}
\phi = (\phi _{11},\phi _{21},\phi _{31},\phi _{41},\phi
_{12},\phi _{22},\phi _{32},\phi _{42},\phi _{13},\phi _{23},\phi _{33},\phi
_{43},\phi _{14},\phi _{24},\phi _{34},
\phi _{44},\phi _{15},\phi _{25},\phi
_{35},\phi _{45})
\end{array}
$$
The table of conformal weights for the $(A_{4},A_{5})$ system is given
by the following table which contains $4 \times 5 = 20$ entries but
only $10 = 20/2$ distinct weights.  Only those weights $h_{r,s}$ such
that $s$ belongs to the set of exponents of $D_{4}$ are conformal
weights for the (undeformed) $(A_{4},D_{4})$ model.  Exponents of
$D_{4}$ are $1,3, 5$ and $3$, as it is well known (or calculate the
adjacency matrix of the diagram and use section 2.3).  The conformal
weights obeying this criteria are typed in bold in the following
table; there are only 6 of them.  In the Virasoro minimal models
language, these states (called ``Potts model states'') correspond
only to a subset of conformal primary fields and are closed under
fusion rules.  Introduction of twists in general involves states
  of the $(A_{4},A_{5})$ system which are different from the usual
  Potts states; the usual identification, stemming from the $Z_{2}$
  symmetry, of course still holds.
\begin{center}
\begin{tabular}{|l|l|}
\hline
$\mathbf{h}_{11}\mathbf{=h}_{45}\mathbf{=0}$ & $\mathbf{h}_{21}\mathbf{=h}
_{35}\mathbf{=2/5}$ \\ \hline
$h_{12}=h_{44}=1/8\;$ & $h_{22}=h_{34}=1/40$ \\ \hline
$\mathbf{h}_{13}\mathbf{=h}_{43}\mathbf{=2/3}$ & $\mathbf{h}_{23}\mathbf{=h}
_{33}\mathbf{=1/15} $ \\ \hline
$h_{14}=h_{42}=13/8$ & $h_{24}=h_{32}=21/40$ \\ \hline
$\mathbf{h}_{15}\mathbf{=h}_{41}\mathbf{=3}$ & $\mathbf{h}_{25}\mathbf{=h}
_{31}\mathbf{=7/5}$ \\ \hline
\end{tabular}
\end{center}

In order to build the fundamental toric matrices for this model, we
need to use those corresponding to the Ocneanu graphs associated with
diagrams $A_{4}$ and $D_{4}$.  The first is easy: $Oc(A_{4})=A_{4}$.
The corresponding toric matrices are
{\scriptsize
$ W_{0}({A_{4}})=\left(
\begin{array}{llll}
1 & 0 & 0 & 0 \\
0 & 1 & 0 & 0 \\
0 & 0 & 1 & 0 \\
0 & 0 & 0 & 1
\end{array}
\right) \;, \; $ $W_{1}({A_{4}})=\left(
\begin{array}{llll}
0 & 1 & 0 & 0 \\
1 & 0 & 1 & 0 \\
0 & 1 & 0 & 1 \\
0 & 0 & 1 & 0
\end{array}
\right) ,\;$
}
and
{\scriptsize
$\; W_{2}({A_{4}})=\left(
\begin{array}{llll}
0 & 0 & 1 & 0 \\
0 & 1 & 0 & 1 \\
1 & 0 & 1 & 0 \\
0 & 1 & 0 & 0
\end{array}
\right) \;, \;
W_{3}({A_{4}})=\left(
\begin{array}{llll}
0 & 0 & 0 & 1 \\
0 & 0 & 1 & 0 \\
0 & 1 & 0 & 0 \\
1 & 0 & 0 & 0
\end{array}
\right) \;.\;$
}

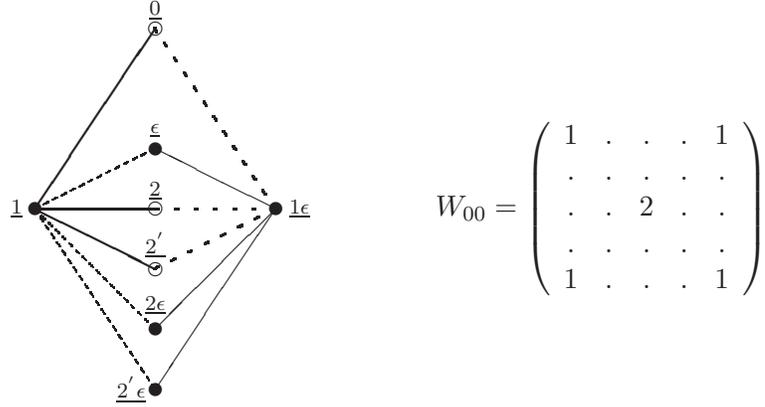
\begin{figure}[hhh]
\unitlength 0.8mm
\begin{center}
\begin{picture}(120,70)
\multiput(25,5)(0,10){2}{\circle*{2}}
\multiput(25,25)(0,10){2}{\circle{2}}
\put(25,45){\circle*{2}}
\put(5,35){\circle*{2}}
\put(25,65){\circle{2}}
\put(45,35){\circle*{2}}

\thicklines
\put(5,35){\line(1,0){20}}
\put(5,35){\line(2,-1){20}}
\put(5,35){\line(2,3){20}}

\thinlines
\put(45,35){\line(-1,-1){20}}
\put(45,35){\line(-2,-3){20}}
\put(45,35){\line(-2,1){20}}

\thicklines
\dashline[50]{1}(45,35)(25,35)
\dashline[50]{1}(45,35)(25,65)
\dashline[50]{1}(45,35)(25,25)

\thinlines
\dashline[50]{1}(5,35)(25,45)
\dashline[50]{1}(5,35)(25,15)
\dashline[50]{1}(5,35)(25,5)

\scriptsize
\put(25,68){\makebox(0,0){$\ud{0}$}}
\put(25,48){\makebox(0,0){$\ud{\epsilon}$}}
\put(25,38){\makebox(0,0){$\ud{2}$}}
\put(25,29){\makebox(0,0){$\ud{2^{'}}$}}
\put(25,19){\makebox(0,0){$\ud{2\epsilon}$}}
\put(21,5){\makebox(0,0){$\ud{2^{'}\epsilon}$}}
\put(49,35){\makebox(0,0){$\ud{1\epsilon}$}}
\put(2,35){\makebox(0,0){$\ud{1}$}}
\normalsize
\put(100,35){\makebox(0,0){$
W_{00}=\left( \begin{array}{ccccc}
1 & . & . & . & 1 \\
. & . & . & . & . \\
. & . & 2 & . & . \\
. & . & . & . & . \\
1 & . & . & . & 1 \\
\end{array} \right)
$
}}
\end{picture}
\caption{The $D_4$ Ocneanu graph and the modular invariant matrix}
\label{grocD4}
\end{center}
\end{figure}

The graph $Oc(D4)$, given in \cite{Ocneanu:paths} has eight points; it
looks like like two ``mixed'' copies of the diagram $D_{4}$ (see fig.
\ref{grocD4}).  A study of the corresponding algebra --- which is non
commutative\footnote{Classical symmetries of the $D_{4}$ diagram
are described by the non commutative group algebra of the permutation
group $S_{3}$, this non commutativity also shows up at the quantum level
in the structure of $Oc(D4)$.}  --- was performed in one of the sections of
\cite{CoqueGil:Tmodular}.  These points are labelled $0,1,2,2'$ and $
\epsilon, 1 \epsilon, 2 \epsilon, 2' \epsilon$.

There are eight generators but it turns out (see also
\cite{CoqueGil:Tmodular}) that there are only five distinct toric
matrices with one twist $W_{x}$ with $x \in \{0,2,\epsilon, 1, 1
\epsilon\}$ (indeed $W_{\epsilon}=W_{2\epsilon}=W_{2'\epsilon}$ and
$W_{2}=W_{2'}$).  We call $\overline{1}\doteq 1 \epsilon$.  The non
commutativity of $Oc(D_{4})$ does not show up in the fundamental
twisted partition functions, indeed, although $2 \epsilon \neq
\epsilon 2$ in this algebra \cite{CoqueGil:ADE} (actually
$\epsilon 2 = 2' \epsilon$), the toric matrices
associated with these two points are the same.
{\scriptsize{
$$ W_{0}(D_{{4}})=\left(
\begin{array}{lllll}
1 & 0 & 0 & 0 & 1 \\
0 & 0 & 0 & 0 & 0 \\
0 & 0 & 2 & 0 & 0 \\
0 & 0 & 0 & 0 & 0 \\
1 & 0 & 0 & 0 & 1
\end{array}
\right) \;,\;W_{2}(D_{{4}})=\left(
\begin{array}{lllll}
0 & 0 & 1 & 0 & 0 \\
0 & 0 & 0 & 0 & 0 \\
1 & 0 & 1 & 0 & 1 \\
0 & 0 & 0 & 0 & 0 \\
0 & 0 & 1 & 0 & 0
\end{array}
\right) \;,\;
W_{\epsilon}(D_{{4}})=\left(
\begin{array}{lllll}
0 & 0 & 0 & 0 & 0 \\
0 & 1 & 0 & 1 & 0 \\
0 & 0 & 0 & 0 & 0 \\
0 & 1 & 0 & 1 & 0 \\
0 & 0 & 0 & 0 & 0
\end{array}
\right) \;,\;
$$
$$
W_{1}(D_{{4}})=\left(
\begin{array}{lllll}
0 & 0 & 0 & 0 & 0 \\
1 & 0 & 2 & 0 & 1 \\
0 & 0 & 0 & 0 & 0 \\
1 & 0 & 2 & 0 & 1 \\
0 & 0 & 0 & 0 & 0
\end{array}
\right) \;,\;W_{\overline{1}}(D_{{4}})=\left(
\begin{array}{lllll}
0 & 1 & 0 & 1 & 0 \\
0 & 0 & 0 & 0 & 0 \\
0 & 2 & 0 & 2 & 0 \\
0 & 0 & 0 & 0 & 0 \\
0 & 1 & 0 & 1 & 0
\end{array}
\right)$$
}}
We expect $\ZZ_{2}$ to act as usual on $A_{4}$ but trivially on
$Oc(D_{4})$; identification is a priori $Z_{(x_{1},0);(x_{2},0)} =
Z_{(3 - x_{1},0);(x_{2},0)}$, \ie $Z_{0;x}=Z_{3;x}$ and $Z_{1;x} =
Z_{2;x}$; this can be checked explicitly.  So the number $4 \times 8 =
32$ of partition functions of this type reduces to $4 \times 5$
because of the accidental degeneracy between the toric matrices of
$D_{4}$ (only five cases) and actually to $2 \times 5 = 10$ because of
the $\ZZ_{2}$ identifications.
In the following we list these partition functions: we have $20$
fundamental toric matrices, but $10$ distinct partition functions (and
only four among them involving the usual Potts' model states):

\begin{eqnarray*}
  Z_{00} &=&\frac{1}{2}\overline{\phi }\left( W_{0}({A_{4}})\otimes W_{0}(D_{{
4 }})\right) \phi \\
&=&\frac{1}{2}\overline{\phi }\left(
{\scriptsize\begin{tabular}{llllllllllllllllllll}
1 & . & . & . & \multicolumn{1}{|l}{.} & . & . & . & \multicolumn{1}{|l}{.}
& . & . & . & \multicolumn{1}{|l}{.} & . & . & . & \multicolumn{1}{|l}{1} & .
& . & . \\
.. & 1 & . & \multicolumn{1}{l|}{.} & . & . & . & \multicolumn{1}{l|}{.} & .
& . & . & \multicolumn{1}{l|}{.} & . & . & . & \multicolumn{1}{l|}{.} & . & 1
& . & . \\
.. & . & 1 & \multicolumn{1}{l|}{.} & . & . & . & \multicolumn{1}{l|}{.} & .
& . & . & \multicolumn{1}{l|}{.} & . & . & . & \multicolumn{1}{l|}{.} & . & .
& 1 & . \\
.. & . & . & \multicolumn{1}{l|}{1} & . & . & . & \multicolumn{1}{l|}{.} & .
& . & . & \multicolumn{1}{l|}{.} & . & . & . & \multicolumn{1}{l|}{.} & . & .
& . & 1 \\ \hline
.. & . & . & \multicolumn{1}{l|}{.} & . & . & . & \multicolumn{1}{l|}{.} & .
& . & . & \multicolumn{1}{l|}{.} & . & . & . & \multicolumn{1}{l|}{.} & . & .
& . & . \\
.. & . & . & \multicolumn{1}{l|}{.} & . & . & . & \multicolumn{1}{l|}{.} & .
& . & . & \multicolumn{1}{l|}{.} & . & . & . & \multicolumn{1}{l|}{.} & . & .
& . & . \\
.. & . & . & \multicolumn{1}{l|}{.} & . & . & . & \multicolumn{1}{l|}{.} & .
& . & . & \multicolumn{1}{l|}{.} & . & . & . & \multicolumn{1}{l|}{.} & . & .
& . & . \\
.. & . & . & \multicolumn{1}{l|}{.} & . & . & . & \multicolumn{1}{l|}{.} & .
& . & . & \multicolumn{1}{l|}{.} & . & . & . & \multicolumn{1}{l|}{.} & . & .
& . & . \\ \hline
.. & . & . & \multicolumn{1}{l|}{.} & . & . & . & \multicolumn{1}{l|}{.} & 2
& . & . & \multicolumn{1}{l|}{.} & . & . & . & \multicolumn{1}{l|}{.} & . & .
& . & . \\
.. & . & . & \multicolumn{1}{l|}{.} & . & . & . & \multicolumn{1}{l|}{.} & .
& 2 & . & \multicolumn{1}{l|}{.} & . & . & . & \multicolumn{1}{l|}{.} & . & .
& . & . \\
.. & . & . & \multicolumn{1}{l|}{.} & . & . & . & \multicolumn{1}{l|}{.} & .
& . & 2 & \multicolumn{1}{l|}{.} & . & . & . & \multicolumn{1}{l|}{.} & . & .
& . & . \\
.. & . & . & \multicolumn{1}{l|}{.} & . & . & . & \multicolumn{1}{l|}{.} & .
& . & . & \multicolumn{1}{l|}{2} & . & . & . & \multicolumn{1}{l|}{.} & . & .
& . & . \\ \hline
.. & . & . & \multicolumn{1}{l|}{.} & . & . & . & \multicolumn{1}{l|}{.} & .
& . & . & \multicolumn{1}{l|}{.} & . & . & . & \multicolumn{1}{l|}{.} & . & .
& . & . \\
.. & . & . & \multicolumn{1}{l|}{.} & . & . & . & \multicolumn{1}{l|}{.} & .
& . & . & \multicolumn{1}{l|}{.} & . & . & . & \multicolumn{1}{l|}{.} & . & .
& . & . \\
.. & . & . & \multicolumn{1}{l|}{.} & . & . & . & \multicolumn{1}{l|}{.} & .
& . & . & \multicolumn{1}{l|}{.} & . & . & . & \multicolumn{1}{l|}{.} & . & .
& . & . \\
.. & . & . & \multicolumn{1}{l|}{.} & . & . & . & \multicolumn{1}{l|}{.} & .
& . & . & \multicolumn{1}{l|}{.} & . & . & . & \multicolumn{1}{l|}{.} & . & .
& . & . \\ \hline
1 & . & . & \multicolumn{1}{l|}{.} & . & . & . & \multicolumn{1}{l|}{.} & .
& . & . & \multicolumn{1}{l|}{.} & . & . & . & \multicolumn{1}{l|}{.} & 1 & .
& . & . \\
.. & 1 & . & \multicolumn{1}{l|}{.} & . & . & . & \multicolumn{1}{l|}{.} & .
& . & . & \multicolumn{1}{l|}{.} & . & . & . & \multicolumn{1}{l|}{.} & . & 1
& . & . \\
.. & . & 1 & \multicolumn{1}{l|}{.} & . & . & . & \multicolumn{1}{l|}{.} & .
& . & . & \multicolumn{1}{l|}{.} & . & . & . & \multicolumn{1}{l|}{.} & . & .
& 1 & . \\
.. & . & . & 1 & \multicolumn{1}{|l}{.} & . & . & . & \multicolumn{1}{|l}{.}
& . & . & . & \multicolumn{1}{|l}{.} & . & . & . & \multicolumn{1}{|l}{.} & .
& . & 1
\end{tabular}}
\right) \phi \\
&=&\frac{1}{2}\left[ \sum\limits_{r=1}^{4}\left( 2\left| \phi _{r,3}\right|
^{2}+\left| \phi _{r,1}+\phi _{r,5}\right| ^{2}\right) \right]
=\sum\limits_{r=1}^{2}\left( 2\left| \phi _{r,3}\right| ^{2}+\left| \phi
_{r,1}+\phi _{r,5}\right| ^{2}\right)
\end{eqnarray*}

\begin{eqnarray*}
  Z_{02} &=&\frac{1}{2}\overline{\phi }\left( W_{0}({A_{4}})\otimes 
W_{2}(D_{{4}})\right) \phi \\
&=&\frac{1}{2}\overline{\phi }\left(
{\scriptsize\begin{tabular}{llllllllllllllllllll}
.. & . & . & . & \multicolumn{1}{|l}{.} & . & . & . & \multicolumn{1}{|l}{1}
& . & . & . & \multicolumn{1}{|l}{.} & . & . & . & \multicolumn{1}{|l}{.} & .
& . & . \\
.. & . & . & \multicolumn{1}{l|}{.} & . & . & . & \multicolumn{1}{l|}{.} & .
& 1 & . & \multicolumn{1}{l|}{.} & . & . & . & \multicolumn{1}{l|}{.} & . & .
& . & . \\
.. & . & . & \multicolumn{1}{l|}{.} & . & . & . & \multicolumn{1}{l|}{.} & .
& . & 1 & \multicolumn{1}{l|}{.} & . & . & . & \multicolumn{1}{l|}{.} & . & .
& . & . \\
.. & . & . & \multicolumn{1}{l|}{.} & . & . & . & \multicolumn{1}{l|}{.} & .
& . & . & \multicolumn{1}{l|}{1} & . & . & . & \multicolumn{1}{l|}{.} & . & .
& . & . \\ \hline
.. & . & . & \multicolumn{1}{l|}{.} & . & . & . & \multicolumn{1}{l|}{.} & .
& . & . & \multicolumn{1}{l|}{.} & . & . & . & \multicolumn{1}{l|}{.} & . & .
& . & . \\
.. & . & . & \multicolumn{1}{l|}{.} & . & . & . & \multicolumn{1}{l|}{.} & .
& . & . & \multicolumn{1}{l|}{.} & . & . & . & \multicolumn{1}{l|}{.} & . & .
& . & . \\
.. & . & . & \multicolumn{1}{l|}{.} & . & . & . & \multicolumn{1}{l|}{.} & .
& . & . & \multicolumn{1}{l|}{.} & . & . & . & \multicolumn{1}{l|}{.} & . & .
& . & . \\
.. & . & . & \multicolumn{1}{l|}{.} & . & . & . & \multicolumn{1}{l|}{.} & .
& . & . & \multicolumn{1}{l|}{.} & . & . & . & \multicolumn{1}{l|}{.} & . & .
& . & . \\ \hline
1 & . & . & \multicolumn{1}{l|}{.} & . & . & . & \multicolumn{1}{l|}{.} & 1
& . & . & \multicolumn{1}{l|}{.} & . & . & . & \multicolumn{1}{l|}{.} & 1 & .
& . & . \\
.. & 1 & . & \multicolumn{1}{l|}{.} & . & . & . & \multicolumn{1}{l|}{.} & .
& 1 & . & \multicolumn{1}{l|}{.} & . & . & . & \multicolumn{1}{l|}{.} & . & 1
& . & . \\
.. & . & 1 & \multicolumn{1}{l|}{.} & . & . & . & \multicolumn{1}{l|}{.} & .
& . & 1 & \multicolumn{1}{l|}{.} & . & . & . & \multicolumn{1}{l|}{.} & . & .
& 1 & . \\
.. & . & . & \multicolumn{1}{l|}{1} & . & . & . & \multicolumn{1}{l|}{.} & .
& . & . & \multicolumn{1}{l|}{1} & . & . & . & \multicolumn{1}{l|}{.} & . & .
& . & 1 \\ \hline
.. & . & . & \multicolumn{1}{l|}{.} & . & . & . & \multicolumn{1}{l|}{.} & .
& . & . & \multicolumn{1}{l|}{.} & . & . & . & \multicolumn{1}{l|}{.} & . & .
& . & . \\
.. & . & . & \multicolumn{1}{l|}{.} & . & . & . & \multicolumn{1}{l|}{.} & .
& . & . & \multicolumn{1}{l|}{.} & . & . & . & \multicolumn{1}{l|}{.} & . & .
& . & . \\
.. & . & . & \multicolumn{1}{l|}{.} & . & . & . & \multicolumn{1}{l|}{.} & .
& . & . & \multicolumn{1}{l|}{.} & . & . & . & \multicolumn{1}{l|}{.} & . & .
& . & . \\
.. & . & . & \multicolumn{1}{l|}{.} & . & . & . & \multicolumn{1}{l|}{.} & .
& . & . & \multicolumn{1}{l|}{.} & . & . & . & \multicolumn{1}{l|}{.} & . & .
& . & . \\ \hline
.. & . & . & \multicolumn{1}{l|}{.} & . & . & . & \multicolumn{1}{l|}{.} & 1
& . & . & \multicolumn{1}{l|}{.} & . & . & . & \multicolumn{1}{l|}{.} & . & .
& . & . \\
.. & . & . & \multicolumn{1}{l|}{.} & . & . & . & \multicolumn{1}{l|}{.} & .
& 1 & . & \multicolumn{1}{l|}{.} & . & . & . & \multicolumn{1}{l|}{.} & . & .
& . & . \\
.. & . & . & \multicolumn{1}{l|}{.} & . & . & . & \multicolumn{1}{l|}{.} & .
& . & 1 & \multicolumn{1}{l|}{.} & . & . & . & \multicolumn{1}{l|}{.} & . & .
& . & . \\
.. & . & . & . & \multicolumn{1}{|l}{.} & . & . & . & \multicolumn{1}{|l}{.}
& . & . & 1 & \multicolumn{1}{|l}{.} & . & . & . & \multicolumn{1}{|l}{.} & .
& . & .
\end{tabular}}
\right) \phi \\
&=&\frac{1}{2}\left\{ \sum\limits_{r=1}^{4}\left[ \left| \phi _{r,3}\right|
^{2}+\left( \overline{\phi _{r,3}}\left( \phi _{r,1}+\phi _{r,5}\right)
+hc\right) \right] \right\} =\sum\limits_{r=1}^{2}\left[ \left| \phi
_{r,3}\right| ^{2}+\left( \overline{\phi _{r,3}}\left( \phi _{r,1}+\phi
_{r,5}\right) +hc\right) \right]
\end{eqnarray*}

 From now on, we no longer display the tensor product of
matrices.

\begin{eqnarray*}
  Z_{0\epsilon} &=&\frac{1}{2}\overline{\phi }\left( 
W_{0}({A_{4}})\otimes W_{\epsilon}(D_{4})\right) \phi \\
&=&\frac{1}{2}\left[ \sum\limits_{r=1}^{4}\left| \phi _{r,2}+\phi
_{r,4}\right| ^{2}\right] =\sum\limits_{r=1}^{2}\left| \phi _{r,2}+\phi
_{r,4}\right| ^{2}
\\
  Z_{01} &=&\frac{1}{2}\overline{\phi }\left( W_{0}({A_{4}})\otimes 
W_{1}(D_{4}))\right) \phi \\
&=&\frac{1}{2}\left[ \sum\limits_{r=1}^{4}\left( \overline{\phi _{r,2}}+
\overline{\phi _{r,4}}\right) \left( \phi _{r,1}+\phi _{r,5}+2\phi
_{r,3}\right) \right] =\sum\limits_{r=1}^{2}\left[ \left( \overline{\phi
_{r,2}}+\overline{\phi _{r,4}}\right) \left( \phi _{r,1}+\phi _{r,5}+2\phi
_{r,3}\right) \right]
\\
  Z_{0\overline{1}} &=&\frac{1}{2}\overline{\phi }\left( 
W_{0}({A_{4}})\otimes W_{\overline{1}}(D_{4}))\right) \phi \\
&=&\frac{1}{2}\left[ \sum\limits_{r=1}^{4}\left( \phi _{r,2}+\phi
_{r,4}\right) \left( \overline{\phi _{r,1}}+\overline{\phi _{r,5}}+2
\overline{\phi _{r,3}}\right) \right] =\sum\limits_{r=1}^{2}\left[ \left(
\phi _{r,2}+\phi _{r,4}\right) \left( \overline{\phi _{r,1}}+\overline{\phi
_{r,5}}+2\overline{\phi _{r,3}}\right) \right] \\
&=&\overline{Z_{03}}
\end{eqnarray*}
\begin{eqnarray*}
  Z_{10} &=&\frac{1}{2}\overline{\phi }\left( W_{1}({A_{4}})\otimes 
W_{0}(D_{{4}})\right) \phi  \\
&=&2\left| \phi _{2,3}\right| ^{2}+\left| \phi _{2,1}+\phi _{2,5}\right|
^{2}+\left[ \left( \overline{\phi _{2,1}}+\overline{\phi _{2,5}}\right)
\left( \phi _{1,1}+\phi _{1,5}\right) +2\overline{\phi _{2,3}}\phi
_{1,3}+hc\right]
\\
  Z_{12} &=&\frac{1}{2}\overline{\phi }\left( W_{1}({A_{4}})\otimes 
W_{2}(D_{4}))\right) \phi  \\
&=&\left[ \phi _{1,3}\left( \overline{\phi _{2,1}}+\overline{\phi _{2,5}}+
\overline{\phi _{2,3}}\right) +\sum\limits_{r=1}^{2}\phi _{2,3}\left(
\overline{\phi _{r,1}}+\overline{\phi _{r,5}}\right) +hc\right] +\left| \phi
_{2,3}\right| ^{2}
\\
  Z_{1\epsilon} &=&\frac{1}{2}\overline{\phi }\left( 
W_{1}({A_{4}})\otimes W_{\epsilon}(D_{4})\right) \phi \\
\\ {}
&=&\left| \phi _{2,2}+\phi _{2,4}\right| ^{2}+\left[ \left( \phi _{2,2}+\phi
_{2,4}\right) \left( \overline{\phi _{1,2}}+\overline{\phi _{1,4}}\right)
+hc\right]
\\
  Z_{11} &=&\frac{1}{2}\overline{\phi }\left( W_{1}({A_{4}})\otimes 
W_{1}(D_{4}))\right) \phi \\
&=&\sum\limits_{r=1}^{2}\left[ \left( \overline{\phi _{r,2}}+\overline{\phi
_{r,4}}\right) \left( \phi _{2,1}+\phi _{2,5}+2\phi _{2,3}\right) \right]
+\left( \overline{\phi _{2,2}}+\overline{\phi _{2,4}}\right) \left( \phi
_{1,1}+\phi _{1,5}+2\phi _{1,3}\right)
\\
  Z_{1\overline{1}} &=&\frac{1}{2}\overline{\phi }\left( 
W_{1}({A_{4}})\otimes W_{\overline{1}}(D_{4}))\right) \phi = 
\overline{Z_{13}}
\end{eqnarray*}
$$
\begin{array}{ccc}
Z_{20} = \frac{1}{2}\overline{\phi }\left( W_{2}({A_{4}})\otimes 
W_{0}(D_{{4}})\right) \phi
= Z_{10}
&{\; ,}&
  Z_{22} = \frac{1}{2}\overline{\phi }\left( W_{2}({A_{4}})\otimes 
W_{2}(D_{4}))\right) \phi
  = Z_{12}
  \\
  Z_{2\epsilon} =\frac{1}{2}\overline{\phi }\left( 
W_{2}({A_{4}})\otimes W_{\epsilon}(D_{4})\right) \phi
  = Z_{1\epsilon}
&{\; ,}&
  Z_{21} =\frac{1}{2}\overline{\phi }\left( W_{2}({A_{4}})\otimes 
W_{1}(D_{4}))\right) \phi
= Z_{11}
\\
  Z_{2\overline{1}} =\frac{1}{2}\overline{\phi }\left( 
W_{2}({A_{4}})\otimes W_{\overline{1}}(D_{4}))\right) \phi
  = Z_{1\overline{1}}
&{\; ,}&
  Z_{30} =\frac{1}{2}\overline{\phi }\left( W_{3}({A_{4}})\otimes 
W_{0}(D_{{4}})\right) \phi
= Z_{00}
\\
  Z_{32} =\frac{1}{2}\overline{\phi }\left( W_{3}({A_{4}})\otimes 
W_{2}(D_{4}))\right) \phi
= Z_{02}
&{\; ,}&
  Z_{3\epsilon} =\frac{1}{2}\overline{\phi }\left( 
W_{3}({A_{4}})\otimes W_{\epsilon}(D_{4})\right) \phi
  = Z_{0\epsilon}
\\
  Z_{31} =\frac{1}{2}\overline{\phi }\left( W_{3}({A_{4}})\otimes 
W_{1}(D_{4}))\right) \phi
  = Z_{01}
&{\; ,}&
  Z_{3\overline{1}} =\frac{1}{2}\overline{\phi }\left( 
W_{3}({A_{4}})\otimes W_{\overline{1}}(D_{4}))\right) \phi
= Z_{0\overline{1}}
\end{array}
$$


The results are summarized in the following table:

\begin{center}
\begin{tabular}{|l|}
\hline
$\mathbf{ Z}_{00}\mathbf{=Z}_{30}\mathbf{=}\sum\limits_{r=1}^{2}
\left( 2\left| \phi _{r,3}\right| ^{2}+\left| \phi _{r,1}+\phi _{r,5}\right|
^{2}\right) $ \\ \hline
$\mathbf{ Z}_{02}\mathbf{=Z}_{32}\mathbf{=}\sum\limits_{r=1}^{2}
\left[ \left| \phi _{r,3}\right| ^{2}+\left( \overline{\phi _{r,3}}\left(
\phi _{r,1}+\phi _{r,5}\right) +hc\right) \right] $ \\ \hline
$ Z_{0\epsilon}=Z_{3\epsilon}=\sum\limits_{r=1}^{2}\left| \phi _{r,2}+\phi
_{r,4}\right| ^{2}$ \\ \hline
$ Z_{01}=Z_{31}=\sum\limits_{r=1}^{2}\left[ \left( \overline{\phi
_{r,2}}+\overline{\phi _{r,4}}\right) \left( \phi _{r,1}+\phi _{r,5}+2\phi
_{r,3}\right) \right] $ \\ \hline
$ 
Z_{0\overline{1}}=Z_{3\overline{1}}=\overline{Z_{01}}=\overline{Z_{31}}$ 
\\ \hline
$\mathbf{ Z}_{10}\mathbf{=Z}_{20}\mathbf{=2}\left| \phi
_{2,3}\right| ^{2}\mathbf{+}\left| \phi _{2,1}+\phi _{2,5}\right| ^{2}
\mathbf{+}\left[ \left( \overline{\phi _{2,1}}+\overline{\phi _{2,5}}\right)
\left( \phi _{1,1}+\phi _{1,5}\right) +2\overline{\phi _{2,3}}\phi
_{1,3}+hc\right] $ \\ \hline
$\mathbf{ Z}_{12}\mathbf{=Z}_{22}\mathbf{=}\left[ \phi
_{1,3}\left( \overline{\phi _{2,1}}+\overline{\phi _{2,5}}+\overline{\phi
_{2,3}}\right) +\sum\limits_{r=1}^{2}\phi _{2,3}\left( \overline{\phi _{r,1}}
+\overline{\phi _{r,5}}\right) +hc\right] \mathbf{+}\left| \phi
_{2,3}\right| ^{2}$ \\ \hline
$ Z_{1\epsilon}=Z_{2\epsilon}=\left| \phi _{2,2}+\phi _{2,4}\right| ^{2}+\left[
\left( \phi _{2,2}+\phi _{2,4}\right) \left( \overline{\phi _{1,2}}+
\overline{\phi _{1,4}}\right) +hc\right] $ \\ \hline
$ Z_{11}=Z_{21}=\sum\limits_{r=1}^{2}\left[ \left( \overline{\phi
_{r,2}}+\overline{\phi _{r,4}}\right) \left( \phi _{2,1}+\phi _{2,5}+2\phi
_{2,3}\right) \right] +\left( \overline{\phi _{2,2}}+\overline{\phi _{2,4}}
\right) \left( \phi _{1,1}+\phi _{1,5}+2\phi _{1,3}\right) $ \\ \hline
$ 
Z_{1\overline{1}}=Z_{2\overline{1}}=\overline{Z_{11}}=\overline{Z_{21}}$ 
\\ \hline
\end{tabular}
\end{center}

Only the partition functions written in bold letters involve the subset
of states corresponding to the undeformed three states Potts model;
we rewrite them by using conformal weights as field subscripts.
We set $\mathbf{Z}_{0}=Z_{00}, \mathbf{Z}_{1}=Z_{02},
\mathbf{Z}_{2}=Z_{10}, \mathbf{Z}_{3}=Z_{12}$.

\begin{center}
\begin{tabular}{|l|}
\hline
$\mathbf{ Z}_{0}\mathbf{=}2\left( \left| \phi _{2/3}\right|
^{2}+\left| \phi _{1/15}\right| ^{2}\right) +\left| \phi _{0}+\phi
_{3}\right| ^{2}+\left| \phi _{2/5}+\phi _{7/5}\right| ^{2}$ \\ \hline
$\mathbf{ Z}_{1}\mathbf{=}\left| \phi _{2/3}\right| ^{2}+\left|
\phi _{1/15}\right| ^{2}+\left( \overline{\phi _{2/3}}\left( \phi _{0}+\phi
_{3}\right) +\overline{\phi _{1/15}}\left( \phi _{2/5}+\phi _{7/5}\right)
+hc\right) $ \\ \hline
$\mathbf{ Z}_{2}\mathbf{=2}\left| \phi _{1/15}\right| ^{2}\mathbf{
+}\left| \phi _{2/5}+\phi _{7/5}\right| ^{2}\mathbf{+}\left[ \left(
\overline{\phi _{2/5}}+\overline{\phi _{7/5}}\right) \left( \phi _{0}+\phi
_{3}\right) +2\overline{\phi _{1/15}}\phi _{2/3}+hc\right] $ \\ \hline
$\mathbf{ Z}_{3}\mathbf{=}\left[ \phi _{2/3}\left( \overline{\phi
_{2/5}}+\overline{\phi _{7/5}}+\overline{\phi _{1/15}}\right) +\phi
_{1/15}\left( \overline{\phi _{0}}+\overline{\phi _{3}}\right) +\phi
_{1/15}\left( \overline{\phi _{2/5}}+\overline{\phi _{7/5}}\right)
+hc\right] \mathbf{+}\left| \phi _{1/15}\right| ^{2}$ \\ \hline
\end{tabular}
\end{center}

When no twisted boundary conditions are imposed, we recover the usual
modular invariant partition function $\mathbf{ Z}_{0}$.

\subsubsection{The $A_{10} - E_{6} $ example}

Finally we consider the model $ (A_{10} - E_{6})$ with dual Coxeter
numbers $(\kappa_{A_{10}}=11,\, \kappa_{E_6}=12)$ and central charge
$c=\frac{21}{22}$.  First notice that ${\cal A}(E_{6}) = A_{11}$ so
that conformal weights $h_{r,s}$ of this model (which is unitary since
$11=10+1$) is a subset of the set of conformal weights for
$(A_{10},A_{11})$.  Index $r$ stands for $A_{10}$ and $s$ for
$A_{11}$.  A priori there are $10\times 11 = 110$ possibilities, but
because of the $Z_{2}$ symmetry of the Kac table
($h_{r,s}=h_{11-r,12-s}; \;1\leq r\leq 10\;;\;1\leq s\leq 11$), there
are only half of them, so $55$.  The following table lists the
conformal dimensions associated to the primary fields of this model.
Only those columns $h_{r,s}$ such that $s$ belongs to the set of
exponents of $E_{6}$ are conformal weights for the usual (undeformed)
$(A_{10},E_{6})$ model.  The exponents of diagram $E_{6}$ are
$1,4,5,7,8,11$ (cf section 2.3).  Columns $1,4,5,7,8,11$ of the
following table are in boldface.

\begin{center}
\begin{tabular}{|l|l|l|l|l|l|l|l|l|l|l|l|}
\hline
$h_{rs}$ & ${\bf s=1}$ & ${2}$ & ${3}$ & ${\bf 4}$ & ${\bf 5}$ & ${6}$ & $
{\bf 7}$ & ${\bf 8}$ & ${9}$ & ${10}$ & ${\bf 11}$ \\ \hline
${r=1}$ & ${\bf 0}$ & $\frac{3}{16}$ & $\frac{5}{6}$ & ${\bf \frac{31}{16}}$
& ${\bf \frac{7}{2}}$ & $\frac{265}{48}$ & ${\bf 8}$ & ${\bf \frac{175}{16}}$
& $\frac{43}{3}$ & $\frac{291}{16}$ & ${\bf \frac{45}{2}}
\rule[-0.20cm]{0cm}{0.6cm}$  \\ \hline
${2}$ & ${\bf \frac{7}{22}}$ & $\frac{1}{176}$ & $\frac{5}{33}$ & ${\bf
\frac{133}{176}}$ & ${\bf \frac{20}{11}}$ & $\frac{1763}{528}$ & ${\bf \frac{
117}{22}}$ & ${\bf \frac{1365}{176}}$ & $\frac{703}{66}$ & $\frac{2465}{176}$
& ${\bf \frac{196}{11}}$ \rule[-0.2cm]{0.0cm}{0.6cm} \\ \hline
${3}$ & ${\bf \frac{13}{11}}$ & $\frac{65}{176}$ & $\frac{1}{66}$ & ${\bf
\frac{21}{176}}$ & ${\bf \frac{15}{22}}$ & $\frac{899}{528}$ & ${\bf \frac{35
}{11}}$ & ${\bf \frac{901}{176}}$ & $\frac{248}{66}$ & $\frac{1825}{176}$ & $
{\bf \frac{301}{22}}$ \rule[-0.2cm]{0.0cm}{0.6cm} \\ \hline
${4}$ & ${\bf \frac{57}{22}}$ & $\frac{225}{176}$ & $\frac{14}{33}$ & ${\bf
\frac{5}{176}}$ & ${\bf \frac{1}{11}}$ & $\frac{323}{528}$ & ${\bf \frac{35}{
22}}$ & ${\bf \frac{533}{176}}$ & $\frac{325}{66}$ & $\frac{1281}{176}$ & $
{\bf \frac{111}{11}}$ \rule[-0.2cm]{0.0cm}{0.6cm} \\ \hline
${5}$ & ${\bf \frac{50}{11}}$ & $\frac{481}{176}$ & $\frac{91}{66}$ & ${\bf
\frac{85}{176}}$ & ${\bf \frac{1}{22}}$ & $\frac{35}{528}$ & ${\bf \frac{6}{
11}}$ & ${\bf \frac{261}{176}}$ & $\frac{95}{33}$ & $\frac{833}{176}$ & $
{\bf \frac{155}{22}}$ \rule[-0.2cm]{0.0cm}{0.6cm} \\ \hline
\end{tabular}
\end{center}

In order to build the fundamental toric matrices for this model we
need to use the toric matrices (with one twist) associated with
diagrams $A_{10}$ and $E_{6}$.  The more general toric matrices of the
model can be gotten from the multiplication table of $Oc(A_{10})$
which coincides with $A_{10}$ itself (so it is easy) and the
multiplication table of $Oc(E_{6})$ was explicitly studied in a
previous section.  We have twelve toric matrices $ W_{i}(E_{6})$ which
are of size $11 \times 11$ and ten toric matrices $W_{i}({A_{10}})$
which are of size $10\times 10$.  Partition functions are then
associated with matrices $Z_{(x_{1},0);(x_{2},0)} = 1/2 \,
W_{x_{1},0}(A_{10}) \otimes W_{x_{2},0}(E_{6})$ (of size $110\times
110$).  A priori, we have $10 \times 12 = 120$ of them, but because of
$Z_{2}$ identifications only half
of them, so $60$, will be distinct.
It is therefore enough to consider partition
functions of the type $Z_{(x_{1},0);(x_{2},0)}$ for $x_{1}=0,1,2,3,4
\in A_{10}$ and $x_{2}$, a point of $Oc(E_{6})$, so $x_{2}$ is a
member of the list: $\{(0\otimesdot 0, 0 \otimesdot 3, 0 \otimesdot
4), (1 \otimesdot 0, 2 \otimesdot 0, 5 \otimesdot 0), (0 \otimesdot 1,
0 \otimesdot 2, 0 \otimesdot 5), (1 \otimesdot 1, 2 \otimesdot 1, 5
\otimesdot 1)\}$.  Since the $12$ points $x_{2}$ of $Oc(E_{6})$ belong
to four distinct subsets of three points each (ambichiral, left
chiral, right chiral, or complementary), it is natural to decompose
our $60$ candidates into four subsets of $15$ each, labelled in the
same way.  The corresponding table of results is quite long\ldots so
we shall only give the fifteen twisted partition functions that belong
to the first family: they are of the type $Z_{(x_{1},0);(x_{2},0)}$
for $x_{1} = 0,1,2,3,4$ and $x_{2} = \{0\otimesdot 0, 0 \otimesdot 3,
0 \otimesdot 4\}$.  These partition functions are the ones that
involve only the combination of characters $(1,7),(4,8),(5,11)$, \ie
the adapted vector $w$ of section 3.1.5. The symmetry relations for this
family will read: $Z_{(x_{1},0);(0\otimesdot 0 ,0)} = Z_{(9 -
x_{1},0);(0\otimesdot 4,0)}$ and $Z_{(x_{1},0);(0\otimesdot 3 ,0)} =
Z_{(9 - x_{1},0);(0\otimesdot 3,0)} $. Similar expressions are obtained for
the other three families. Here are the explicit results.

Calling $\phi$ the characters vector of $110$ components (only $55$
distinct conformal weight), the $15$ twisted partition functions
involving the characters of type $\phi _{i1}+\phi _{i7}$, $\phi
_{i4}+\phi _{i8}$, $\phi _{i5}+\phi _{i11}$ take the following general
form:

\[
Z_{ij}=\frac{1}{2} \, \overline{\phi }\left( W_{i}({A_{10}})\otimes
W_{j}(E_{6})\right) \phi
\]
with $i=0,...,9$ the labels of the vertices of the $A_{10}$ diagram
and $j=0,1,2 $ corresponding respectively to $\{0\otimesdot 0,
0 \otimesdot 3, 0 \otimesdot 4\}$ vertices of $Oc(E_{6})$.

\begin{center}
\begin{tabular}{|l|}
\hline
$ Z_{00}=Z_{92}=\sum\limits_{r=1}^{5}\left| \phi _{r,1}+\phi
_{r,7}\right| ^{2}+\left| \phi _{r,4}+\phi _{r,8}\right| ^{2}+\left| \phi
_{r,5}+\phi _{r,11}\right| ^{2}$ \\ \hline
$ Z_{10}=Z_{82}=\left[ \sum\limits_{r=1}^{4}(\overline{\phi _{r,1}
}+\overline{\phi _{r,7}})(\phi _{r+1,1}+\phi _{r+1,7})+(\overline{\phi _{r,4}
}+\overline{\phi _{r,8}})(\phi _{r+1,4}+\phi _{r+1,8})+\right. $ \\
$\left. (\overline{\phi _{r,5}}+\overline{\phi _{r,11}})+(\phi _{r+1,5}+\phi
_{r+1,11})+(\overline{\phi _{5,5}}+\overline{\phi _{5,11}})(\phi _{5,1}+\phi
_{5,7})+hc\right] +\left| \phi _{5,4}+\phi _{5,8}\right| ^{2}$ \\ \hline
$ Z_{20}=Z_{72}=\sum\limits_{r=2}^{5}\left| \phi _{r,1}+\phi
_{r,7}\right| ^{2}+\left| \phi _{r,4}+\phi _{r,8}\right| ^{2}+\left| \phi
_{r,5}+\phi _{r,11}\right| ^{2}$ \\
$+\sum\limits_{r=1}^{3}(\overline{\phi _{r,1}}+\overline{\phi _{r,7}})(\phi
_{r+2,1}+\phi _{r+2,7})+(\overline{\phi _{r,4}}+\overline{\phi _{r,8}})(\phi
_{r+2,4}+\phi _{r+2,8})+$ \\
$(\overline{\phi _{r,5}}+\overline{\phi _{r,11}})(\phi _{r+2,5}+\phi
_{r+2,11})+(\overline{\phi _{5,1}}+\overline{\phi _{5,7}})(\phi _{4,1}+\phi
_{4,7})+(\overline{\phi _{5,4}}+\overline{\phi _{5,8}})(\phi _{4,4}+\phi
_{4,8})$ \\
$+(\overline{\phi _{5,5}}+\overline{\phi _{5,11}})(\phi _{4,5}+\phi
_{4,11})+hc$ \\ \hline
$ Z_{30}=Z_{62}=\sum\limits_{r=0}^{2}(\overline{\phi _{4,1}}+
\overline{\phi _{4,7}})(\phi _{2r+1,1}+\phi _{2r+1,7})+(\overline{\phi _{4,4}
}+\overline{\phi _{4,8}})(\phi _{2r+1,4}+\phi _{2r+1,8})+$ \\
$(\overline{\phi _{4,5}}+\overline{\phi _{4,11}})(\phi _{2r+1,5}+\phi
_{2r+1,11})+\sum\limits_{r=1}^{2}(\overline{\phi _{2,1}}+\overline{\phi
_{2,7}})(\phi _{2r+1,1}+\phi _{2r+1,7})+ $\\
$
(\overline{\phi _{2,4}}+\overline{\phi _{2,8}})(\phi _{2r+1,4}+\phi
_{2r+1,8})+(\overline{\phi _{2,5}}+\overline{\phi _{2,11}})(\phi
_{2r+1,5}+\phi _{2r+1,11})+$ \\
$(\overline{\phi _{5,1}}+\overline{\phi _{5,7}})(\phi _{3,5}+\phi _{3,11})+(
\overline{\phi _{5,4}}+\overline{\phi _{5,8}})(\phi _{3,4}+\phi _{3,8})+(
\overline{\phi _{5,5}}+\overline{\phi _{5,11}})(\phi _{3,1}+\phi _{3,7})$ \\
$+\sum\limits_{r=4}^{5}(\overline{\phi _{r,1}}+\overline{\phi _{r,7}})(\phi
_{r,5}+\phi _{r,11})+hc+\sum\limits_{r=4}^{5}\left| \phi _{r,4}+\phi
_{r,8}\right| ^{2}$ \\ \hline
$ Z_{40}=Z_{52}=\sum\limits_{r=3}^{5}\left| \phi _{r,1}+\phi
_{r,7}\right| ^{2}+\left| \phi _{r,4}+\phi _{r,8}\right| ^{2}+\left| \phi
_{r,5}+\phi _{r,11}\right| ^{2}$ \\
$+\sum\limits_{r=1,3}(\overline{\phi _{5,1}}+\overline{\phi _{5,7}})(\phi
_{r,1}+\phi _{r,7})+(\overline{\phi _{5,4}}+\overline{\phi _{5,8}})(\phi
_{r,4}+\phi _{r,8})+(\overline{\phi _{5,5}}+\overline{\phi _{5,11}})(\phi
_{r,5}+\phi _{r,11})$ \\
$+(\overline{\phi _{4,1}}+\overline{\phi _{4,7}})(\phi _{2,1}+\phi _{2,7})+(
\overline{\phi _{4,4}}+\overline{\phi _{4,8}})(\phi _{2,4}+\phi _{2,8})+(
\overline{\phi _{4,5}}+\overline{\phi _{4,11}})(\phi _{2,5}+\phi _{2,11})$
\\
$+\sum\limits_{r=2}^{4}(\overline{\phi _{5,1}}+\overline{\phi _{5,7}})(\phi
_{r,5}+\phi _{r,11})+(\overline{\phi _{5,4}}+\overline{\phi _{5,8}})(\phi
_{r,4}+\phi _{r,8})+(\overline{\phi _{5,5}}+\overline{\phi _{5,11}})(\phi
_{r,1}+\phi _{r,7})+hc$ \\ \hline
$ Z_{50}=Z_{42}=\sum\limits_{r=3}^{5}\left| \phi _{r,4}+\phi
_{r,8}\right| ^{2}+\left( (\overline{\phi _{r,1}}+\overline{\phi _{r,7}}
)(\phi _{r,5}+\phi _{r,11})+hc\right) $ \\
$+\sum\limits_{r=2,4}(\overline{\phi _{5,1}}+\overline{\phi _{5,7}})(\phi
_{r,1}+\phi _{r,7})+(\overline{\phi _{5,4}}+\overline{\phi _{5,8}})(\phi
_{r,4}+\phi _{r,8})+(\overline{\phi _{5,5}}+\overline{\phi _{5,11}})(\phi
_{r,5}+\phi _{r,11})$ \\
$+\sum\limits_{r=1,3}(\overline{\phi _{5,1}}+\overline{\phi _{5,7}})(\phi
_{r,5}+\phi _{r,11})+(\overline{\phi _{5,4}}+\overline{\phi _{5,8}})(\phi
_{r,4}+\phi _{r,8})+(\overline{\phi _{5,5}}+\overline{\phi _{5,11}})(\phi
_{r,1}+\phi _{r,7})$ \\
$+(\overline{\phi _{4,5}}+\overline{\phi _{4,11}})\left( (\phi _{2,1}+\phi
_{2,7})+(\phi _{3,5}+\phi _{3,11})\right) +(\overline{\phi _{4,1}}+\overline{
\phi _{4,7}})\left( (\phi _{2,5}+\phi _{2,11})+(\phi _{3,1}+\phi
_{3,7})\right) $ \\
$\sum\limits_{r=1,3}(\overline{\phi _{4,4}}+\overline{\phi _{4,8}})(\phi
_{r,4}+\phi _{r,8})+hc$ \\ \hline
\end{tabular}
\\[0pt]

\begin{tabular}{|l|}
\hline
$ Z_{60}=Z_{32}=\sum\limits_{r=4}^{5}\left| \phi _{r,1}+\phi
_{r,7}\right| ^{2}+\left| \phi _{r,4}+\phi _{r,8}\right| ^{2}+\left| \phi
_{r,5}+\phi _{r,11}\right| ^{2}$ \\
$+(\overline{\phi _{5,1}}+\overline{\phi _{5,7}})(\phi _{3,1}+\phi _{3,7})+(
\overline{\phi _{5,4}}+\overline{\phi _{5,8}})(\phi _{3,4}+\phi _{3,8})+(
\overline{\phi _{5,5}}+\overline{\phi _{5,11}})(\phi _{3,5}+\phi _{3,11})$
\\
$+\sum\limits_{r=1,3,5}(\overline{\phi _{4,1}}+\overline{\phi _{4,7}})(\phi
_{r,5}+\phi _{r,11})+(\overline{\phi _{4,4}}+\overline{\phi _{4,8}})(\phi
_{r,4}+\phi _{r,8})+(\overline{\phi _{4,5}}+\overline{\phi _{4,11}})(\phi
_{r,1}+\phi _{r,7})+hc$ \\ \hline
$ Z_{70}=Z_{22}=\sum\limits_{r=2}^{5}\left| \phi _{r,4}+\phi
_{r,8}\right| ^{2}+(\overline{\phi _{r,1}}+\overline{\phi _{r,7}})(\phi
_{r,5}+\phi _{r,11})+$ \\
$+\sum\limits_{r=1}^{3}(\overline{\phi _{r,1}}+\overline{\phi _{r,7}})(\phi
_{r+2,5}+\phi _{r+2,11})+(\overline{\phi _{r,4}}+\overline{\phi _{r,8}}
)(\phi _{r+2,4}+\phi _{r+2,8})+(\overline{\phi _{r,5}}+\overline{\phi _{r,11}
})(\phi _{r+2,1}+\phi _{r+2,7})$ \\
$+(\overline{\phi _{5,1}}+\overline{\phi _{5,7}})(\phi _{4,1}+\phi _{4,7})+(
\overline{\phi _{5,4}}+\overline{\phi _{5,8}})(\phi _{4,4}+\phi _{4,8})+(
\overline{\phi _{5,5}}+\overline{\phi _{5,11}})(\phi _{4,5}+\phi _{4,11})+hc$
\\ \hline
$ Z_{80}=Z_{12}=\sum\limits_{r=1}^{4}(\overline{\phi _{r,1}}+
\overline{\phi _{r,7}})(\phi _{r+1,5}+\phi _{r+1,11})+(\overline{\phi _{r,4}}
+\overline{\phi _{r,8}})(\phi _{r+1,4}+\phi _{r+1,8})+$ \\
$(\overline{\phi _{r,5}}+\overline{\phi _{r,11}})(\phi _{r+1,1}+\phi
_{r+1,7})+hc+\left| \phi _{5,1}+\phi _{5,7}\right| ^{2}+\left| \phi
_{5,4}+\phi _{5,8}\right| ^{2}+\left| \phi _{5,5}+\phi _{5,11}\right| ^{2}$
\\ \hline
$ Z_{90}=Z_{02}=\sum\limits_{r=1}^{5}\left| \phi _{r,4}+\phi
_{r,8}\right| ^{2}+\left( (\overline{\phi _{r,1}}+\overline{\phi _{r,7}}
)(\phi _{r,5}+\phi _{r,11})+hc\right) $ \\ \hline
\end{tabular}
\\[0pt]
\end{center}

\begin{center}
\begin{tabular}{|l|}
\hline
$ Z_{01}=Z_{91}=\sum\limits_{r=1}^{5}(\overline{\phi _{r,4}}+
\overline{\phi _{r,8}})(\phi _{r,5}+\phi _{r,11}+\phi _{r,1}+\phi _{r,7})+hc$
\\ \hline
$ Z_{11}=Z_{81}=\sum\limits_{r=1}^{4}(\overline{\phi _{r,4}}+
\overline{\phi _{r,8}})(\phi _{r+1,5}+\phi _{r+1,11}+\phi _{r+1,1}+\phi
_{r+1,7})+(r\leftrightarrow r+1)+hc$ \\
$+(\overline{\phi _{5,4}}+\overline{\phi _{5,8}})(\phi _{5,5}+\phi
_{5,11}+\phi _{5,1}+\phi _{5,7})+hc$ \\ \hline
$ Z_{21}=Z_{71}=\sum\limits_{r=2}^{5}(\overline{\phi _{r,4}}+
\overline{\phi _{r,8}})(\phi _{r,5}+\phi _{r,11}+\phi _{r,1}+\phi _{r,7})+hc$
\\
$+\sum\limits_{r=1}^{3}(\overline{\phi _{r,4}}+\overline{\phi _{r,8}})(\phi
_{r+2,5}+\phi _{r+2,11}+\phi _{r+2,1}+\phi _{r+2,7})+(r\leftrightarrow
r+2)+hc$ \\
$+(\overline{\phi _{3,4}}+\overline{\phi _{3,8}})(\phi _{4,5}+\phi
_{4,11}+\phi _{4,1}+\phi _{4,7})+(3\leftrightarrow 4)+hc$ \\ \hline
$ Z_{31}=Z_{61}=\sum\limits_{r=4}^{5}(\overline{\phi _{r,4}}+
\overline{\phi _{r,8}})(\phi _{r,5}+\phi _{r,11}+\phi _{r,1}+\phi _{r,7})+hc$
\\
$+\sum\limits_{r=2,4}\sum\limits_{s=1,2}(\overline{\phi _{r,4}}+\overline{
\phi _{r,8}})(\phi _{2s+1,5}+\phi _{2s+1,11}+\phi _{2s+1,1}+\phi
_{2s+1,7})+(r\leftrightarrow 2s+1)+hc$ \\
$+(\overline{\phi _{3,4}}+\overline{\phi _{3,8}})(\phi _{5,5}+\phi
_{5,11}+\phi _{5,1}+\phi _{5,7})+(3\leftrightarrow 5)+hc$ \\
$+(\overline{\phi _{1,4}}+\overline{\phi _{1,8}})(\phi _{4,5}+\phi
_{4,11}+\phi _{4,1}+\phi _{4,7})+(1\leftrightarrow 4)+hc$ \\ \hline
$ Z_{41}=Z_{51}=\sum\limits_{r=3}^{5}(\overline{\phi _{r,4}}+
\overline{\phi _{r,8}})(\phi _{r,5}+\phi _{r,11}+\phi _{r,1}+\phi _{r,7})+hc$
\\
$+\sum\limits_{r=1}^{4}(\overline{\phi _{5,4}}+\overline{\phi _{5,8}})(\phi
_{r,5}+\phi _{r,11}+\phi _{r,1}+\phi _{r,7})+(5\leftrightarrow r)+hc$ \\
$+\sum\limits_{r=2}^{3}(\overline{\phi _{4,4}}+\overline{\phi _{4,8}})(\phi
_{r,5}+\phi _{r,11}+\phi _{r,1}+\phi _{r,7})+(4\leftrightarrow r)+hc$ \\
\hline
\end{tabular}
\end{center}

\subsection{Examples from higher Coxeter - Dynkin system}

In general a pair of generalized Dynkin diagrams (Di Francesco-Zuber
diagrams in the case of the $SU(3)$ system) and of levels $k_1$ and
$k_2$ can be associated with a conformal theory whose central charge
was recalled in a previous section.  For the $SU(3)$ case, the
corresponding generalized dual Coxeter numbers or altitudes are
obtained from the relation $ k=\kappa -3$.  Taking $k_1 +1=k_2$, like
in the Virasoro minimal models, leads to a series of unitary ${\cal
W}_{3}$ minimal models with central charges ${\frac{4}{5},
\frac{6}{5},\frac{10}{7},\frac{ 11}{7},\frac{5}{3},\frac{26}{15},
\frac{98}{55},\frac{20}{11},\frac{24}{13} ...}$.  In what follows we
discuss two of these unitary theories corresponding to the $\left(
{\cal A}_{1},{\cal A}_{2}\right)$ and the $ \left( {\cal A}_{4},{\cal
E}_{5}\right)$ pairs.

\subsubsection{The $\left( {\cal A}_{1},{\cal A}_{2}\right) $ model}
\begin{figure}[hhh]
\begin{center}
\unitlength 0.30mm
\par
\begin{picture}(300,160)(0,-15)

\put(0,0){\begin{picture}(60,60)
\put(0,0){\color{green} \circle*{5}}
\put(40,0){\color{blue} \circle*{5}}
\put(20,30){\color{red} \circle*{5}}
\put(0,0){\vector(1,0){21}}
\put(20,0){\line(1,0){20}}
\put(40,0){\vector(-2,3){11.5}}
\put(30,15){\line(-2,3){10}}
\put(20,30){\vector(-2,-3){11.5}}
\put(10,15){\line(-2,-3){10}}\end{picture}}

\put(-5,-10){\makebox(0,0){(0,0)}}
\put(40,-10){\makebox(0,0){(1,0)}}
\put(20,40){\makebox(0,0){(0,1)}}

\put(160,0){\begin{picture}(180,100)

\put(0,0){\begin{picture}(40,40)
\put(0,0){\color{green} \circle*{5}}
\put(40,0){\color{blue} \circle*{5}}
\put(20,30){\color{red} \circle*{5}}
\put(0,0){\vector(1,0){21}}
\put(20,0){\line(1,0){20}}
\put(40,0){\vector(-2,3){11.5}}
\put(30,15){\line(-2,3){10}}
\put(20,30){\vector(-2,-3){11.5}}
\put(10,15){\line(-2,-3){10}}\end{picture}}

\put(40,0){\begin{picture}(40,40)
\put(40,0){\color{red} \circle*{5}}
\put(20,30){\color{green} \circle*{5}}
\put(0,0){\vector(1,0){21}}
\put(20,0){\line(1,0){20}}
\put(40,0){\vector(-2,3){11.5}}
\put(30,15){\line(-2,3){10}}
\put(20,30){\vector(-2,-3){11.5}}
\put(10,15){\line(-2,-3){10}}\end{picture}}

\put(20,30){\begin{picture}(40,40)
\put(20,30){\color{blue} \circle*{5}}
\put(0,0){\vector(1,0){21}}
\put(20,0){\line(1,0){20}}
\put(40,0){\vector(-2,3){11.5}}
\put(30,15){\line(-2,3){10}}
\put(20,30){\vector(-2,-3){11.5}}
\put(10,15){\line(-2,-3){10}}\end{picture}}

\put(-5,-10){\makebox(0,0){(0,0)}}
\put(40,-10){\makebox(0,0){(1,0)}}
\put(3,30){\makebox(0,0){(0,1)}}
\put(80,30){\makebox(0,0){$(1,1)$}}
\put(90,-10){\makebox(0,0){$(2,0)$}}
\put(40,70){\makebox(0,0){$(0,2)$}}

\end{picture}}

\end{picture}
\par
\end{center}
\caption{The $\mathcal{A}_1$ and $\mathcal{A}_2$ generalized Dynkin diagrams}
\end{figure}
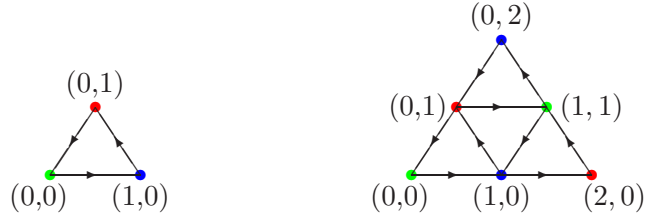

\paragraph{Toric matrices and partition functions.}
The first member of this series, $c=4/5$, corresponds to the $\left(
{\cal A}_{1},{\cal A} _{2}\right) $ pair of diagrams with generalized
Coxeter numbers $(\kappa _{ {\cal A}_{1}}=4,\kappa _{{\cal
A}_{2}=5})$.  We know that there are already two minimal (\ie ${\cal
W}_{2}$ - minimal) models associated with this value of central
charge, the diagonal $(A_{4},A_{5})$ theory and the three states Potts
model $(A_{4},D_{4})$ already discussed in the previous section.  We
start considering the toric matrices of type $W_{{\cal A}
_{k}}(\lambda ,00)=W_{{\cal A}_{k}}(\lambda )$ with $k=1,2$ and
$\lambda $ the weight of the representation (also index of the
vertices of the diagram), $\lambda _{{\cal A}_{1}}=\{(00),(10),(01)\}$
and $\lambda _{ {\cal A}_{2}}=\{(00),(11),(02),(10),(01),(20)\}.$
Triality $\theta$ is obviously defined on these two diagrams in the
following way: ${\cal A}_{1}: \theta(00)=0, \theta(10)=1,\theta(01)=2$
and ${\cal A}_{2}: \theta(00)=\theta(10)=0, \theta(01)=\theta(20)=2,
\theta(10)=\theta(02)=1$.  We have three toric matrices (also graph
matrices in this case) for ${\cal A}_{1}$:
$$
\begin{array}{cc}
W_{{\cal A}_{1}}(00)\equiv W_{{\cal A}_{1}}[1]=\left(
\begin{array}{lll}
1 & 0 & 0 \\
0 & 1 & 0 \\
0 & 0 & 1
\end{array}
\right),
&
W_{{\cal A}_{1}}(10)\equiv W_{{\cal A}_{1}}[2]=\left(
\begin{array}{lll}
0 & 0 & 1 \\
1 & 0 & 0 \\
0 & 1 & 0
\end{array}
\right),
\\
{} & {} \\
W_{{\cal A}_{1}}(01)\equiv W_{{\cal A}_{1}}[3]=\left(
\begin{array}{lll}
0 & 1 & 0 \\
0 & 0 & 1 \\
1 & 0 & 0
\end{array}
\right),
& {}
\end{array}
$$
\noindent
and six toric matrices (also graph matrices in this case) for ${\cal
A}_{2}$:
$$
\begin{array}{cc}
W_{{\cal A}_{2}}(00)\equiv W_{{\cal A}_{2}}[1]=\left(
\begin{array}{llllll}
1 & 0 & 0 & 0 & 0 & 0 \\
0 & 1 & 0 & 0 & 0 & 0 \\
0 & 0 & 1 & 0 & 0 & 0 \\
0 & 0 & 0 & 1 & 0 & 0 \\
0 & 0 & 0 & 0 & 1 & 0 \\
0 & 0 & 0 & 0 & 0 & 1
\end{array}
\right) , &
W_{{\cal A}_{2}}(10)\equiv W_{{\cal A}_{2}}[2]=\left(
\begin{array}{llllll}
0 & 0 & 0 & 0 & 1 & 0 \\
0 & 0 & 0 & 0 & 1 & 1 \\
0 & 1 & 0 & 0 & 0 & 0 \\
1 & 1 & 0 & 0 & 0 & 0 \\
0 & 0 & 1 & 1 & 0 & 0 \\
0 & 0 & 0 & 1 & 0 & 0
\end{array}
\right),
\end{array}
$$

$$
\begin{array}{cc}
W_{{\cal A}_{2}}(01)\equiv W_{{\cal A}_{2}}[3]=\left( W_{
{\cal A}_{2}}(10)\right) ^{T}, &
W_{{\cal A}_{2}}(20)\equiv W_{{\cal A}
_{2}}[4]=W_{{\cal A}_{2}}(10).W_{{\cal A}_{2}}(10)-W_{{\cal A}
_{2}}(01),\\
W_{{\cal A}_{2}}(02)\equiv W_{{\cal A}_{2}}[5]=\left( W_{
{\cal A}_{2}}(20)\right) ^{T}, &
W_{{\cal A}_{2}}(11)\equiv W_{
{\cal A}_{2}}[6]=W_{{\cal A}_{2}}(10).W_{{\cal A}_{2}}(01)-W_{{\cal A
}_{2}}(00)
\end{array}
$$

The fundamental twisted partition functions are
$
Z(\lambda ,\nu )=\frac{1}{3}\overline{\chi }\, W^{({\cal 
A}_{1}{\cal,A}_{2})} \, (\lambda
,\nu )\chi,$
with
$W^{({\cal A}_{1}{\cal A}
_{2})}[\lambda ,\nu ]=W^{({\cal A}_{1})}[\lambda ]\otimes W^{(
{\cal A}_{2})}[\nu ]$, a matrix of of dimension $(3\times 6)^{2}$, and
where
$\chi =\chi [\lambda ^{{\cal A}_{1}},\mu ^{{\cal A}_{2}}]$ denotes the
basis
$$
\begin{array}{l}
     \{\chi[00,00],\chi [10,00],\chi [01,00],\chi [00,11],\chi [10,11],\chi
[01,11],
\chi [00,02],\chi [10,02],\chi [01,02],
\\
\chi [00,10], \chi [10,10],\chi [01,10],\chi
[00,01],\chi [10,01],
\chi [01,01],\chi [00,20],\chi [10,20],\chi [01,20]\}
\end{array}
$$
The matrix $W^{({\cal A}_{1}{\cal A}_{2})}(00,00) \equiv W^{({\cal
A}_{1}{\cal A}_{2})}[1,1]$ $W^{({\cal A}_{1})}[1]\otimes W^{( {\cal
A}_{2})}[1]$ is the modular invariant; in the base $\{\chi\}$, it is
just the identity matrix of size $(18,18)$.  We give another example, 
$W^{({\cal
A}_{1},{\cal A} _{2})}(00,10) \equiv W^{({\cal A}_{1}{\cal
A}_{2})}[1,2] \doteq W^{({\cal A}_{1})}[1]\otimes W^{({\cal
A}_{2})}[2]$ which is one of the twisted mass matrices:

\[
W^{({\cal A}_{1}{\cal,A}_{2})}[1,2]=\left(
\begin{tabular}{lll|lll|lll|lll|lll|lll}
. & . & . & . & . & . & . & . & . & . & . & . & 1 & . & . & . & . & . \\
. & . & . & . & . & . & . & . & . & . & . & . & . & 1 & . & . & . & . \\
. & . & . & . & . & . & . & . & . & . & . & . & . & . & 1 & . & . & . \\
\hline
. & . & . & . & . & . & . & . & . & . & . & . & 1 & . & . & 1 & . & . \\
. & . & . & . & . & . & . & . & . & . & . & . & . & 1 & . & . & 1 & . \\
. & . & . & . & . & . & . & . & . & . & . & . & . & . & 1 & . & . & 1 \\
\hline
. & . & . & 1 & . & . & . & . & . & . & . & . & . & . & . & . & . & . \\
. & . & . & . & 1 & . & . & . & . & . & . & . & . & . & . & . & . & . \\
. & . & . & . & . & 1 & . & . & . & . & . & . & . & . & . & . & . & . \\
\hline
1 & . & . & 1 & . & . & . & . & . & . & . & . & . & . & . & . & . & . \\
. & 1 & . & . & 1 & . & . & . & . & . & . & . & . & . & . & . & . & . \\
. & . & 1 & . & . & 1 & . & . & . & . & . & . & . & . & . & . & . & . \\
\hline
. & . & . & . & . & . & 1 & . & . & 1 & . & . & . & . & . & . & . & . \\
. & . & . & . & . & . & . & 1 & . & . & 1 & . & . & . & . & . & . & . \\
. & . & . & . & . & . & . & . & 1 & . & . & 1 & . & . & . & . & . & . \\
\hline
. & . & . & . & . & . & . & . & . & 1 & . & . & . & . & . & . & . & . \\
. & . & . & . & . & . & . & . & . & . & 1 & . & . & . & . & . & . & . \\
. & . & . & . & . & . & . & . & . & . & . & 1 & . & . & . & . & . & .
\end{tabular}
\right)
\]

As in the previous section, we are considering only the ``fundamental
toric matrices'' \ie those of type $W^{({\cal A}_{1},{\cal A}
_{2})}[\lambda ,\nu ]=W^{({\cal A}_{1})}[\lambda,00 ]\otimes W^{(
{\cal A}_{2})}[\nu,00 ]$.  The more general ones would be of the type
$W^{({\cal A}_{1},{\cal A} _{2})}[\lambda ,\nu ,\tau,\mu]=W^{({\cal
A}_{1})}[\lambda,\nu ]\otimes W^{( {\cal A}_{2})}[\tau,\mu ]$.  A
priori, we obtain in this way $3 \times 6 = 18$ toric matrices, but
because of the identifications resulting from the $\ZZ_{3}$ symmetries
of the ${\cal W}_{3}$ Kac table, we obtain, at the end, only $18/3 =
6$ distinct\footnote{For a ${\cal W}_{3}$
minimal model of type $({\cal A}_{k},{\cal A}_{k+1})$, we would obtain
$(k+1)(k+2)^{2}(k+3)/12$ distinct functions.} partition functions $Z[i,j]$.
The group $\ZZ_{3}$ acts on the pairs of vertices belonging to these
two diagrams in a geometrically very intuitive way (counter-clockwise on
both ${\cal A}_{1}$ and ${\cal A}_{2}$) so that
characters are identified as follows:
$$
\begin{array}{cc}
  \chi [00,00]=\chi [10,20]=\chi [01,02]=\chi_{1}, \;
  \chi [00,02]=\chi [10,00]=\chi [01,20]=\chi_{2} \\
  \chi [00,20]=\chi [10,02]=\chi [01,00]=\chi_{3}, \;
  \chi [00,11]=\chi [10,01]=\chi [01,10]=\chi_{4} \\
  \chi [00,10]=\chi [10,11]=\chi [01,01]=\chi_{5}, \;
  \chi [00,01]=\chi [10,10]=\chi [01,11]=\chi_{6} \\
\end{array}
$$
Implementation of this $Z_3$ symmetry over
the characters leads to the
following table where we list the six partition functions of
the form $Z[i,j]$:

\begin{center}
\begin{tabular}{|l|}
\hline
$ Z[1,1]=Z[2,4]=3Z[3,5]=\sum\limits_{i=1}^{6}\left|
\chi _{i}\right| ^{2}$ \\ \hline
$ Z[1,2]=Z[2,6]=3Z[3,3]=\chi _{1}\overline{\chi _{5}} +
\chi _{2}\overline{\chi _{6}}+\chi _{3}\overline{\chi _{4}}+
\chi_{4} \left(\overline{\chi _{2}}+\overline{\chi _{5}}\right)+ $ \\
$\chi_{5} \left(\overline{\chi _{3}}+\overline{\chi _{6}}\right)+
\chi_{6} \left(\overline{\chi _{1}}+\overline{\chi _{4}}\right)$ \\ \hline
$ Z[1,3]=Z[2,2]=3Z[3,6]=\chi _{1}\overline{\chi _{6}} +
\chi _{2}\overline{\chi _{4}}+\chi _{3}\overline{\chi _{5}}+
\chi_{4} \left(\overline{\chi _{3}}+\overline{\chi _{6}}\right)+$ \\
$\chi_{5} \left(\overline{\chi _{1}}+ \overline{\chi _{4}}\right)+
\chi_{6} \left(\overline{\chi _{2}}+\overline{\chi _{5}}\right) $ \\ \hline
$ Z[1,4]=Z[2,5]=3Z[3,1]=\chi _{1}\overline{\chi _{3}} +
\chi _{2}\overline{\chi _{1}}+\chi _{3}\overline{\chi _{2}}+
\chi_{4}\overline{\chi _{6}}+\chi_{5}\overline{\chi 
_{4}}+\chi_{6}\overline{\chi _{5}}$ \\ \hline
$ Z[1,5]=Z[2,1]=3Z[3,4]=\overline{Z}[1,4]$ \\ \hline
$ Z[1,6]=Z[2,3]=3Z[3,2]=\left| \chi _{1}+\chi_{4}\right| ^{2}+
\left| \chi _{2}+\chi_{5}\right| ^{2}+ \left| \chi _{3}+\chi_{6}\right| ^{2}+
\sum\limits_{i=4}^{6}\left|\chi _{i}\right| ^{2}$ \\ \hline
\end{tabular}
\end{center}

\paragraph{The Potts model recovered.}
 From the value of the central charge (4/5) it is expected that the
present model is the Potts model in a new guise.  It is indeed so and
this has been known for quite a while.  However, here we want to show
that not only we recover the usual (undeformed) partition function,
but also the whole set of (four) twisted partition functions that were
determined in section \ref{sec:caracw3} and denoted in boldface.  We first
calculate conformal weights for the $SU(3)$ fields $\chi$ from the
generalized ${\cal W}_{3}$ Rocha - Cariddi recalled in section
\ref{RochaGene}.
Remember that $r,s$ labels are shifted by $(1,1)$ compared with
$(\lambda,\mu)$ labels.
One finds:
$h(\chi[00,00]) = 0$ and this is compatible\footnote{Compatibility of
weights of $SU(2)$ versus $SU(3)$ is only meaningful modulo integers.
} with the $SU(2)$ fields for which $h=0$ or $h=3$, \ie
$\phi_{11}$ and $\phi_{41}$;
$h(\chi[00,11]) = 2/5$, compatible with the $SU(2)$ fields for
which $h=2/5$ or $h=7/5 = 2/5 + 1$, \ie $\phi_{21}$ and $\phi_{31}$;
$h(\chi[00,02]) = 2/3$, compatible with $\phi_{13}$,
and  $h(\chi[00,20]) = 2/3$,  also compatible with $\phi_{13}$;
$h(\chi[00,10]) = 1/15$,  compatible with $\phi_{23}$
and $h(\chi[00,01]) = 1/15$, also compatible with $\phi_{23}$.
It is therefore natural to consider the branching rules:
$$
\begin{array}{cc}
\chi[00,00] \rightarrow \phi_{11} + \phi_{41} &
\chi[00,11] \rightarrow \phi_{21} + \phi_{31} \\
\chi[00,02] \rightarrow \phi_{13} &
\chi[00,20] \rightarrow \phi_{13} \\
\chi[00,10] \rightarrow \phi_{23} &
\chi[00,01] \rightarrow \phi_{23} \\
\end{array}
$$
If we now perform these substitutions in the twisted partition
functions of the previous table, we find:
$$
Z[11] \rightarrow Z_{0},
Z[12] \rightarrow 2 Z_{3},
Z[13] \rightarrow 2 Z_{2},
Z[14] \rightarrow Z1,
Z[15] \rightarrow \overline{Z1},
Z[16] \rightarrow Z0 + Z2
$$
The six twisted partitions functions of this $SU(3)$ minimal model can
therefore be reinterpreted in terms of the four twisted partitions 
functions of the
$SU(2)$ Potts model obtained in section \ref{sec:PottsSU2}.


\subsubsection{The $\left( {\mathcal A}_{4},{\mathcal E}_{5}\right) $ model}
\label{sec:w3minimal}

\begin{figure}[tbp]
\begin{center}
\includegraphics*{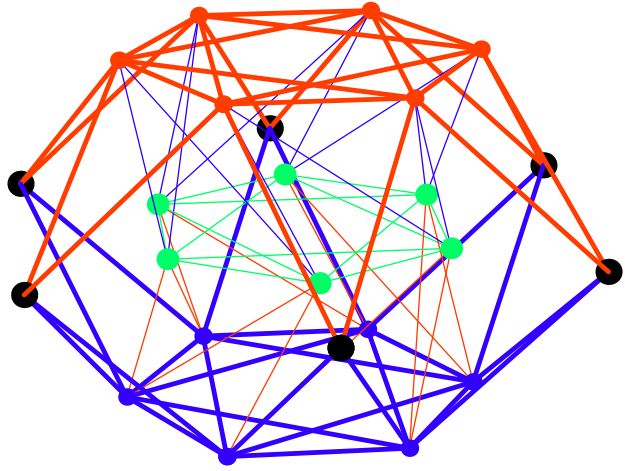}
\end{center}
\caption{Ocneanu graph for $\mathcal{E}_{5}$}
\label{OcGraphE5}
\end{figure}

The $\left( {\mathcal A}_{4},{\mathcal E}_{5}\right) $ pair of diagrams
with generalized Coxeter numbers $(\kappa _{{\mathcal A}_{4}}=7,\kappa
_{{\mathcal E} _{5}}=8)$ corresponds to a ${\cal W}_{3}$ minimal and
unitary conformal model, with central charge $ c=11/7$.  One of these
diagrams (${\mathcal E}_{5}$) was displayed in section
\ref{sec:E5partfun}, and ${\mathcal A}_{4}$ is of course similar to
${\mathcal A}_{5}$, (displayed in the same section) but with only four
levels.  There are several possible theories associated to this value
of central charge, one is the diagonal theory associated to the pair
$\left( {\mathcal A}_{4},{\mathcal A}_{5}\right) $, another is the one
we are considering here.  There are 15 toric matrices (also graph
matrices in this case) of type $W_{{\mathcal A}_{4}}(\lambda ,00)=W_{
{\mathcal A}_{4}}(\lambda )$ with $\lambda $ labelling the vertices of
the generalized Dynkin diagram, $\lambda _{{\mathcal A} _{4}}=
\{(00),(30),(03),(11),(22),(10),(40),(21),(02),(13),$ \\
$(20),(12),(04),(01),(31) \}$ and 24 of type $W_{{\mathcal E}_{5}}(\sigma
,00)=W_{{\mathcal E} _{5}}(\sigma )$ with $\sigma =\rho \times \nu $
labels of the $Oc({\mathcal E} _{5})$ vertices (this graph was obtained
in \cite{CoqueGil:Tmodular} and is recalled below \ref{OcGraphE5}).
Here $\rho ,\nu \in \lambda _{{\mathcal E}_{5}}=
\{1_{0},1_{3},2_{3},2_{0},1_{2},1_{5},2_{2},2_{5},1_{1},1_{4},2_{1},2_{4}\}$
labels the vertices of the generalized $ {\mathcal E}_{5}$ Dynkin
diagram.  The $W_{{\mathcal E}_{5}}(\sigma )$ are matrices of dimension
$21\times 21$ whose entries $(i,j)$ corresponds to the vertices of the
diagram ${\mathcal A}_{5} = {\cal A}({\cal E}_{5})$.

The general twisted partition functions are given by:
\[
Z(\lambda ,\sigma )=\frac{1}{3}\overline{\chi }W^{({\mathcal A}_{4}{\mathcal,E}
_{5})}(\lambda ,\sigma )\chi
\]
where $\chi =\chi [\lambda ^{{\mathcal A}_{4}},\lambda ^{{\mathcal 
A}_{5}}]$ and
$W^{({\mathcal A}_{4}{\mathcal E}_{5})}[\lambda ,\sigma ]=W^{({\mathcal A}
_{4})}[\lambda ]\otimes W^{({\mathcal A}_{5})}[\sigma ]$.

Exponents of ${\cal E}_{5}$ can be read, for example, from the modular
invariant toric matrix given in section \ref{sec:E5partfun}).  These
are particular ${\cal A}_{5}$ vertices $s = (s_{1},s_{2})$  given by the
list
$\{(0,0),(2,2),(0,2),(3,2),(2,0),(2,3),(2,1),(0,5),(3,0),(0,3),(1,2),(5,0)\}$.
The $\ZZ_{3}$ action on ${\cal A}_{4}$ gives, a priori, five
equivalence classes labelled for example by $\{(0,0),(0,1),(0,2),\\
(0,3),(1,1)\}$. All together, the untwisted $({\cal A}_{4}, {\cal E}_{5})$
model will therefore involve in principle $12 \times 5 = 60$ distinct ${\cal W}_{3}$
characters with conformal weights given by the following table where we are 
including the fifteen $r$ vertices of ${\cal A}_{4}$ in a particular order 
$\{(00),(40),(04)\},$ \\
$\{(22),(02),(20)\},\{(03),(10),(31)\},\{(13),(01),(30)\},
\{(11),(12),(21)\}$ to make manifest the ocurrence of the five mentioned 
equivalence classes
(this is only a subset of the Kac table of the pair $({\cal A}_{4}, {\cal A}_{5}))$.

\begin{center}
\begin{tabular}{|l|l|l|l|l|l|l|l|l|l|l|l|l|l|l|l|}
\hline
$s\backslash r$ & $00$ & $40$ & $04$ & $22$ & $02$ & $20$ & $03$ & $10$ & $31
$ & $13$ & $01$ & $30$ & $11$ & $12$ & $21$ \\ \hline
$00$ & $0$ & $\frac{20}{3}$ & $\frac{20}{3}$ & $\frac{36}{7}$ & $\frac{38}{21
}$ & $\frac{38}{21}$ & $\frac{27}{7}$ & $\frac{11}{21}$ & $\frac{116}{21}$ &
$\frac{116}{21}$ & $\frac{11}{21}$ & $\frac{27}{7}$ & $\frac{10}{7}$ & $
\frac{65}{21}$ & $\frac{65}{21}\rule[-0.2cm]{0cm}{0.6cm}$ \\ \hline
$05$ & $\frac{20}{3}$ & $\frac{20}{3}$ & $0$ & $\frac{38}{21}$ & $\frac{38}{
21}$ & $\frac{36}{7}$ & $\frac{11}{21}$ & $\frac{116}{21}$ & $\frac{27}{7}$
& $\frac{11}{21}$ & $\frac{27}{7}$ & $\frac{116}{21}$ & $\frac{65}{21}$ & $
\frac{10}{7}$ & $\frac{65}{21}\rule[-0.2cm]{0cm}{0.6cm}$ \\ \hline
$50$ & $\frac{20}{3}$ & $0$ & $\frac{20}{3}$ & $\frac{38}{21}$ & $\frac{36}{7
}$ & $\frac{38}{21}$ & $\frac{116}{21}$ & $\frac{27}{7}$ & $\frac{11}{21}$ &
$\frac{27}{7}$ & $\frac{116}{21}$ & $\frac{11}{21}$ & $\frac{65}{21}$ & $
\frac{65}{21}$ & $\frac{10}{7}\rule[-0.2cm]{0cm}{0.6cm}$ \\ \hline
$03$ & $\frac{9}{4}$ & $\frac{59}{12}$ & $\frac{11}{12}$ & $\frac{39}{28}$ &
$\frac{5}{84}$ & $\frac{173}{84}$ & $\frac{3}{28}$ & $\frac{149}{84}$ & $
\frac{233}{84}$ & $\frac{65}{84}$ & $\frac{65}{84}$ & $\frac{87}{28}$ & $
\frac{19}{28}$ & $\frac{29}{84}$ & $\frac{113}{84}\rule[-0.2cm]{0cm}{0.6cm}$
\\ \hline
$20$ & $\frac{11}{12}$ & $\frac{9}{4}$ & $\frac{59}{12}$ & $\frac{173}{84}$
& $\frac{39}{28}$ & $\frac{5}{84}$ & $\frac{233}{84}$ & $\frac{3}{28}$ & $
\frac{149}{84}$ & $\frac{87}{28}$ & $\frac{65}{84}$ & $\frac{65}{84}$ & $
\frac{29}{84}$ & $\frac{113}{84}$ & $\frac{19}{28}\rule[-0.2cm]{0cm}{0.6cm}$
\\ \hline
$32$ & $\frac{59}{12}$ & $\frac{11}{12}$ & $\frac{9}{4}$ & $\frac{5}{84}$ & $
\frac{173}{84}$ & $\frac{39}{28}$ & $\frac{149}{84}$ & $\frac{233}{84}$ & $
\frac{3}{28}$ & $\frac{65}{84}$ & $\frac{87}{28}$ & $\frac{65}{84}$ & $\frac{
113}{84}$ & $\frac{19}{28}$ & $\frac{29}{84}\rule[-0.2cm]{0cm}{0.6cm}$ \\
\hline
$30$ & $\frac{9}{4}$ & $\frac{11}{12}$ & $\frac{59}{12}$ & $\frac{39}{28}$ &
$\frac{173}{84}$ & $\frac{5}{84}$ & $\frac{87}{28}$ & $\frac{65}{84}$ & $
\frac{65}{84}$ & $\frac{233}{84}$ & $\frac{149}{84}$ & $\frac{3}{28}$ & $
\frac{19}{28}$ & $\frac{113}{84}$ & $\frac{29}{84}\rule[-0.2cm]{0cm}{0.6cm}$
\\ \hline
$23$ & $\frac{59}{12}$ & $\frac{9}{4}$ & $\frac{11}{12}$ & $\frac{5}{84}$ & $
\frac{39}{28}$ & $\frac{173}{84}$ & $\frac{65}{84}$ & $\frac{87}{28}$ & $
\frac{65}{84}$ & $\frac{3}{28}$ & $\frac{233}{84}$ & $\frac{149}{84}$ & $
\frac{113}{84}$ & $\frac{29}{84}$ & $\frac{19}{28}\rule[-0.2cm]{0cm}{0.6cm}$
\\ \hline
$02$ & $\frac{11}{12}$ & $\frac{59}{12}$ & $\frac{9}{4}$ & $\frac{173}{84}$
& $\frac{5}{84}$ & $\frac{39}{28}$ & $\frac{65}{84}$ & $\frac{65}{84}$ & $
\frac{87}{28}$ & $\frac{149}{84}$ & $\frac{3}{28}$ & $\frac{233}{84}$ & $
\frac{29}{84}$ & $\frac{19}{28}$ & $\frac{113}{84}\rule[-0.2cm]{0cm}{0.6cm}$
\\ \hline
$22$ & $3$ & $\frac{5}{3}$ & $\frac{5}{3}$ & $\frac{1}{7}$ & $\frac{17}{21}$
& $\frac{17}{21}$ & $\frac{6}{7}$ & $\frac{32}{21}$ & $\frac{11}{21}$ & $
\frac{11}{21}$ & $\frac{32}{21}$ & $\frac{6}{7}$ & $\frac{3}{7}$ & $\frac{2}{
21}$ & $\frac{2}{21}\rule[-0.2cm]{0cm}{0.6cm}$ \\ \hline
$12$ & $\frac{5}{3}$ & $3$ & $\frac{5}{3}$ & $\frac{17}{21}$ & $\frac{1}{7}$
& $\frac{17}{21}$ & $\frac{11}{21}$ & $\frac{6}{7}$ & $\frac{32}{21}$ & $
\frac{6}{7}$ & $\frac{11}{21}$ & $\frac{32}{21}$ & $\frac{2}{21}$ & $\frac{2
}{21}$ & $\frac{3}{7}\rule[-0.2cm]{0cm}{0.6cm}$ \\ \hline
$21$ & $\frac{5}{3}$ & $\frac{5}{3}$ & $3$ & $\frac{17}{21}$ & $\frac{17}{21}
$ & $\frac{1}{7}$ & $\frac{32}{21}$ & $\frac{11}{21}$ & $\frac{6}{7}$ & $
\frac{32}{21}$ & $\frac{6}{7}$ & $\frac{11}{21}$ & $\frac{2}{21}$ & $\frac{3
}{7}$ & $\frac{2}{21}\rule[-0.2cm]{0cm}{0.6cm}$ \\ \hline
\end{tabular}
\end{center}

The Ocneanu graph of ${\cal E}_{5}$ has $24$ points and the
intersection of the vector spaces spanned by the $12$ left and the
$12$ right generators -- ambichiral subspace -- is of dimension $6$
(generators are of the type $1_{0} \otimesdot 1_{j} = 1_{j} \otimesdot
1_{0}$). The supplementary subspace has also dimension $6$.
We therefore expect to obtain four sets of $5 \times 6$
twisted fundamental partition functions.

If we further restrict our attention to those fundamental twisted
partition functions which only involve the fields appearing in the
undeformed $({\cal A}_{4}, {\cal E}_{5})$ model (labelled by the above
60 conformal weights), \ie if we only take the ambichiral points
into account, we expect $5 \times 6 = 30$ distinct cases.  Notice that
the description of this ${\cal W}_{3}$ minimal model is very similar
to the one that we made for the Virasoro minimal model of type
$(A_{10},E_{6})$.

We shall not give this full list of  $30$ partition  functions
but only two of them: those associated with
toric matrices
$W_{00,00}({\cal A}_{4}) \otimes W_{1_{0}\otimes 1_{0}}({\cal E}_{5})$
and
$W_{00,00}({\cal A}_{4}) \otimes W_{1_{0}\otimes 1_{3}}({\cal E}_{5})$.

\begin{center}
\begin{tabular}{|l|}
\hline
$ Z[00,1_{0}\times 1_{0}]=\sum\limits_{i}\left| \chi [i,05]+\chi
[i,21]\right| ^{2}+\left| \chi [i,00]+\chi [i,22]\right| ^{2}+\left| \chi
[i,20]+\chi [i,23]\right| ^{2}$ \\
$+\left| \chi [i,03]+\chi [i,30]\right| ^{2}+\left| \chi [i,02]+\chi
[i,32]\right| ^{2}+\left| \chi [i,12]+\chi [i,50]\right| ^{2}$ \\ \hline
$ Z[00,1_{0}\times 1_{3}]=\sum\limits_{i}\left( \chi [i,02]+\chi
[i,32]\right) \left( \overline{\chi }[i,05]+\overline{\chi }[i,21]\right) +$
\\
$\left( \chi [i,03]+\chi [i,30]\right) \left( \overline{\chi }[i,00]+
\overline{\chi }[i,22]\right) +\left( \chi [i,12]+\chi [i,50]\right) \left(
\overline{\chi }[i,20]+\overline{\chi }[i,23]\right) +hc$ \\ \hline
\end{tabular}
\end{center}

where the sums are over $i=(0,0),(0,1),(0,2),(0,3),(1,1)$ vertices of
${\mathcal A_{4}}$.

\label{fig:OcE5}

\section*{Acknowledgments}

Part of this work was performed at
the Centre de Physique Theorique, in Marseille. One of us (M.H.)
wants to acknowledge CNRS (Poste Rose), Fundaci\'on Antorchas, Argentina
and The Abdus Salam ICTP, for financial support.


\end{document}